\documentclass{article}

\usepackage[final]{neurips_2024}

\usepackage[utf8]{inputenc} %
\usepackage[T1]{fontenc}    %
\usepackage{hyperref}       %
\usepackage{url}            %
\usepackage{booktabs}       %
\usepackage{amsfonts}       %
\usepackage{nicefrac}       %
\usepackage{microtype}      %
\usepackage{xcolor}         %

\usepackage[percent]{overpic}
\usepackage{fontawesome}

\usepackage{amsmath}
\usepackage{amssymb}
\usepackage{bm}
\usepackage{bbm}
\usepackage{xspace}
\usepackage{cleveref}
\usepackage{subcaption} %
\usepackage{multirow} 
\usepackage{enumitem} 

\usepackage{natbib}
\setcitestyle{numbers,square,sort}

\newcommand{\search}{\faSearch\@\xspace}

\newcommand*{\ie}{i.e.,\@\xspace}
\newcommand*{\eg}{e.g.,\@\xspace}
\newcommand*{\etc}{etc.\@\xspace}

\usepackage{graphicx}

\title{Combining Observational Data and Language\\ for Species Range Estimation}

\author{%
  \textbf{Max Hamilton}\textsuperscript{1} \quad 
  \textbf{Christian Lange}\textsuperscript{2} \quad 
  \textbf{Elijah Cole}\textsuperscript{3} \quad
  \textbf{Alexander Shepard}\textsuperscript{4} \quad\\
  \\
  \textbf{Samuel Heinrich}\textsuperscript{5}\quad
  \textbf{Oisin Mac Aodha}\textsuperscript{2} \quad
  \textbf{Grant Van Horn}\textsuperscript{1} \quad
  \textbf{Subhransu Maji}\textsuperscript{1}\\
  \\
  \textsuperscript{1}UMass Amherst \quad 
  \textsuperscript{2}University of Edinburgh \quad  
  \textsuperscript{3}Altos Labs \quad\\
  \textsuperscript{4}iNaturalist  \quad
  \textsuperscript{5}Cornell University
  }

\begin{document}

\maketitle

\begin{abstract}
Species range maps (SRMs) are essential tools for research and policy-making in ecology, conservation, and environmental management. However, traditional SRMs rely on the availability of environmental covariates and high-quality species location observation data, both of which can be challenging to obtain due to geographic inaccessibility and resource constraints. We propose a novel approach combining millions of citizen science species observations with textual descriptions from Wikipedia, covering habitat preferences and range descriptions for tens of thousands of species. Our framework maps locations, species, and text descriptions into a common space, facilitating the learning of rich spatial covariates at a global scale and enabling zero-shot range estimation from textual descriptions. Evaluated on held-out species, our zero-shot SRMs significantly outperform baselines and match the performance of SRMs obtained using tens of observations. Our approach also acts as a strong prior when combined with observational data, resulting in more accurate range estimation with less data. We present extensive quantitative and qualitative analyses of the learned representations in the context of range estimation and other spatial tasks, demonstrating the effectiveness of our approach.
\end{abstract}

\section{Introduction}
Collecting sufficient point-based observations of species in the wild allows us to infer species range maps (SRMs), which describe the spatial extent of where a species is likely to occur. These maps are invaluable, enhancing our understanding of natural history and informing land use and conservation decisions. Large-scale citizen science projects like iNaturalist~\cite{iNaturalist} and eBird~\cite{sullivan2009ebird} have recently accelerated SRM generation by systematically consolidating millions of observations across tens of thousands of species. By combining these extensive databases with environmental covariates, we can produce accurate SRMs. However, there remains a 'long tail' of species with few observations, and current methods fall short of producing reliable range maps in such low-data settings.

We propose learning SRMs by combining citizen science observations with text descriptions of species from Wikipedia (see Figure~\ref{fig:motivation_fig}). 
These text descriptions describe a wide array of properties including habitat preferences, range estimates, and visual attributes. Our framework learns to map location embeddings over the Earth's surface and text embeddings from Wikipedia~\cite{wikipedia} articles into a common space based on the species observations (see Figure~\ref{fig:pipeline_fig}). 
This provides a way to ground the information in text describing tens of thousands of species to spatial locations based on millions of observations. 
As shown in Figure~\ref{fig:motivation_fig}, our resulting model allows us to estimate SRMs based on text descriptions that might be known to an ecologist, such as habitat preferences, even when no location observations are available. 

We evaluate our text-driven approach for zero-shot estimation of SRMs for species from the IUCN~\cite{IUCN} and eBird Status and Trends (S\&T)~\cite{fink2020ebird} benchmarks from~\cite{cole2023spatial}, which were excluded from the training data. Our model easily outperforms baselines (Table~\ref{tab:main-results}) and is competitive with SRMs estimated with ten observations for S\&T species (Figure~\ref{fig:few_shot_results}). Additionally, our model can be combined with observational data to achieve strong few-shot performance. Logistic regression models are regularized to be close to the species vector obtained from text descriptions, allowing us to match the performance of SRMs estimated using an order of magnitude more observations (Figure~\ref{fig:few_shot_results}).

While training is based on Wikipedia articles, it is unrealistic to expect a biologist to provide such extensive information for a novel species. Therefore, we evaluate our model based on short summaries, often just a few sentences long, that describe various aspects of the species. We further organize the summaries based on descriptions of the range (\eg where they appear) and habitat preferences. While a biologist might not know the full range of an unknown species, they may be familiar with some of the latter (\eg a tree might be in a tropical forest, or a bird might prefer wetlands). 
However, training on both forms of data allows the location embeddings to capture a wide range of geographic concepts embedded in language, such as countries, continents, climate regions, topology, and biomes, as revealed in our visualizations (Figures~\ref{fig:zero_shot_w_different_text} and ~\ref{fig:text_model_vis}). Even short text summaries provide a significant boost in estimation accuracy over baselines when there are zero or few observations available. 
Our code and data is publicly available at: \url{https://github.com/cvl-umass/le-sinr}
\vspace{-4pt}

\begin{figure}[t]
\centering
\includegraphics[width=0.95\textwidth]{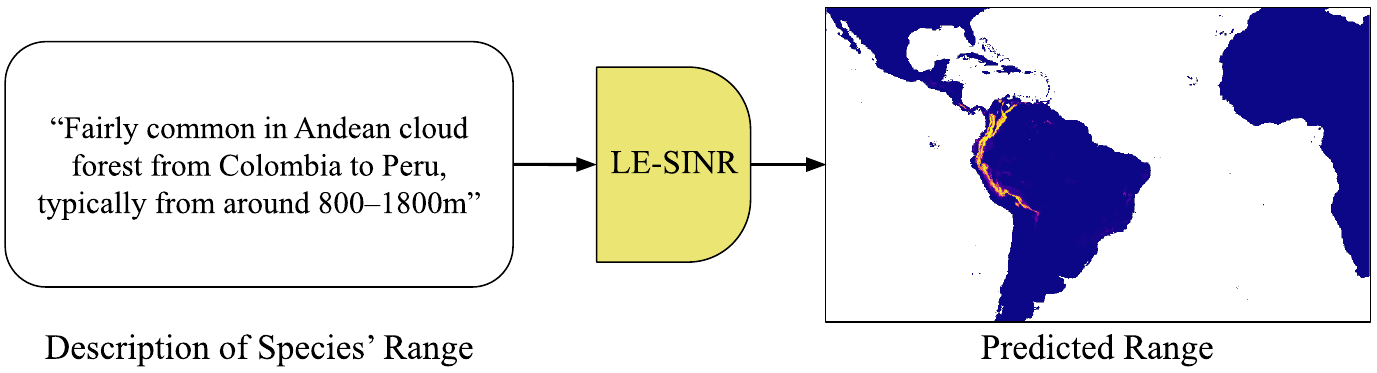} 
\caption{Our LE-SINR model takes as input free-form text describing aspects of a species' preferred habitat or range and geospatially grounds it to generate a plausible range map for that species.}
\label{fig:motivation_fig}
\vspace{-3pt}
\end{figure}

\section{Related work}

\vspace{-3pt}
\paragraph{Species distribution models (SDMs).}
SDMs are a broad family of models, primarily from the ecological statistics literature, that are concerned with modeling and predicting different geospatial properties of a species of interest~\citep{elith2006novel,elith2009species}. 
These spatially varying properties encompass quantities such as occurrence (\ie the presence or absence of a species) through to abundance (\ie a count of the number of individuals from a given species present).  
By integrating occurrence predictions over the earth, we can generate the \emph{range} of a species, which is defined as the geographic area in which a species can be found during its lifetime.   
Existing works can be broadly categorized based on the completeness of the data they are trained on (\eg presence-only vs.~presence-absence data), the number of species they simultaneously model (\eg single vs.~joint methods), or how interpretable they are (\eg machine learning vs.~mechanistic models).    
For an introduction to SDMs, we point interested readers to the following survey~\cite{beery2021species}. 

Most relevant to our work is the growing number of deep SDM approaches. These methods explore the core task from a representation learning perspective by training on raw observation data, typically from multiple species simultaneously, to learn an encoding of geographic space that is more predictive of species presence. \cite{cole2023spatial} demonstrated the advantages of this approach by showing that model performance improves when trained on larger amounts of data, even if that data comes from disjoint species that do not appear in the evaluation set. Recent work has explored various challenges and design decisions related to training these models, addressing topics such as data imbalance~\cite{zbinden2024imbalance}, spatial biases~\cite{chen2019bias}, location encodings~\cite{russwurm2023geographic}, the use of remote sensing data~\cite{deneu2021convolutional,teng2024satbird}, active learning~\cite{lange2024active}, binarizing range maps~\cite{binary_maps_cv4e_2024}, and modeling species co-occurrence~\cite{chen2017deep}. However, the current literature has not extensively investigated the few-shot setting, where very limited or potentially no observation data is available. There are an estimated nine million species on Earth~\cite{mora2011many}, and given that only a limited proportion of these have reliable range estimates, there is a need for methods that can reliably estimate geospatial properties of interest from few observations.

\paragraph{Geospatial data and large language models (LLMs).}
High-capacity transformer-based architectures~\cite{vaswani2017attention}, coupled with large web-sourced text training data, have largely contributed to recent advances in LLMs.   
LLMs have been demonstrated to be effective across a range of language-based reasoning tasks~\cite{zheng2024judging,liang2022holistic,chang2024survey}.    
Inspired by this, recently there have  been multiple attempts to explore what, if any, geospatial information is encoded inside of these models. 
For example, \cite{roberts2023gpt4geo} evaluated a pre-trained closed-source LLM (\ie GPT-4~\cite{achiam2023gpt}) on a range of geospatial tasks such as point based ones (\eg location, distance, and elevation estimation) in addition to more complex path-based ones (\eg geographic and outlines route planning) via carefully designed text prompts. 
They later extended this work to multimodal LLMs that can also take images as input~\cite{roberts2023charting}. 
In both cases, they observed impressive capabilities, but also some notable limitations. 

In \cite{manvi2023geollm} the authors used pre-trained LLMs to map geographic coordinates to continuous geospatial properties (\eg population, house value, \etc). 
They improved upon simple text prompts by engineering a prompt which provides spatial context in terms of relative distance to nearby named locations to the query coordinate of interest. 
By fine-tuning the LLM they demonstrated that their approach works better than the more naive encoding. 
However, their approach is expensive to evaluate at inference time as it requires a full forward pass of the LLM for every geographic coordinate of interest.

Most related to our work is LD-SDM~\cite{sastry2023ld} which also uses an LLM in conjunction with an SDM. 
Their goal is to predict the spatial range of a set of species of interest using location observation data at training time. 
However, instead of simply learning a per-species latent embedding vector as in~\cite{cole2023spatial}, they employ a frozen LLM to map a text string that describes the explicit taxonomic hierarchy (\ie species, genus, family, \etc) of a species of interest to a latent embedding.      
Unlike us, their species text description is not very expressive (\ie it has a very specific hierarchical structure) and thus cannot generalize as effectively to distinct held-out species at evaluation time. 
Instead, in this work, we show that it is possible to predict the range of a previously unseen species from free-form, highly unstructured, internet sourced text that describes its habitat and/or range preferences.    
Furthermore, once trained, we show that our approach is also able to efficiently and densely geospatially ground non-species related text. 
Unlike \cite{manvi2023geollm}, our approach is computationally efficient at inference time, requiring only one LLM forward pass for the text query of a species of interest, as opposed to one forward pass for each location of interest, which can number in the millions depending on the spatial resolution at which the evaluation is performed.

\section{Methods}

\begin{figure}[t]
\centering
\includegraphics[width=1.0\textwidth]{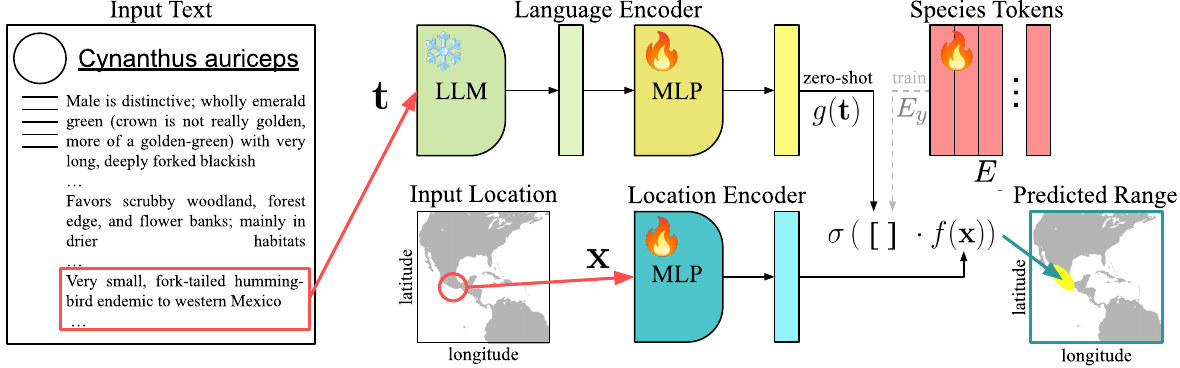} 
\caption{LE-SINR learns to align location and text representations at training time using presence-only observation data and habitat or range descriptions for a set of  species. 
Optionally, we can also include a learnable species token $E_y$ allowing range estimation for seen species $y$ from the training set. 
The model is trained on millions of observations from iNaturalist and language data from Wikipedia articles across thousands of species. 
LE-SINR supports zero-shot range estimation based on text descriptions for novel species and can also be used as a prior for few-shot range estimation.}
\label{fig:pipeline_fig}
\end{figure}

\subsection{Problem Setup}
We focus on the problem of estimating SRMs across multiple species, indicating their presence or absence at each location on Earth. For a given location $\mathbf{x}$, our goal is to predict the probability of each species being present there. Our observation dataset is composed of pairs $\{(\mathbf{x}_i,y_i)\}_{i=1}^N$, where $\mathbf{x}_i = (lat, lon)$ is a geographic location and $y_i \in \{1,\dots,S\}$ is the observed species label. We use the dataset proposed in SINR~\cite{cole2023spatial}, consisting of 35.5 million observations covering 47,375 species observed prior to 2022 on the iNaturalist platform. Included species have at least 50 observations. 
Removing those that are included in the S\&T and IUCN evaluation benchmarks results in 44,181 species, which forms our training set.

We also consider estimating SRMs from text-based descriptions in addition to observational data. To this end we curate a text dataset $\{(\mathbf{t}_i, y_i)\}$, where $\mathbf{t}_i$ is a text description sourced from Wikipedia pertaining to species $y_i$. For a particular species $y_i$ and corresponding text description $\mathbf{t}_i$, we aim to predict whether $y_i$ is present at some location $\mathbf{x}_i$ given $\mathbf{t}_i$. It is important to emphasize that this prediction only depends on the text description $\mathbf{t}_i$ and query location $\mathbf{x}_i$. Thus, we are not constrained to only making range predictions for species that have been observed during training. This allows for zero-shot species range estimation, where we can generate a range map of a species not in our observation dataset by utilizing a text description for it.

\subsection{Text Data}
We source our text descriptions from Wikipedia~\cite{wikipedia}. For a particular species, we search for its Wikipedia article using its scientific name. We then extract the text and divide it into chunks by section. Every article starts with a lead section, which typically provides an overview, followed by a varying number of body sections. These body sections can describe a diverse range of attributes such as taxonomy, description, habitat, behavior, diet, \etc To improve data quality, we remove sections with ``References,'' ``Links,'' and ``Bibliography'' in their names. Since each species has multiple sections, at training time we randomly sample one section per iteration. Overall, our text dataset contains 127,484 sections from 37,889 species' articles. Note, not all species in our observation dataset have an associated text description. 

\subsection{Language Enhanced SINR (LE-SINR) Architecture}

We utilize the Spatial Implicit Neural Representation (SINR)~\cite{cole2023spatial} framework for our approach, which models the probability of presence for a species $y$ at given location $\mathbf{x}$ as $\sigma (f_\theta(\mathbf{x}) \cdot E_y)$, where $f_\theta(\mathbf{x})$ is a location encoding, $E_y$ is an embedding of species $y$ and $\sigma$ is the sigmoid function. Our architecture, shown in Figure~\ref{fig:pipeline_fig}, is similarly composed of two branches: one for representing locations and one for representing species. Each branch outputs a 256-dimensional feature vector. The probability of occurrence is then estimated by the sigmoid of their dot product.

The location branch is a location encoder model, $f_{\theta}(\mathbf{x})$, with parameters $\theta$, which takes a position embedding $\mathbf{x}$ (\eg a location denoted by latitude and longitude) as input. The species branch is composed of two models, allowing us to generate species embeddings in two different ways. The first is a text-based species encoder, $g_{\phi}(\mathbf{t})$, with parameters $\phi$, which takes text $\mathbf{t}$  from our Wikipedia text data as input. The second species representation is a batch of species tokens  optimized directly, $E \in \mathbb{R}^{S \times 256}$.  Given a known species $y \in \{1,\dots,S\}$ in the observation dataset, we can generate its representation with a simple lookup, $E_y$. 
Since we learn a unique species token for each species in the training set, these species tokens cannot be used in the zero-shot setting. 
However, we are able to maintain the ability to have true supervised evaluation on species seen at training time. Additionally, the species tokens are used when a species has no text description.

Our text-based species encoder has two parts: a frozen LLM used to extract text embeddings and a learned fully connected network. As mentioned previously, to perform zero-shot SDM, we cannot directly utilize the species tokens. Instead, we input segments of text sourced from Wikipedia articles. Since these text segments can be as long as multiple paragraphs, we utilize a pretrained Large Language Model, GritLM~\cite{muennighoff2024generative}, to create a fixed-length text embedding. GritLM is a recent language model with strong performance on text embedding tasks like document classification and retrieval. Due to the length of text in our data, it is much easier to work with a fixed-length embedding than with per-token embeddings from generative models like Llama3~\cite{llama3}. Given this text embedding, we then learn a three-layer fully connected language encoder network that outputs the final species embedding. 
As our approach incorporates language within the SINR framework for range estimation, we call it: \textbf{Language Enhanced SINR (LE-SINR)}.

\subsection{Training LE-SINR}
During training we only have access to presence observations, \ie locations where species have been observed, and do not have any absence observations, \ie locations where species have been confirmed to be absent. 
As a result, we train our model with a modified version of $\mathcal{L}_{\textsc{AN-full}}$ from SINR~\cite{cole2023spatial}, which was one of the best-performing losses in their experiments. It minimizes binary cross-entropy with positives sampled from the observation dataset and negatives (\ie `pseudo absences') selected to be all other species at the same observation location, as well as all species at a uniformly sampled random location.

This loss is computationally expensive with our model, as it requires computing a species embedding for every species and sample in the batch. To reduce this computation, we perform an approximation by selecting $M-1$ random negative species at the observation location and $M$ random negatives from the random location, where $M \ll S$. We also weight the negatives by the inverse of the proportion selected. This ensures that the expectation over the randomly selected negative species equals the original loss. Our modified loss is given by: 
\begin{equation}
\mathcal{L}_{\text{AN-full}}' = -\frac{1}{S}\sum_{j=1}^M[\mathbbm{1}_{[z_j=1]}\lambda \log(\hat{y}_j) + \mathbbm{1}_{[z_j\neq 1]}\frac{S-1}{M-1}\log(1-\hat{y}_j) + \frac{S}{M}\log(1-\hat{y}'_j)].
\end{equation}

Here,  $\hat{\mathbf{y}} \in \mathbb{R}^M$ are predictions at one location for the ground truth species and $M-1$ random other species, $\mathbf{z} \in \mathbb{R}^M$ is the corresponding one-hot label, and $\hat{\mathbf{y}}' \in \mathbb{R}^M$ are predictions for a random species at a random location uniformly sampled over earth. For all models, we set $M=192$ based on memory and compute considerations. We also tried values as large as 2,048 but saw significantly slower training times with no effect on zero-shot performance.

During training, we first use the location encoder to generate location features. We then use these to make two predictions: one using the species tokens and the other from the text-based species embeddings. Finally, we apply $\mathcal{L}'_{\textsc{AN-full}}$ to both of these predictions independently. We do not explicitly encourage the species tokens and text-based species embeddings to be close. In our preliminary experiments, this seemed to be too restrictive and hurt performance.

\subsection{Evaluation} \label{sec:eval}
Similar to SINR, we evaluate our model using expert-derived range maps from the eBird Status \& Trends (S\&T) dataset~\cite{fink2020ebird}, which covers 535 bird species with a focus on North America, as well as range maps for 2,418 species from the International Union for Conservation of Nature (IUCN) Red List~\cite{IUCN}. During training, we exclude observations for these species to assess zero-shot and few-shot performance, measured using mean average precision (MAP), \ie average precision (AP) averaged across all species in the set. SINR models trained with target species' observations provide an ``upper bound" on performance.

\paragraph{Zero-shot evaluation.}\label{sec:zero-shot}
Our model naturally supports zero-shot evaluation by providing a text prompt from species not in the training data to the species model. The output species embedding can then be multiplied with position features to generate the probability of occurrence. We can then compute a precision-recall curve by varying the threshold over the score to generate the range map given an expert derived range map. We again report the mean average precision (MAP) across species in the S\&T and IUCN datasets.

An important choice here is which text we provide for the evaluation. The Wikipedia section names are not consistent across articles and often the same information can appear under different headings. At the same time, it is unlikely that one can provide such detailed text for novel species. To standardize the information for a realistic evaluation, we use the open-source Llama-3 model \cite{llama3} to generate two short summaries: a habitat description and a range description from the article text. The range description is most informative as it typically lists specific countries or regions where the species can be found. In practice, such a rich description might not be available, so we also generate a habitat description. During evaluation, we use these summaries instead of the Wikipedia text. 
Figure~\ref{fig:zero_shot_w_different_text} shows some example summary texts along with zero-shot range predictions from LE-SINR. Further examples can be found in the Appendix.

\paragraph{Few-shot evaluation.}\label{sec:few-shot}
While the original SINR model was trained with observations from the target species, we also consider a setting where position features are used to derive SRMs from sparse observational data. We achieve this by performing logistic regression with $L_2$ normalization on the position features to predict presence or absence. For each species, we sample $n_p$ positives from the observation dataset and $n_n=20,000$ negatives. To mimic the $\mathcal{L}_{\textsc{AN-full}}$  loss, we sample 10,000 negatives uniformly across the Earth and 10,000 negatives randomly from the training dataset species locations. 
Our logistic regression loss is given by, 
\begin{equation}
L_{\text{reg}} = -\frac{1}{n_p}\sum_{i=1}^{n_p} -\log\left[\sigma(\mathbf{w}^\top f(\mathbf{x}_i))\right]-\frac{1}{n_n}\sum_{i=1}^{n_n}\log\left[1-\sigma(\mathbf{w}^\top f(\mathbf{n}_i))\right] + \frac{\lambda}{n_pd}||\mathbf{w}||_2^2, 
\end{equation}
where $\mathbf{w} \in \mathbb{R}^{256}$ is the species parameter being optimized, $\lambda$ is the regularization strength, $d=256$, $\sigma$ is the sigmoid function, $f(\mathbf{x}_i)$ is the output of the position branch for the $i$-th positive observation, and $f(\mathbf{n}_i)$ is the output of the position branch for the $i$-th negative observation. We use this loss to estimate a SRM when text is not provided at test time. The learned weights can then be applied to all positions to derive a range map.

To incorporate text at test time along with observational data, we modify the regularization to be the distance from the predicted text-based species embedding, resulting in the following loss, 
\begin{equation}
L_{\text{reg-tx}} = -\frac{1}{n_p}\sum_{i=1}^{n_p} -\log\left[\sigma(\mathbf{w}^\top f(\mathbf{x}_i))\right]-\frac{1}
{n_n}\sum_{i=1}^{n_n}\log\left[1-\sigma(\mathbf{w}^\top f(\mathbf{n}_i))\right] + \frac{\lambda}{n_pd}||\mathbf{w}-\mathbf{w}_{\text{tx}}||_2^2, 
\end{equation}
where $\mathbf{w}_{\text{tx}}$ is the output of the text-based species encoder when provided a text summary. 
This encourages the learned species weight vector to be similar to the text derived one. 
This approach is simple to implement, and our experiments indicate that it is also effective.

\subsection{Implementation Details}\label{sec:training_details}
We closely follow the hyperparameters from SINR  for a fair comparison. We train with the Adam optimizer for 10 epochs with a learning rate of 0.0005. The species network has three linear layers with ReLU activation. The input text embedding dimension is 4,096, the hidden dimension is 512, and the output species embedding dimension is 256. During training, at each iteration, we choose a random Wikipedia section to generate each species embedding. For logistic regression in the later few-shot evaluation experiments, we use a regularization strength $\lambda=20$. 
Training a single LE-SINR model from scratch using all the text and observational data takes about 10 hours on a single NVIDIA RTX 2080ti GPU occupying about 10GB of VRAM. Wikipedia text embeddings and their summaries were generated once using a distributed GPU cluster.

\begin{table}
  \caption{\textbf{Zero-shot Range Estimation}. Our LE-SINR approach enables zero-shot range estimation given a range or habitat description for a particular species. All models were trained with a maximum of 1,000 observations per species and uniformly sampled negatives, using the AN\_full loss. `+Env' indicates models trained with additional environmental features as input, as in SINR~\cite{cole2023spatial}. '+Eval Sp.' indicates models that had observed data from the evaluation species during training. LE-SINR matches Oracle SINR performance when explicit species tokens are available, \ie the model is trained with evaluation species, and we also show the performance on the observed species using the text-based species encoders for completeness in the Appendix.
  }
  \vspace{5pt}
  \centering
  \resizebox{0.72\columnwidth}{!}{
  \begin{tabular}{l|l|c|c|c|c}
        & Method & +Env & +Eval Sp. & IUCN & S\&T \\\hline
        \multirow{5}{*}{Oracle} & SINR &  & \checkmark & \color{gray}0.67 & \color{gray}0.77\\
        & LE-SINR (Species Token) &  & \checkmark & \color{gray}0.69 & \color{gray}0.78\\
        & SINR & \checkmark & \checkmark & \color{gray}0.76 & \color{gray}0.80\\
        & LE-SINR (Species Token) & \checkmark & \checkmark & \color{gray}0.75 & \color{gray}0.80 \\
        \hline
        \multirow{2}{*}{Baselines} & Constant Prediction & &\checkmark & 0.01 & 0.22\\
        & Model Mean & & \checkmark & 0.09 &0.35\\
        \hline
        \multirow{4}{*}{Zero-shot} & LE-SINR (Habitat Text)&  & & 0.28 & 0.51\\
        & LE-SINR (Range Text) &  & & 0.47 & 0.61\\
        & LE-SINR (Habitat Text) & \checkmark & & 0.32 & 0.52 \\
        & LE-SINR (Range Text) & \checkmark & & 0.53 & 0.64         
  \end{tabular}
  }
    \label{tab:main-results}

\caption{\textbf{LE-SINR Ablations}. Impact of various design choices, such as species description (range and habitat vs. taxonomy in LD-SDM~\cite{sastry2023ld}), location encoders, and the loss function on the zero-shot performance of LE-SINR. All models were trained without `Eval Sp' and using 'Env' when appropriate. LD-SDM$^\dagger$ is our model and text encoder but trained with taxonomic text as in~\cite{sastry2023ld}.}
  \vspace{5pt}
  \centering
  \resizebox{0.9\columnwidth}{!}{
  \begin{tabular}{l|l|c|c|c}
        Ablation Type & Method & +Env & IUCN & S\&T \\\hline
        \multirow{2}{*}{Text} & LD-SDM$^\dagger$~\cite{sastry2023ld} (Genus Text) & \checkmark & 0.19 & 0.30\\
        & LD-SDM$^\dagger$ (Species Text) & \checkmark & 0.21 & 0.33\\\hline
        \multirow{4}{*}{Position Encoder} & GeoCLIP Backbone~\cite{cepeda2023geoclip} (Habitat text) & &  0.23 & 0.49\\
        & GeoCLIP Backbone (Range Text) & &  0.41 & 0.58\\
        & Spherical Harmonics Encoding~\cite{russwurm2023geographic} (Habitat Text) & \checkmark &  0.31 & 0.52\\
        & Spherical Harmonics Encoding (Range Text)  & \checkmark &  \bf 0.53 & 0.63\\\hline
        \multirow{2}{*}{Loss Function} & SatCLIP Loss~\cite{klemmer2023satclip} (Habitat Text) & \checkmark &  0.26 & 0.52\\
        & SatCLIP Loss (Range Text)  & \checkmark &  0.46 & 0.63\\\hline
        & Ours (Habitat Text) & \checkmark &  0.32 & 0.52\\
        & Ours (Range Text) & \checkmark & \bf 0.53 & \bf 0.64
  \end{tabular}
  }
  \label{tab:ablations}
  \vspace{-10pt}
\end{table}

\section{Results}\label{sec:results}
\subsection{Zero-shot Range Estimation} 
In Table \ref{tab:main-results}, we compare against several baseline methods to establish performance lower and upper bounds as zero-shot range prediction is a novel task. The constant prediction baseline assumes a species is present everywhere, while model mean predicts the mean species distribution map for all species. We also show the `oracle' performance obtained by training a SINR which sees observations from evaluation species at training time. All models were trained with observations capped at $1,000$ with uniform negatives, using the AN\_full loss and `+Env' indicates models that are trained using extra environmental features as input, as in \cite{cole2023spatial}. We compare these to zero-shot range estimates obtained using various LE-SINR models and input text. 

The results show that the zero-shot range estimates comfortably outperform the baselines, achieving non-trivial performance with no observations. Range text works the best, but even habitat text achieves strong performance on both IUCN and S\&T species. Similar to SINR, we observe a boost when including environmental covariates.  
LE-SINR with explicit species tokens matches the Oracle SINR performance, \ie when evaluation species are used during training, suggesting that LE-SINR does not lose performance on observed species. Therefore, LE-SINR can be used both to explicitly model observed species and for zero-shot prediction from text for novel species.

\begin{figure}[t!]
\centering
\includegraphics[width=\linewidth]{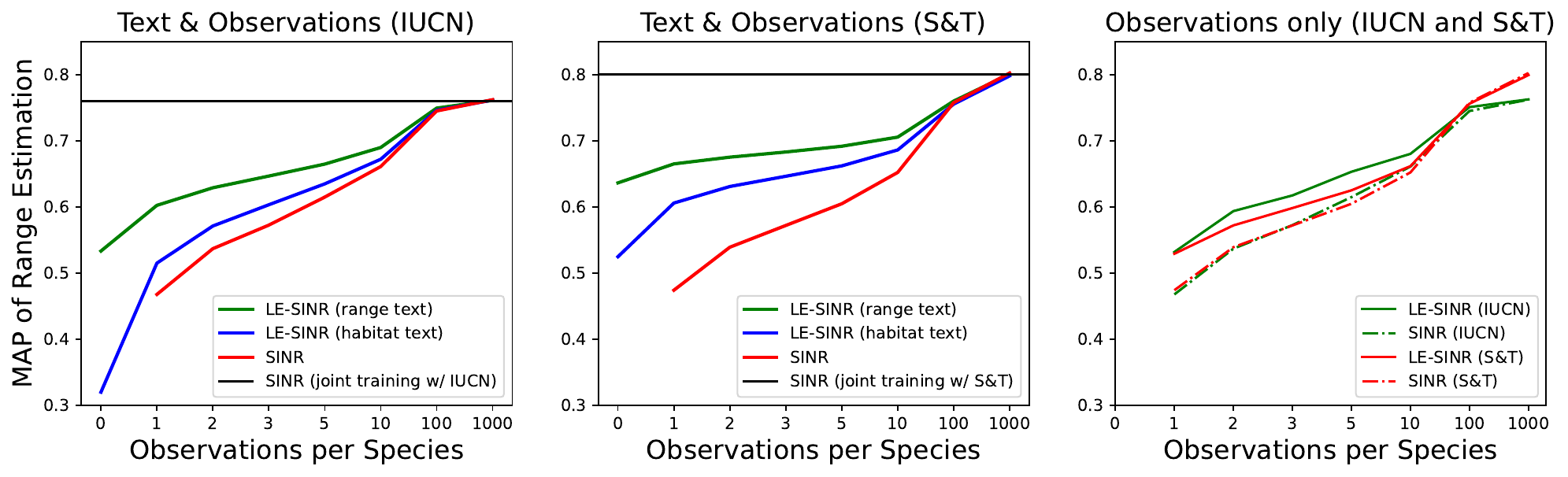}
\caption{\textbf{Range Estimation from Text and Observations}. (Left) IUCN and (Middle) S\&T results for zero-shot range estimation based on text, and few-shot estimation based on the text-driven prior. Both range and habitat texts improve few-shot performance over baseline SINR. (Right) Comparison of the position branch of SINR and LE-SINR for range estimation using a few examples. Language-driven covariates learned by LE-SINR lead to better generalization when observations are limited. We report the MAP for range estimation for species in the S\&T and IUCN test sets. 
}
\vspace{-10pt}
\label{fig:few_shot_results}
\end{figure}

\begin{figure}[t]
    \centering

    \begin{minipage}{0.48\textwidth}
        \includegraphics[trim={0 4cm 0 0},clip,width=\linewidth]{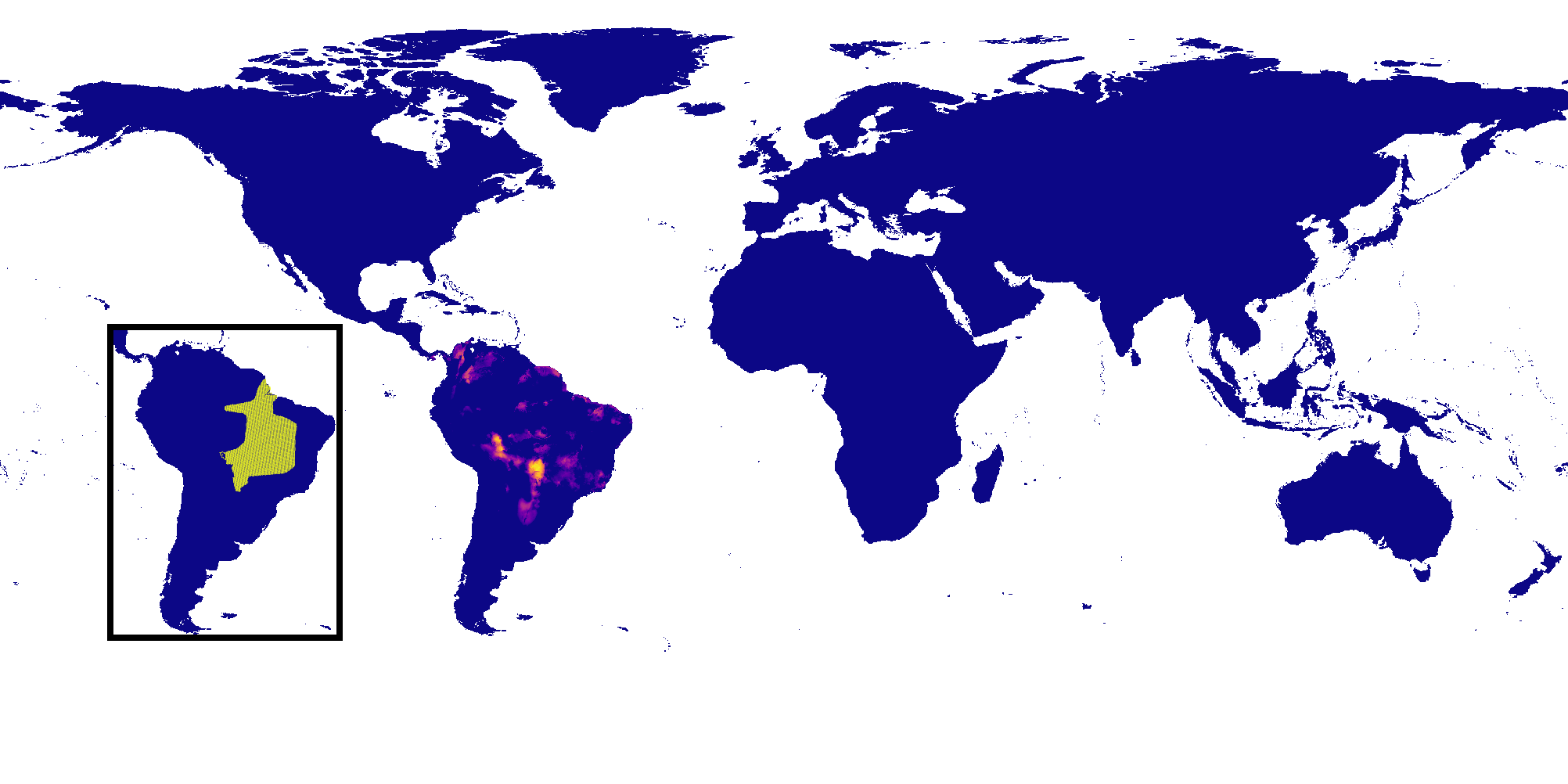}
        \subcaption*{\scriptsize The hyacinth macaw prefers semi-open, somewhat wooded habitats. It usually avoids dense, humid forest, and in regions dominated by such habitats, it is generally restricted to the edge or relatively open sections (\eg along major rivers). In different areas of their range, these parrots are found in savannah grasslands, in dry thorn forests known as caatinga, and in palm stands or swamps, particularly the moriche palm (Mauritia flexuosa).}
    \end{minipage} \hspace{1em}
    \begin{minipage}{0.48\textwidth}
        \includegraphics[trim={0 4cm 0 0},clip,width=\linewidth]{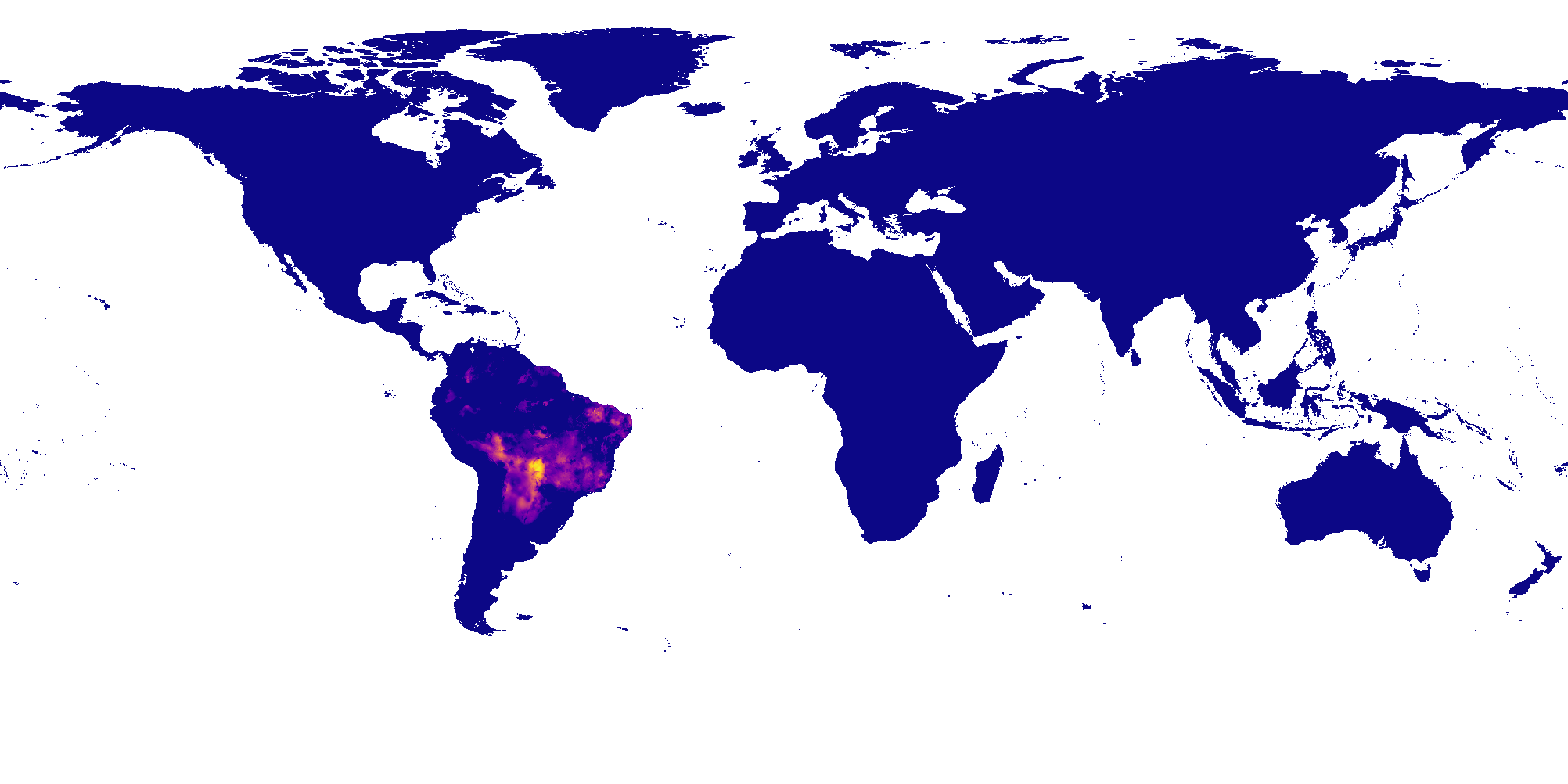}
        \subcaption*{\scriptsize The hyacinth macaw occurs today in three main areas in South America: In the Pantanal region of Brazil, and adjacent eastern Bolivia and northeastern Paraguay, in the cerrado regions of the eastern interior of Brazil (Maranhão, Piauí, Bahia, Tocantins, Goiás, Mato Grosso, Mato Grosso do Sul, and Minas Gerais), and in the relatively open areas associated with the Tocantins River, Xingu River, Tapajós River, and the Marajó island in the eastern Amazon Basin of Brazil.}
    \end{minipage}

    \begin{minipage}{0.48\textwidth}
        \includegraphics[trim={0 4cm 0 0},clip,width=\linewidth]{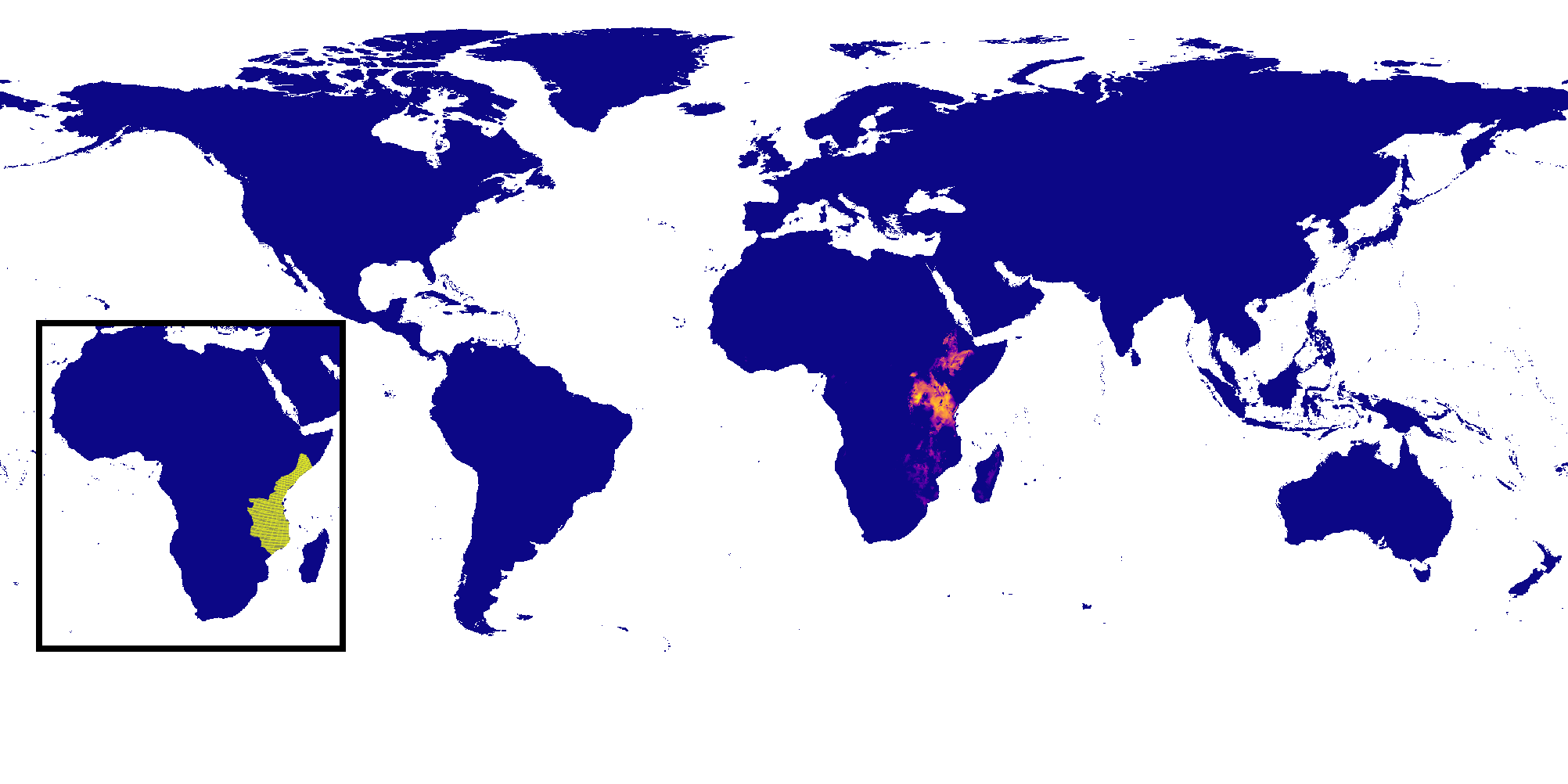}
        \subcaption*{\scriptsize They are diurnal, terrestrial, and live in complex, mixed-gender social groups of 8 to 200 individuals per troop. They prefer savannas and light forests with a climate that is suitable for their omnivorous diet.}
            \centering{\small {\it Habitat text description}}
    \end{minipage}\hspace{1em}
    \begin{minipage}{0.48\textwidth}
        \includegraphics[trim={0 4cm 0 0},clip,width=\linewidth]{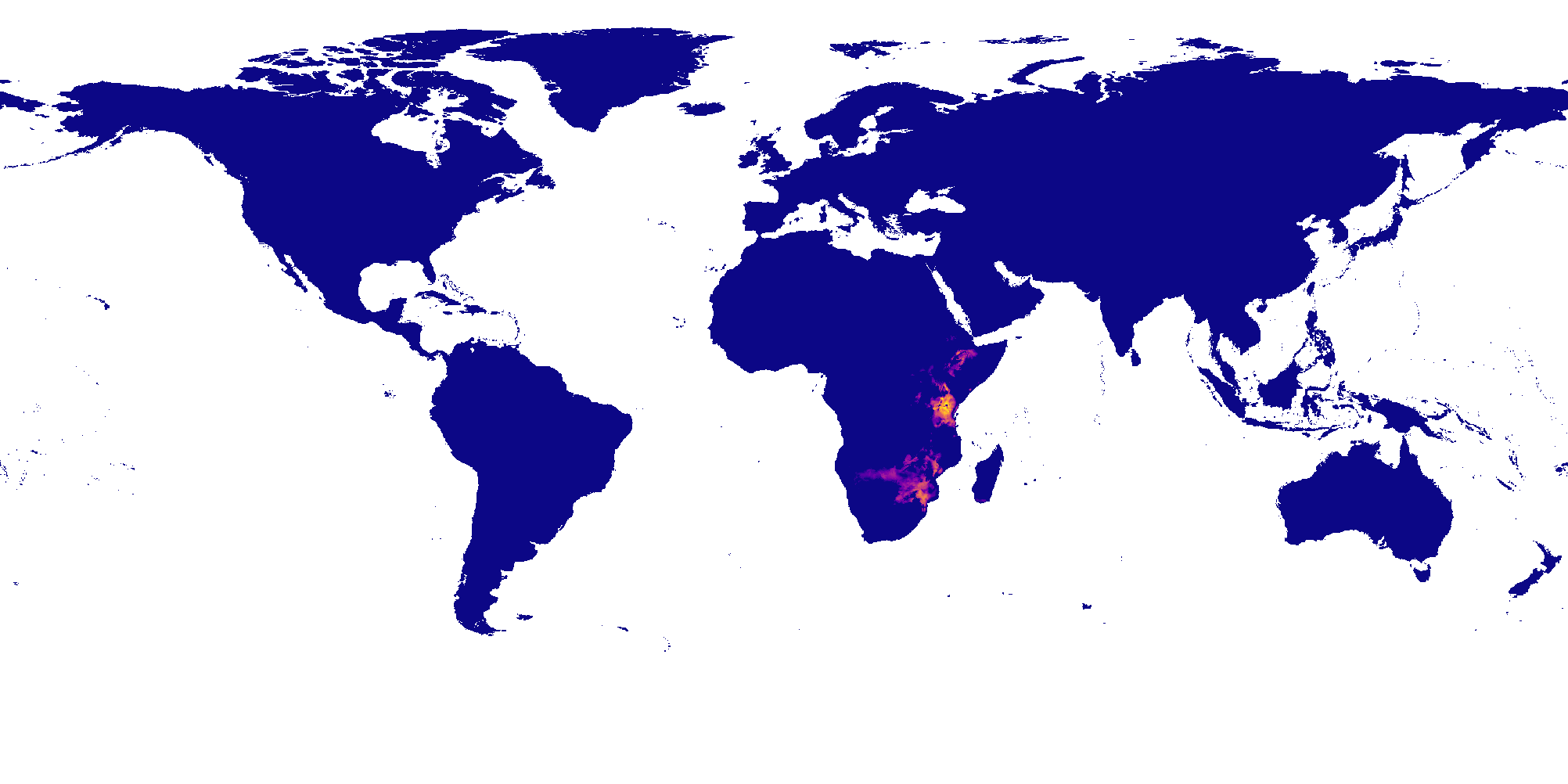}
        \subcaption*{\scriptsize Yellow baboons inhabit savannas and light forests in eastern Africa, from Kenya and Tanzania to Zimbabwe and Botswana.\\}
        \centering{\small {\it Range text description}}
    \end{minipage}
    
    \caption{\textbf{Zero-Shot Range Estimation.} Here we show the `Habitat' and `Range' text descriptions and corresponding zero-shot range maps for the \texttt{Hyacinth Macaw} (top) and the \texttt{Yellow Baboon} (bottom), with expert derived range maps inset.}
    \vspace{-8pt}
    \label{fig:zero_shot_w_different_text}
\end{figure}

Figure~\ref{fig:few_shot_results} compares our zero-shot performance with few-shot range estimation using logistic regression models trained with varying numbers of observations. On S\&T the zero-shot performance with habitat text is worth two observations, while range text is worth ten. Figures~\ref{fig:zero_shot_w_different_text} and \ref{fig:qualitative_low_shot} show range maps generated from text inputs for various species. %

In Table~\ref{tab:ablations} we compare against other training text, position encodings, and loss functions. For the LD-SDM \cite{sastry2023ld} baseline we use our LE-SINR architecture but replace the Wikipedia-derived text with LD-SDM taxonomic classification strings. Unlike LD-SDM's autoregressive LLM, which generates an embedding for each token, this baseline, like our main method, encodes text using a model trained specifically for fixed-size text embeddings. For each species, we construct a string representing different levels of the taxonomic hierarchy, including class, order, family, genus, and species. During training, we randomly select one of these strings in each iteration. We then perform zero-shot evaluations at each taxonomic level. Performance decreases as we move up the hierarchy with species-level text performing the best. For the position encoder baselines we first replace our position encoder with the pre-trained GeoCLIP \cite{cepeda2023geoclip} model. The weights are then fine-tuned during training. We also try the Spherical Harmonics \cite{russwurm2023geographic} coordinate encoding to encode the latitude and longitude. We experimented with multiple values for $L$ as well as the SIREN network architecture and found $L=10$ with our original MLP architecture worked best for this experiment. For the SatCLIP \cite{klemmer2023satclip} baseline, we use our model but replace the $AN_{\text{full}}$ loss with their contrastive loss applied to the position and text embeddings. Looking at Table~\ref{tab:ablations}, none of these experiments perform better than our method.

\vspace{-8pt}
\subsection{Text-driven Priors for Few-shot Range Estimation} 
\vspace{-8pt} 
In addition to zero-shot range map generation, we can also use LE-SINR in the few-shot setting, \ie when there are a small number of location observations available for a previously unseen test species. 
Figure~\ref{fig:few_shot_results} shows few-shot results obtained using the text-driven prior from LE-SINR, as described in Section~\ref{sec:eval}, on IUCN (left) and S\&T (middle) species. Logistic regression  models regularized toward species weights obtained from range and habitat text provide a significant boost over vanilla logistic models regularized toward zero. The gap is significant when the training data is limited, \ie fewer than 100 observations. Range text priors reduce the need for observations on IUCN and S\&T by a factor of 3$\times$ and 10$\times$, respectively.  Figure~\ref{fig:qualitative_low_shot} in the Appendix shows some qualitative examples of how range estimates change with different numbers of observations.

\vspace{-8pt}
\subsection{Evaluation of Learned Positional Embeddings}
\vspace{-8pt}
Prior work on generating SRMs often rely on carefully-selected spatial covariates such as temperature, elevation, precipitation, land cover, \etc, to predict the distribution of species based on a few observations. Our approach provides a way to learn a rich set of spatial covariates based on language data.
Figure~\ref{fig:few_shot_results} (right) compares the position embeddings of SINR and LE-SINR as spatial covariates on the task of few-shot range estimation. In both cases, we use a model whose location branch is trained with position only (\ie only latitude and longitude) as input to avoid conflation of the learned covariates with input ones. We find that the intermediate position embeddings learned when trained using language lead to better generalization, especially when training data is limited. Note that here we do not use any language input at test time, as both models are trained simply using observation data, same as the logistic regression baselines.

\begin{figure}[t]
\centering

\begin{overpic}[width=0.49\textwidth]{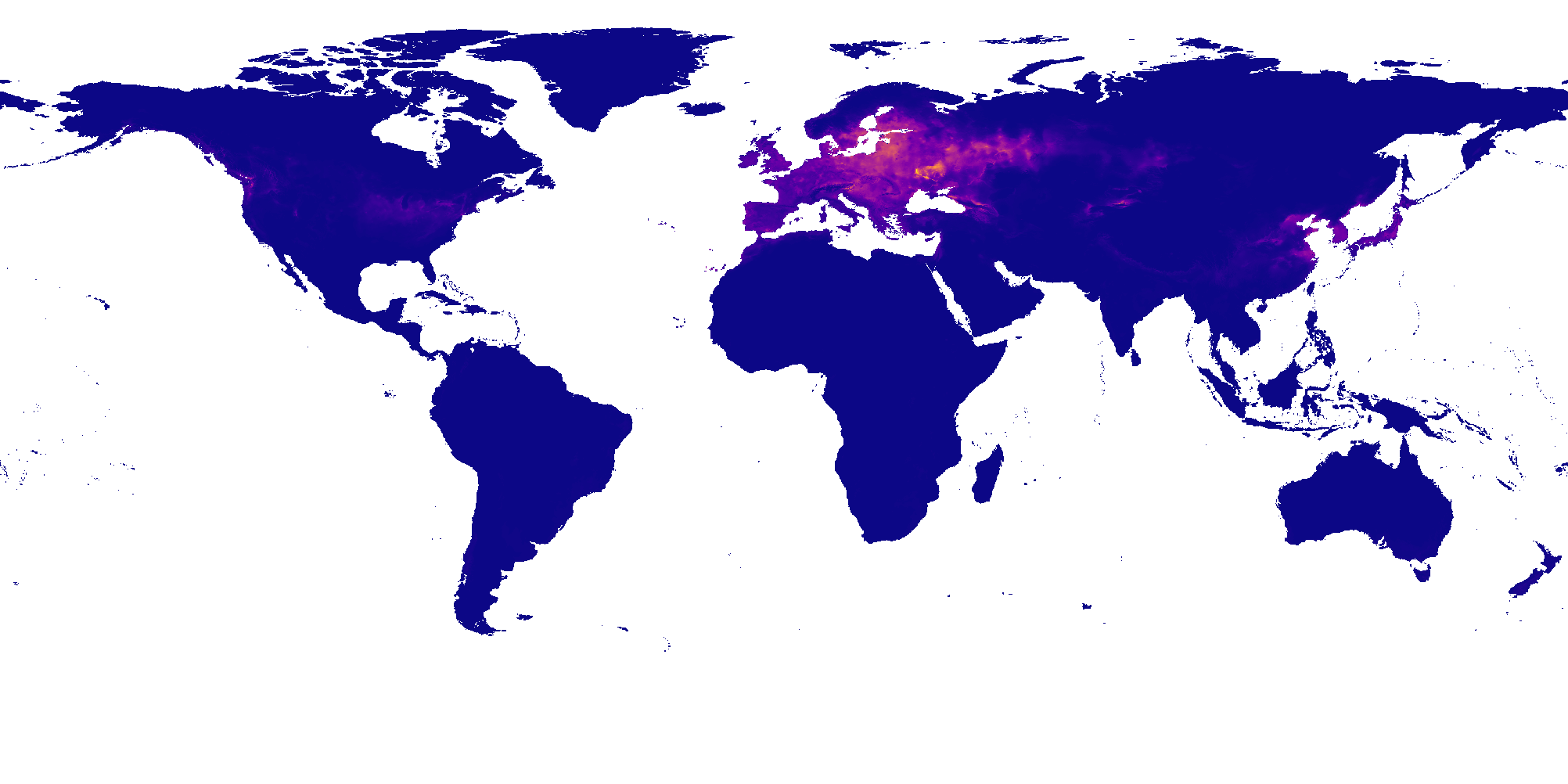}
 \put(0,4){
    \fcolorbox{black}{gray!30}{\strut\normalsize  \search ``europe''}
  }
\end{overpic}\hspace{5pt}
\begin{overpic}[width=0.49\textwidth]{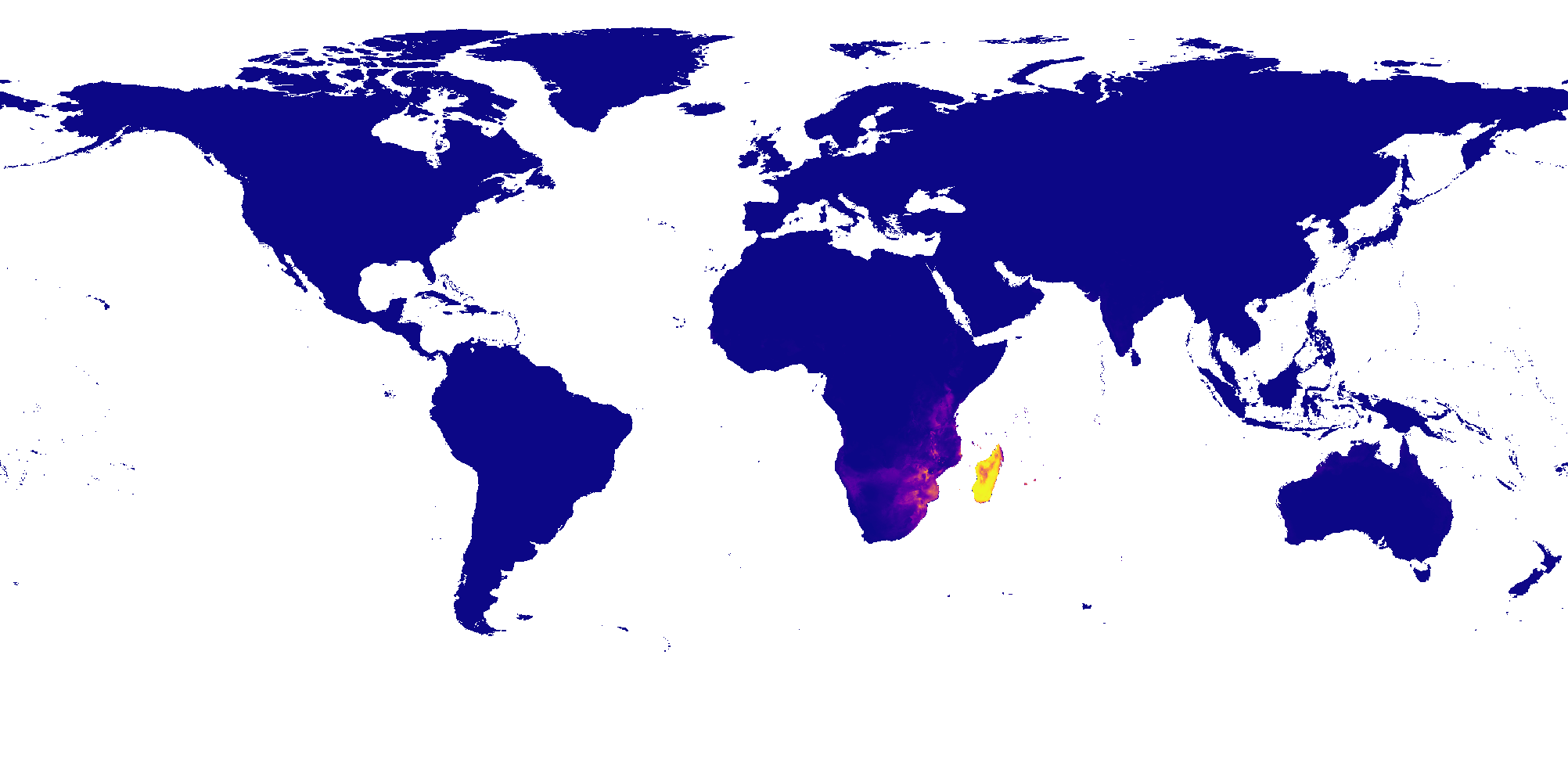}
 \put(0,4){
    \fcolorbox{black}{gray!30}{\strut\normalsize  \search ``madagascar''}
  }
\end{overpic} \\

\begin{overpic}[width=0.49\textwidth]{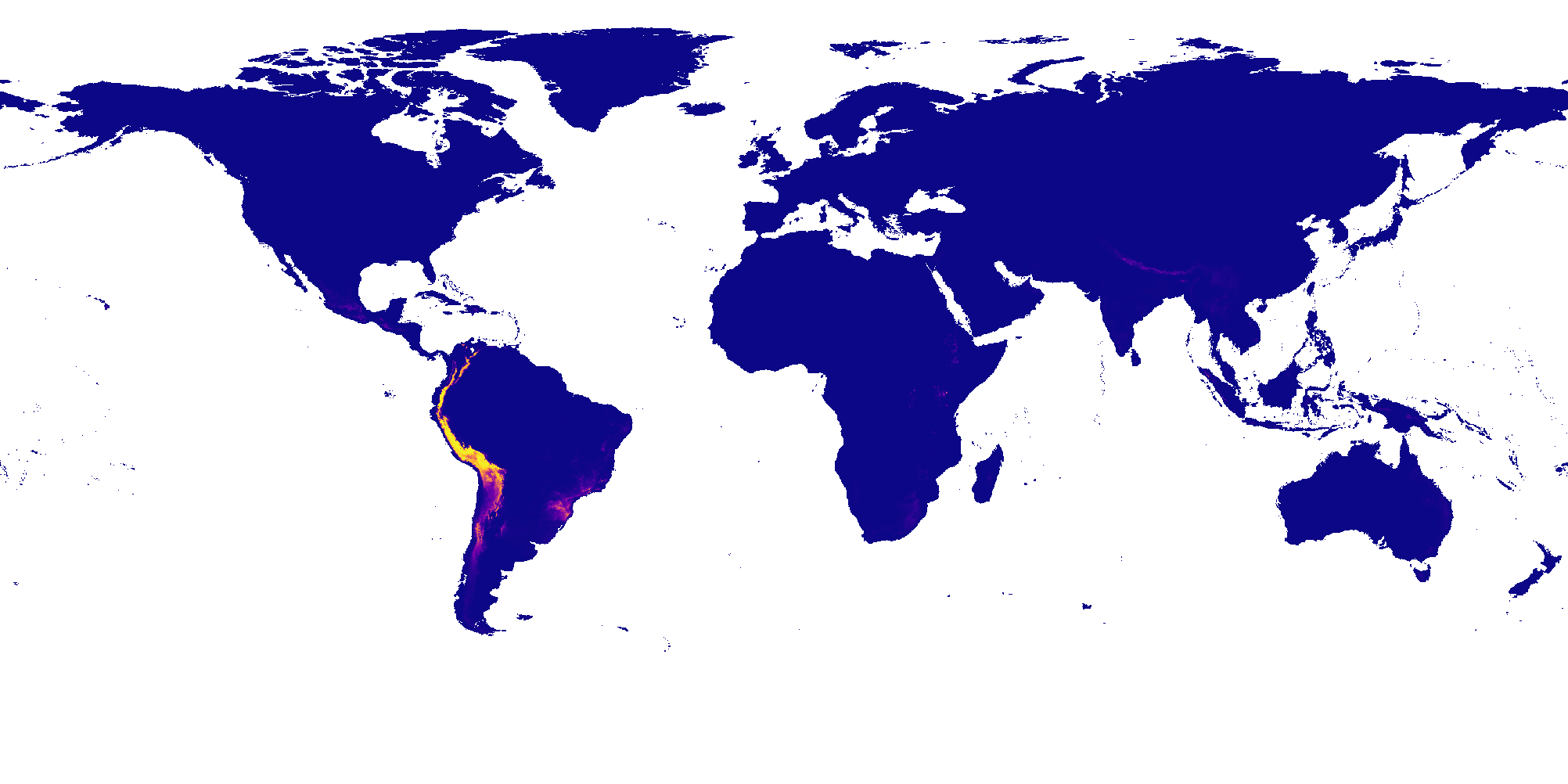}
 \put(0,4){
    \fcolorbox{black}{gray!30}{\strut\normalsize  \search ``Andes''}
  }
\end{overpic}\hspace{5pt}
\begin{overpic}[width=0.49\textwidth]{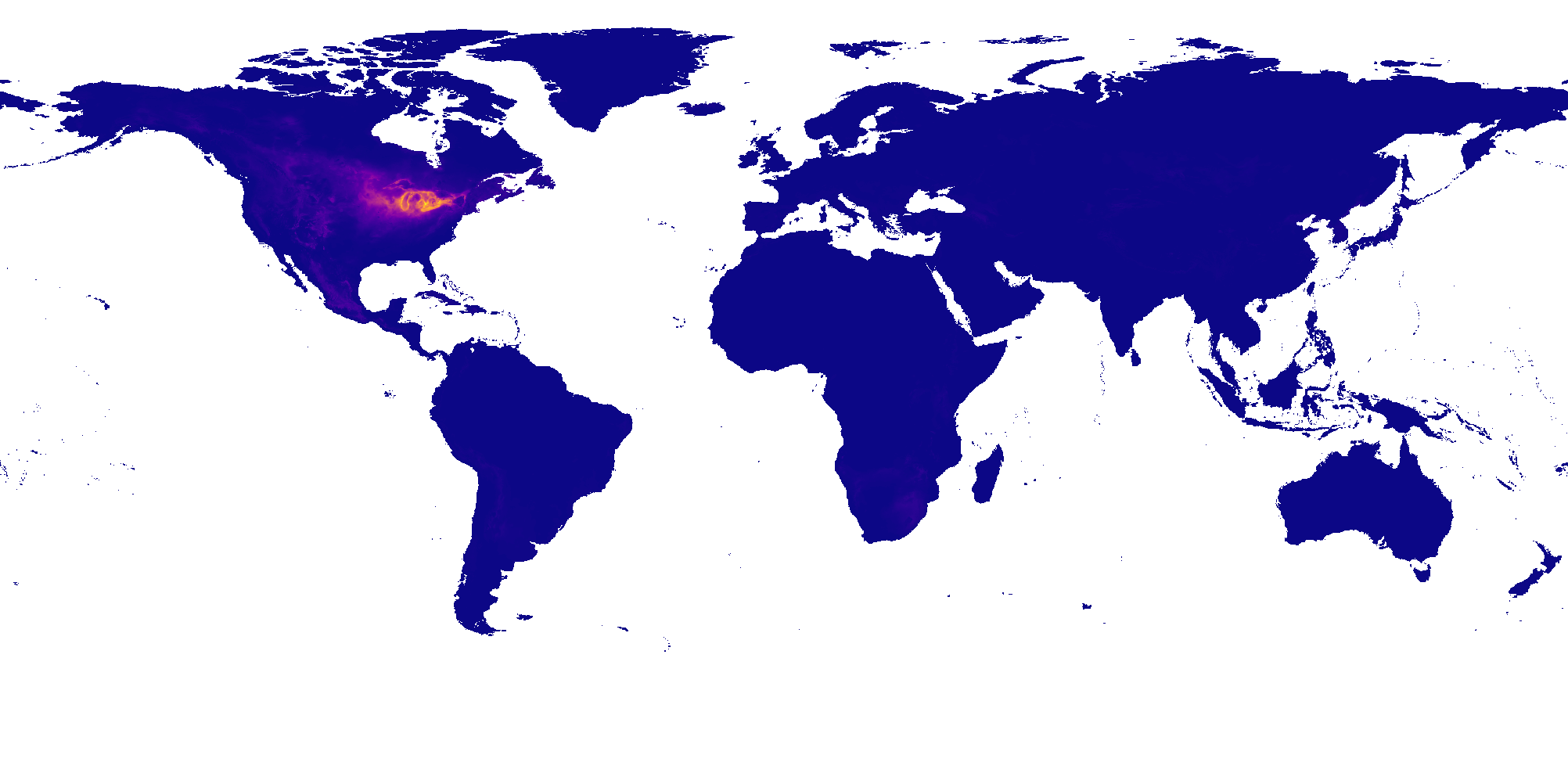}
 \put(0,4){
    \fcolorbox{black}{gray!30}{\strut\normalsize  \search ``great lakes''}
  }
\end{overpic} \\

\begin{overpic}[width=0.49\textwidth]{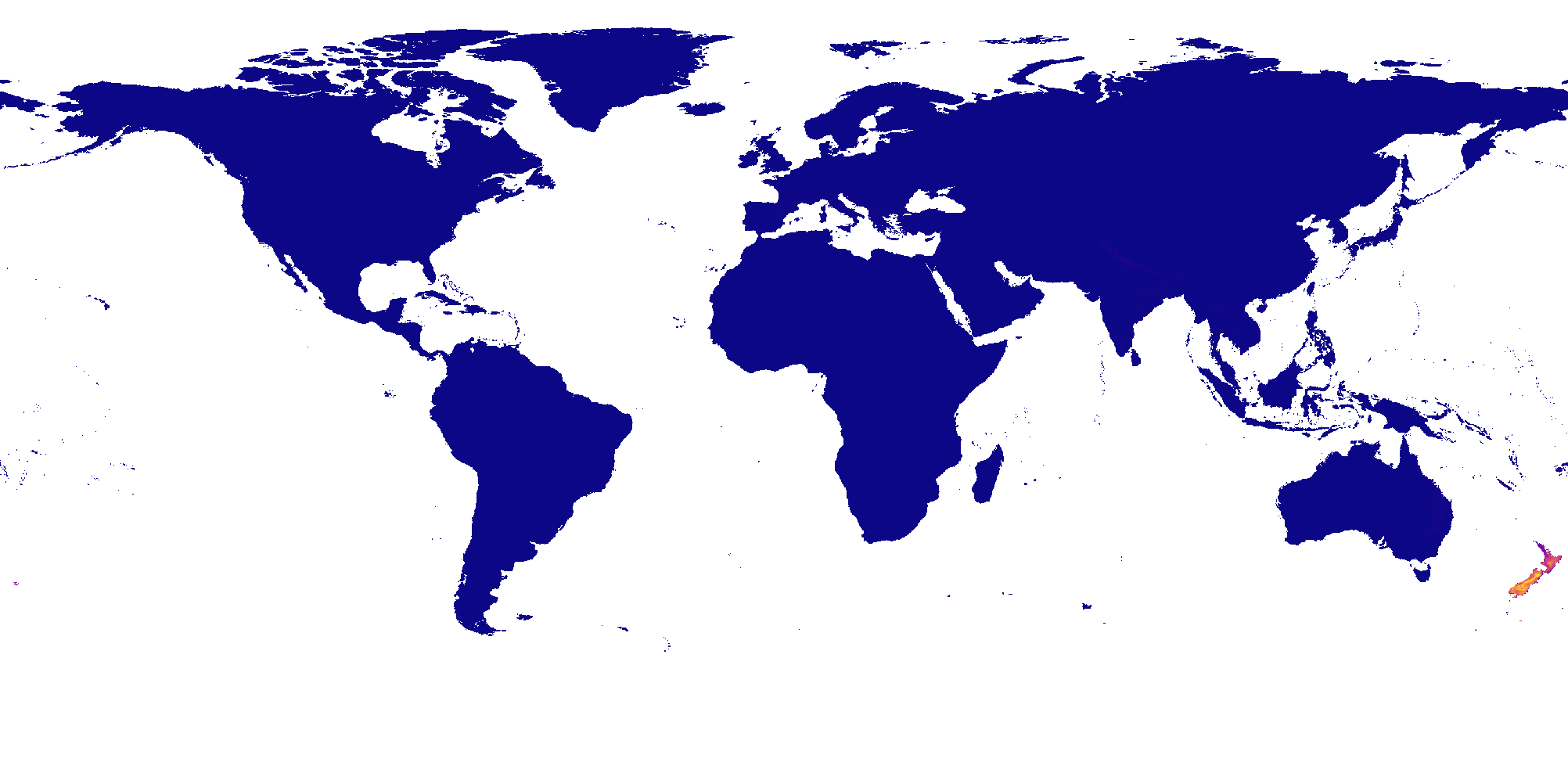}
 \put(0,4){
    \fcolorbox{black}{gray!30}{\strut\normalsize  \search ``Treaty of Waitangi''}
  }
\end{overpic}\hspace{5pt}
\begin{overpic}[width=0.49\textwidth]{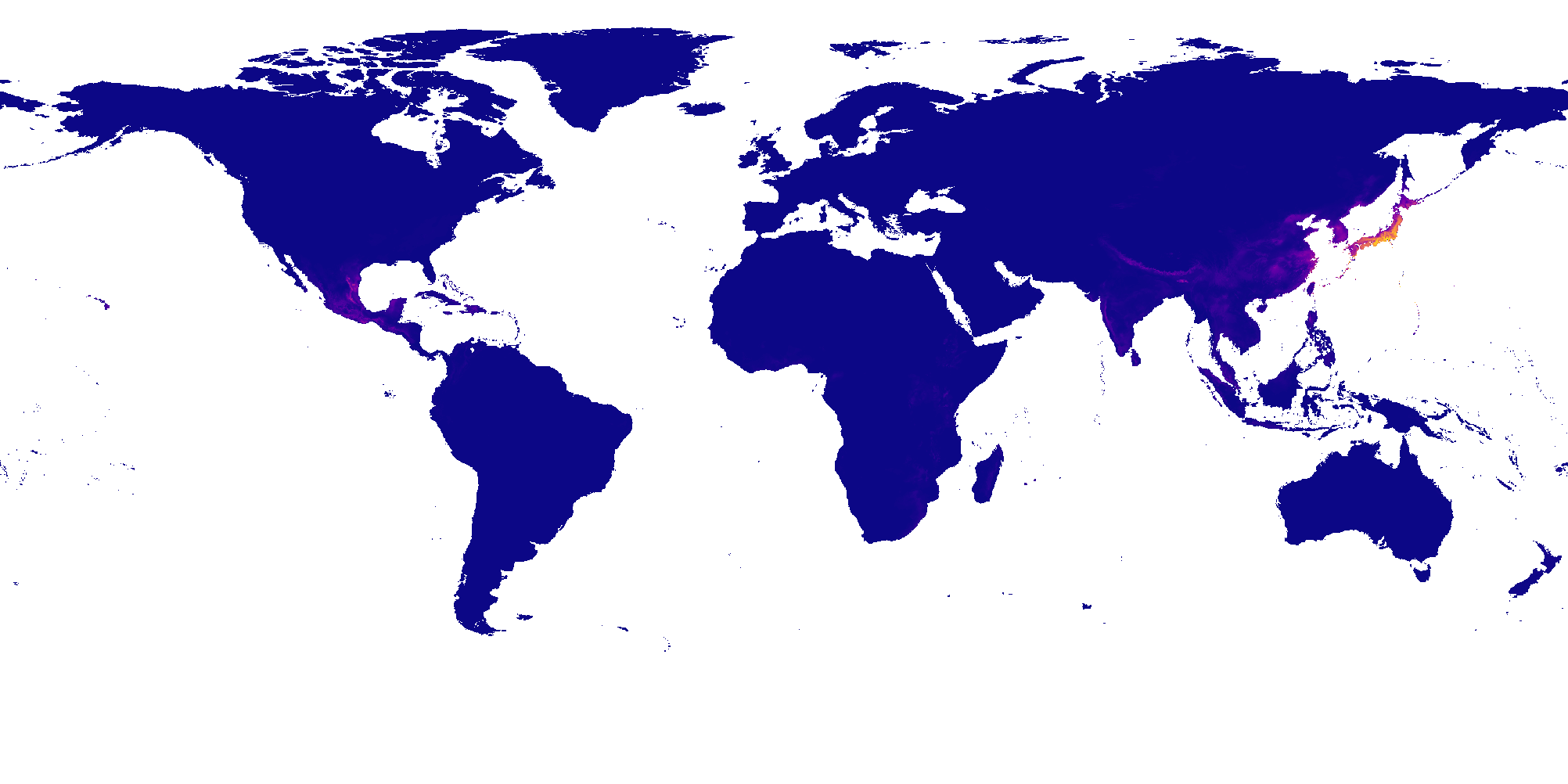}
 \put(0,4){
    \fcolorbox{black}{gray!30}{\strut\normalsize \search ``hello kitty''}
  }
\end{overpic} \\

\begin{overpic}[width=0.49\textwidth]{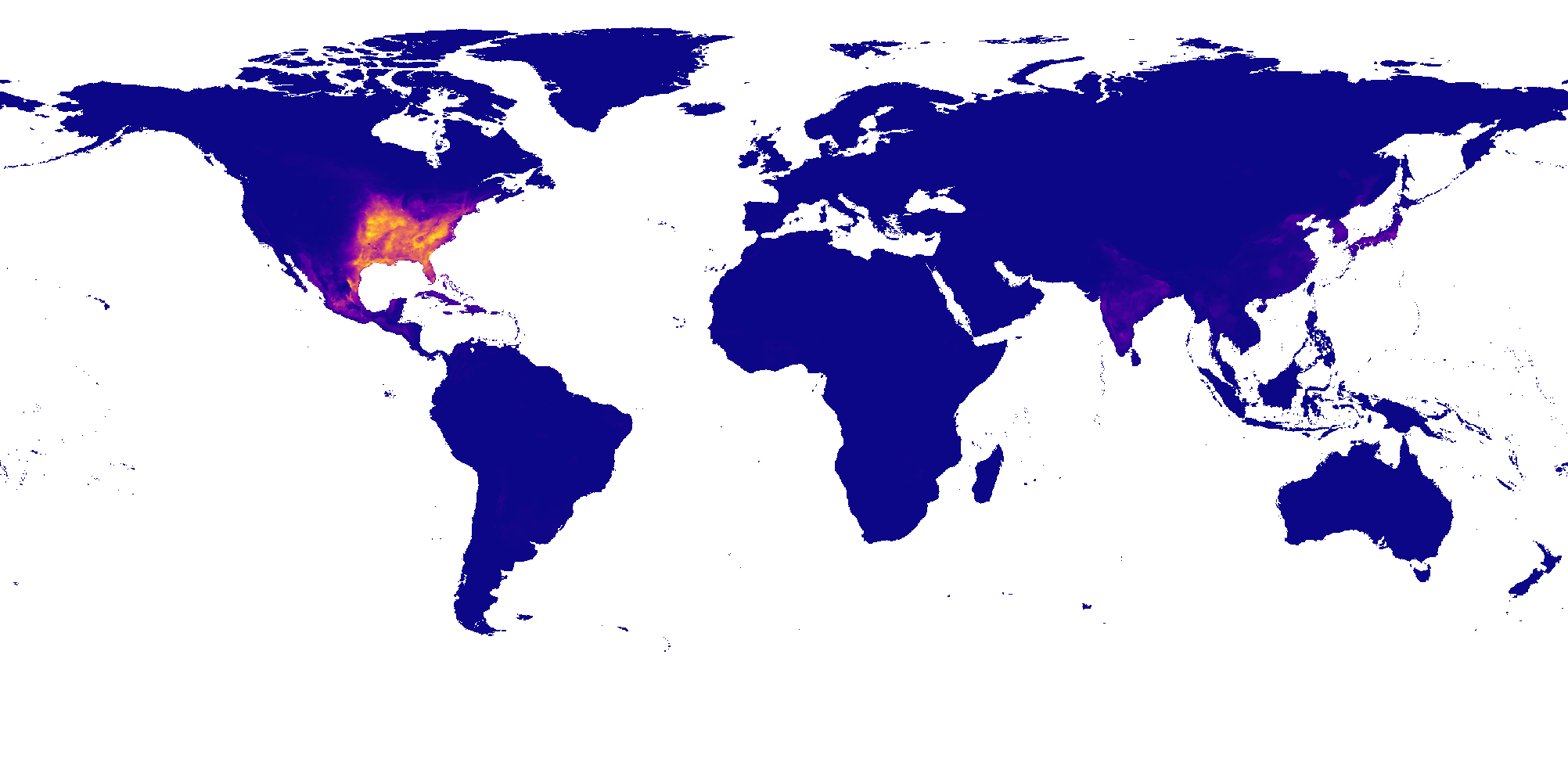}
 \put(0,4){
    \fcolorbox{black}{gray!30}{\strut\normalsize  \search ``Babe Ruth''}
  }
\end{overpic}\hspace{5pt}
\begin{overpic}[width=0.49\textwidth]{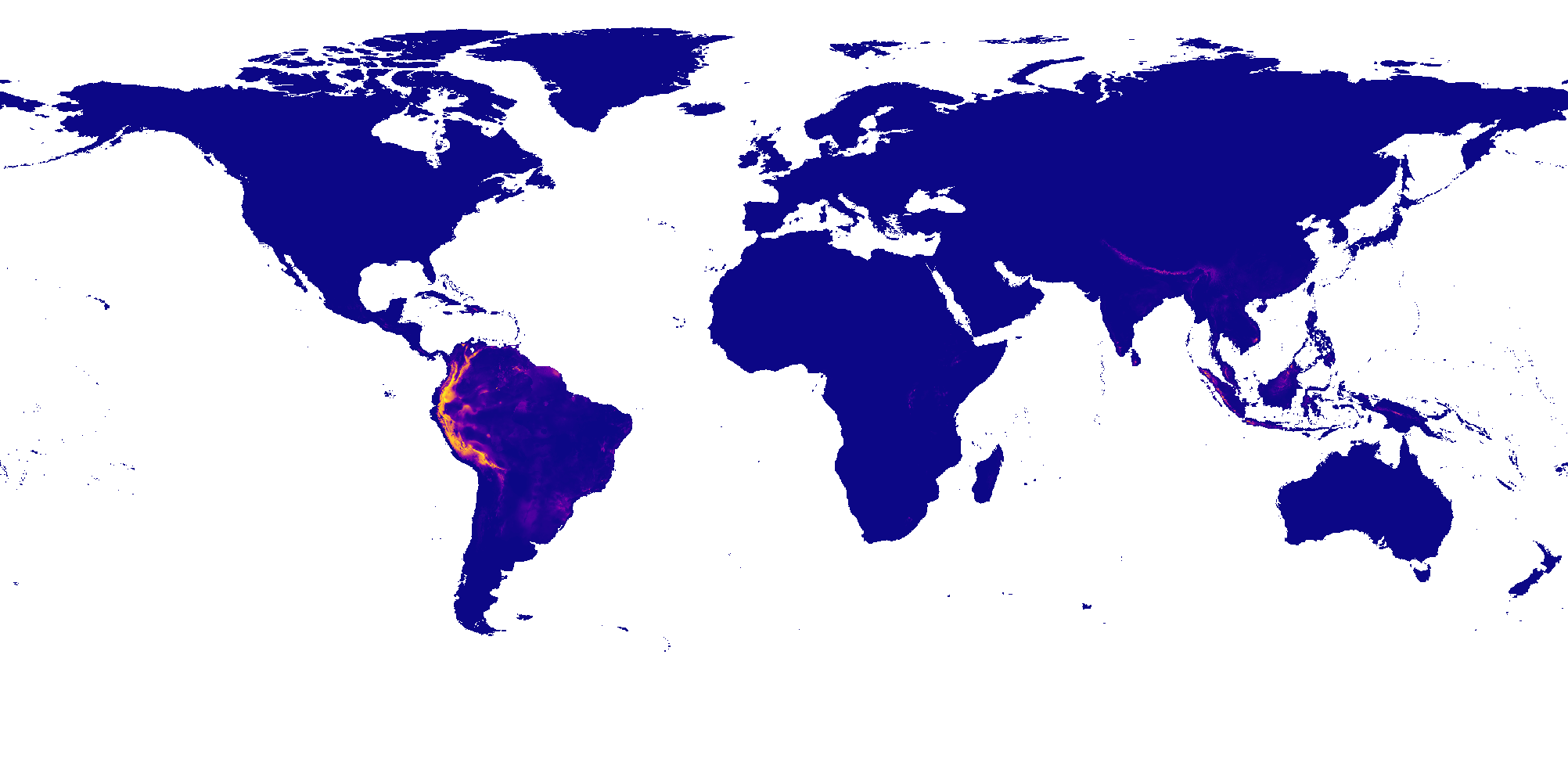}
 \put(0,4){
    \fcolorbox{black}{gray!30}{\strut\normalsize \search ``Simón Bolívar''}
  }
\end{overpic}

\begin{overpic}[width=0.49\textwidth]{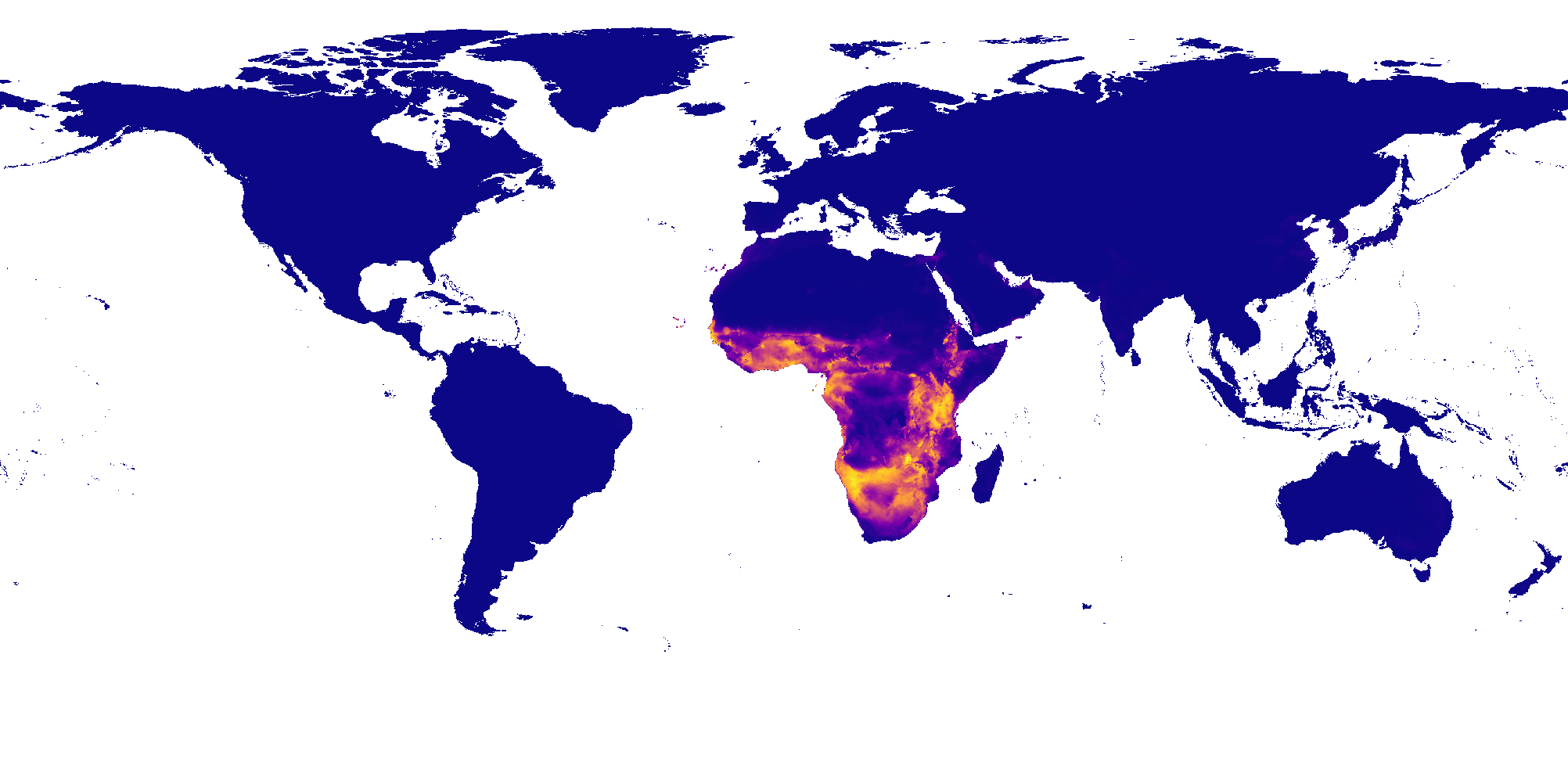}
 \put(0,4){
    \fcolorbox{black}{gray!30}{\strut\normalsize  \search ``africa''}
  }
\end{overpic}\hspace{5pt}
\begin{overpic}[width=0.49\textwidth]{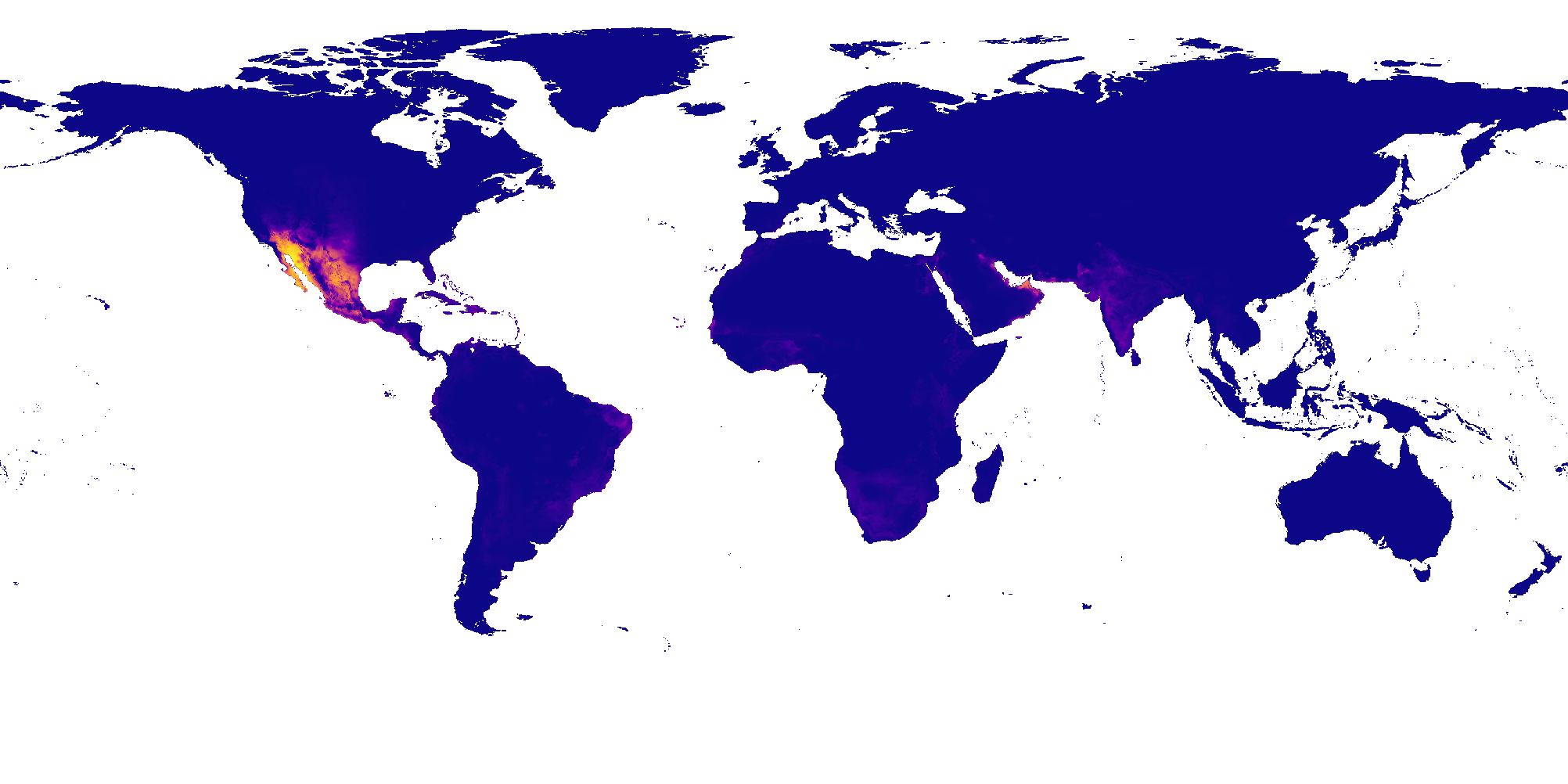}
 \put(0,4){
    \fcolorbox{black}{gray!30}{\strut\normalsize \search ``desert''}
  }
\end{overpic}\vspace{0pt}
\caption{ \textbf{Geospatial Grounding of Non-Species Concepts.}
LE-SINR is able to geographically ground text prompts to locations on the earth. 
Here we display the inner product between the location encoder's features and the language model's encoding of the text displayed in the bottom left of each panel.
This includes coarse concepts such as continents and countries (top row), geographic features such as specific lakes and mountain ranges (second row), in addition to concepts that do not appear in our species text training data but are likely already represented in the language model (third and fourth row). 
We do, however, observe some limitations resulting from the biases in our training data  which favors North America, Europe, and Australasia (final row).  
Please zoom in to see more detail.}
\vspace{-10pt}
\label{fig:text_model_vis}
\end{figure}

Figure~\ref{fig:ica_embedding} in the Appendix shows that the position embeddings of LE-SINR have a richer spatial structure than SINR. This figure was obtained by projecting the learned position embeddings to three dimensions using Independent Component Analysis and visualizing them as color in RGB space. 
Figure~\ref{fig:text_model_vis} visualizes a variety of maps generated from natural language. LE-SINR has learned about geographical regions, climate zones, and even abstract non-species concepts by aligning text representations with geographic locations through species observations. For example, the presence of a species such as the `Fennec Fox' allows the model to learn that the species is associated with concepts such as the deserts of North Africa, particularly the Sahara Desert, sandy environments, and extreme temperatures based on information in Wikipedia text.

\vspace{-8pt}
\subsection{Limitations and Broader Impacts}\label{sec:limitations}
\vspace{-8pt}
While we show that language and location observations can be combined to estimate range maps from a few examples, we do not compare to other few-shot methods such as those based on meta-learning~\cite{snell2017prototypical, lee2019meta, finn2017model}. However, recent results for non-species range estimation tasks indicate that a good underlying representation is a key component to superior few-shot performance, even outperforming more advanced methods~\cite{hu2022pushing}.

Another limitation is that our models are most applicable when range and habitat descriptions are available with very few observations, which is not common. While we evaluate our approach based on Wikipedia, a more realistic scenario involves obtaining descriptions of rarely observed species from domain experts. However, we note that many species have substantial Wikipedia articles despite having fewer than 50 observations on iNaturalist.

Our models rely on text embeddings and summaries generated from LLMs, and thus may inherit the biases contained within them. 
For example, our model could further amplify biases in the data by incorrectly spatially localizing specific text terms inappropriately. 
Both Wikipedia text and observational data from iNaturalist are biased toward the United States and Western Europe. 
As a result, our models may not generalize as well to other geographical regions.
There could also be potential negative consequences associated with using the species range predictions from our model to inform conservation or policy decisions. While our results are promising, they may still fall short of the quality needed for such high-stakes use cases. Therefore, caution is encouraged when making decisions based on the model's predictions. Another risk is that the model could be used to locate threatened or endangered species. To mitigate this, we only train on publicly available observation data that has been deemed safe to redistribute by the iNaturalist platform.

\vspace{-8pt}
\section{Conclusion}
\vspace{-8pt}

The generation of detailed species range maps is often constrained by the need for extensive location observations, which can be expensive and time-consuming to collect. Our LE-SINR approach mitigates this issue by mapping species observations and text descriptions into the same space, enabling zero-shot range map generation from text alone. Unlike other methods that make use of text, we can generate a global range map for a species with a single forward pass of our language model, significantly increasing usability for potential downstream users. 

Our extensive evaluation shows that zero-shot range maps produced by LE-SINR, derived solely from text descriptions, can outperform those produced by state-of-the-art models trained on tens of observations. Additionally, using LE-SINR as a prior also significantly enhances few-shot performance. By learning to relate text descriptions of species with locations where that species has been observed, we show that LE-SINR develops an understanding of a wide range of geographical and environmental features, as well as unrelated concepts not seen in the training data such as historical events and aspects of culture.

\clearpage 
\begin{ack}
\vspace{-7pt}
We thank the iNaturalist community for providing the data used for training our models. 
OMA was in part supported by a Royal Society Research Grant. SM and MH are supported by grant \#2329927 from the National Science Foundation.
\end{ack}

\bibliography{main}

\clearpage
\appendix
\setcounter{table}{0}   
\renewcommand{\thetable}{A\arabic{table}}
\setcounter{figure}{0}
\renewcommand{\thefigure}{A\arabic{figure}}
\noindent{\LARGE Appendix}
\section{Additional Results}

In Table ~\ref{tab:text_oracle} we evaluate our oracle model, that was trained on evaluation species observations, on the habitat and range text summaries. These results show that some of the performance gap between the zero-shot and supervised methods can be explained by inherent ambiguities of our evaluation text descriptions to describe the range.

\begin{figure}[h]
\centering
\begin{overpic}[width=0.87\textwidth]{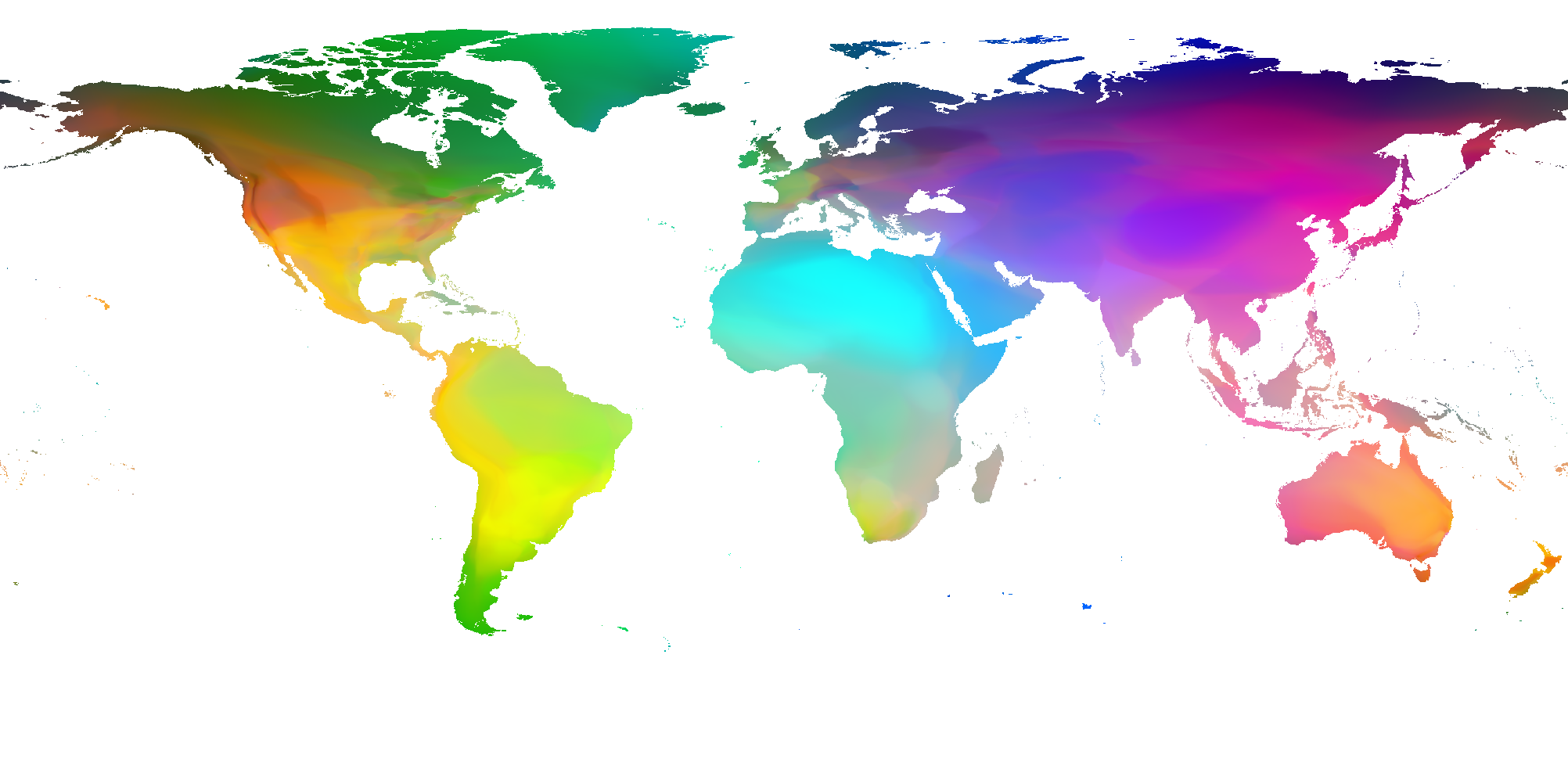} 
    \put(3,5){\large SINR}
\end{overpic}
\begin{overpic}[width=0.87\textwidth]{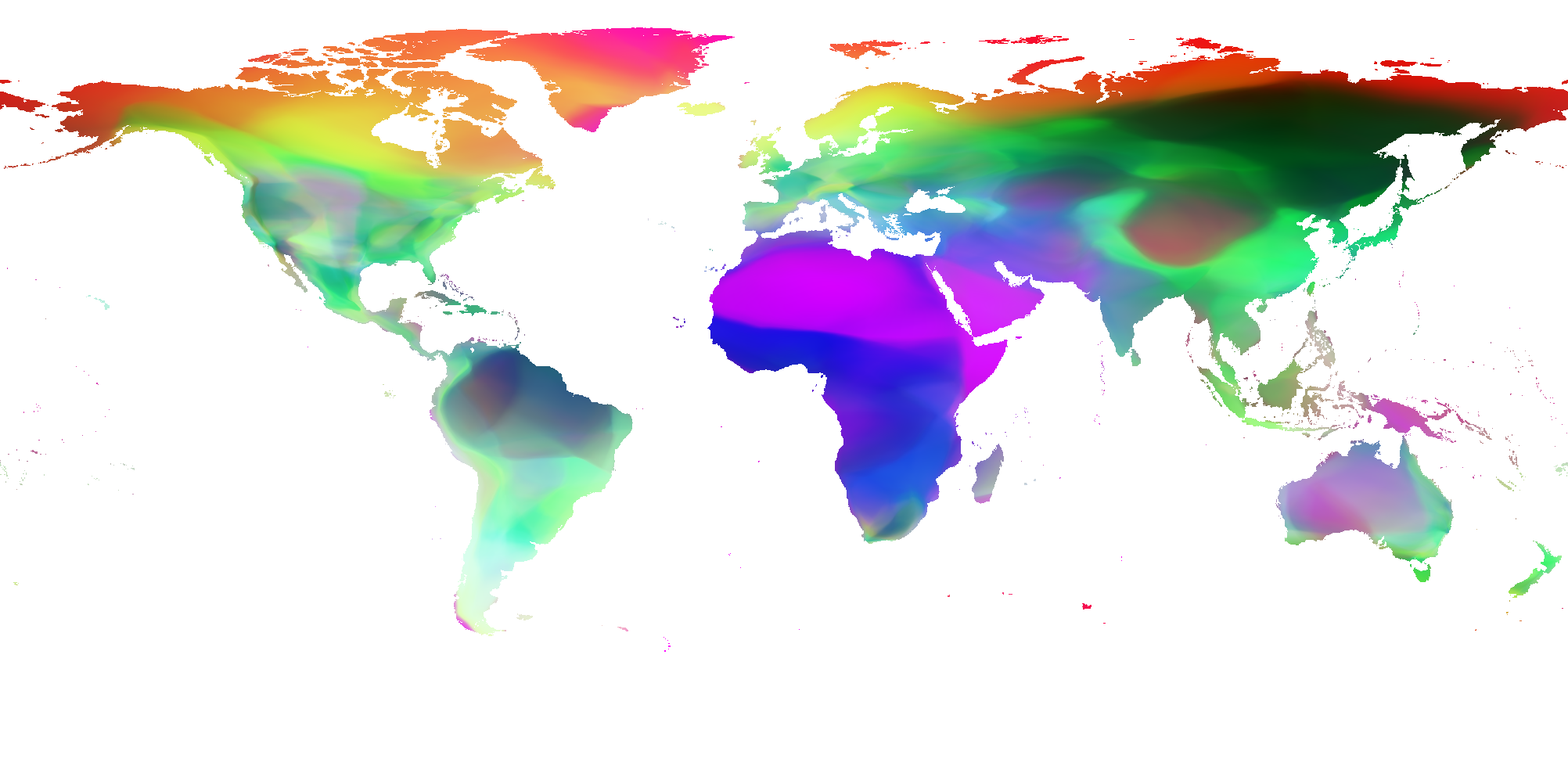} 
    \put(3,5){\large LE-SINR}
\end{overpic}
\vspace{-10pt}
\caption{Visualization of the intermediate position representation learned by the network projected to three dimensions using Independent Component Analysis. 
(Top) Standard SINR model. (Bottom) Our LE-SINR model. Both models were trained with position features as the only input.
}
\label{fig:ica_embedding}
\vspace{-10pt}
\end{figure}

In Figure~\ref{fig:ica_embedding} we display the intermediate representations learned by our location encoder and compare it to the standard SINR model that does not make use of any language information.  
There appears to be more higher frequency spatial information encoded in our representations which is consistent with the improved  few-shot performance we observe when using these features in conjunction with a logistic regression classifier in  Figure~\ref{fig:few_shot_results} from the main paper.  

\begin{figure}
    \centering
    \rotatebox{90}{\hspace{-15pt}\parbox{20mm}{\small\centering\texttt{zero-shot}}}
    \begin{minipage}{0.31\textwidth}
        \includegraphics[width=\linewidth]{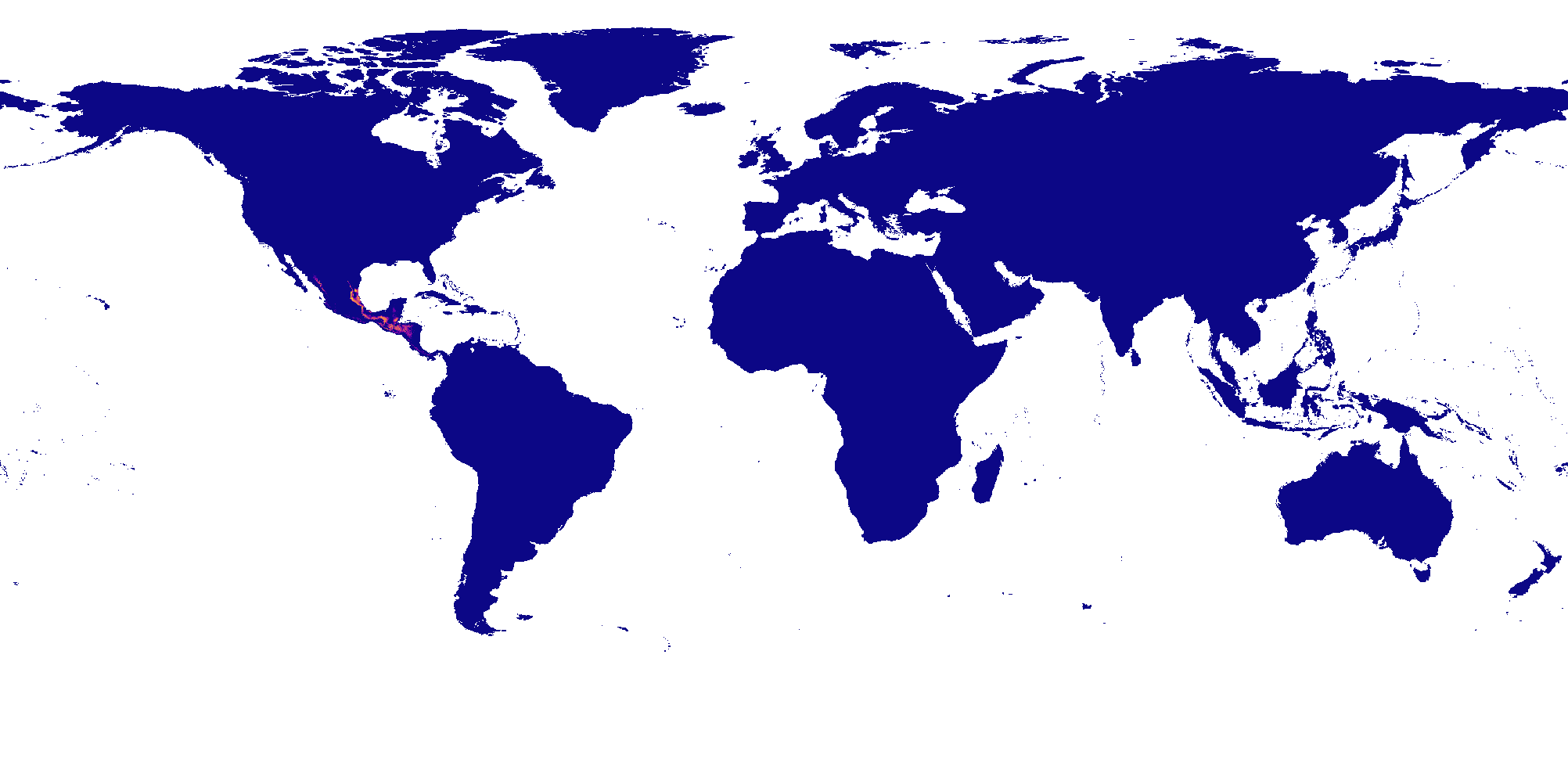}
    \end{minipage}
    \begin{minipage}{0.31\textwidth}
        \includegraphics[width=\linewidth]{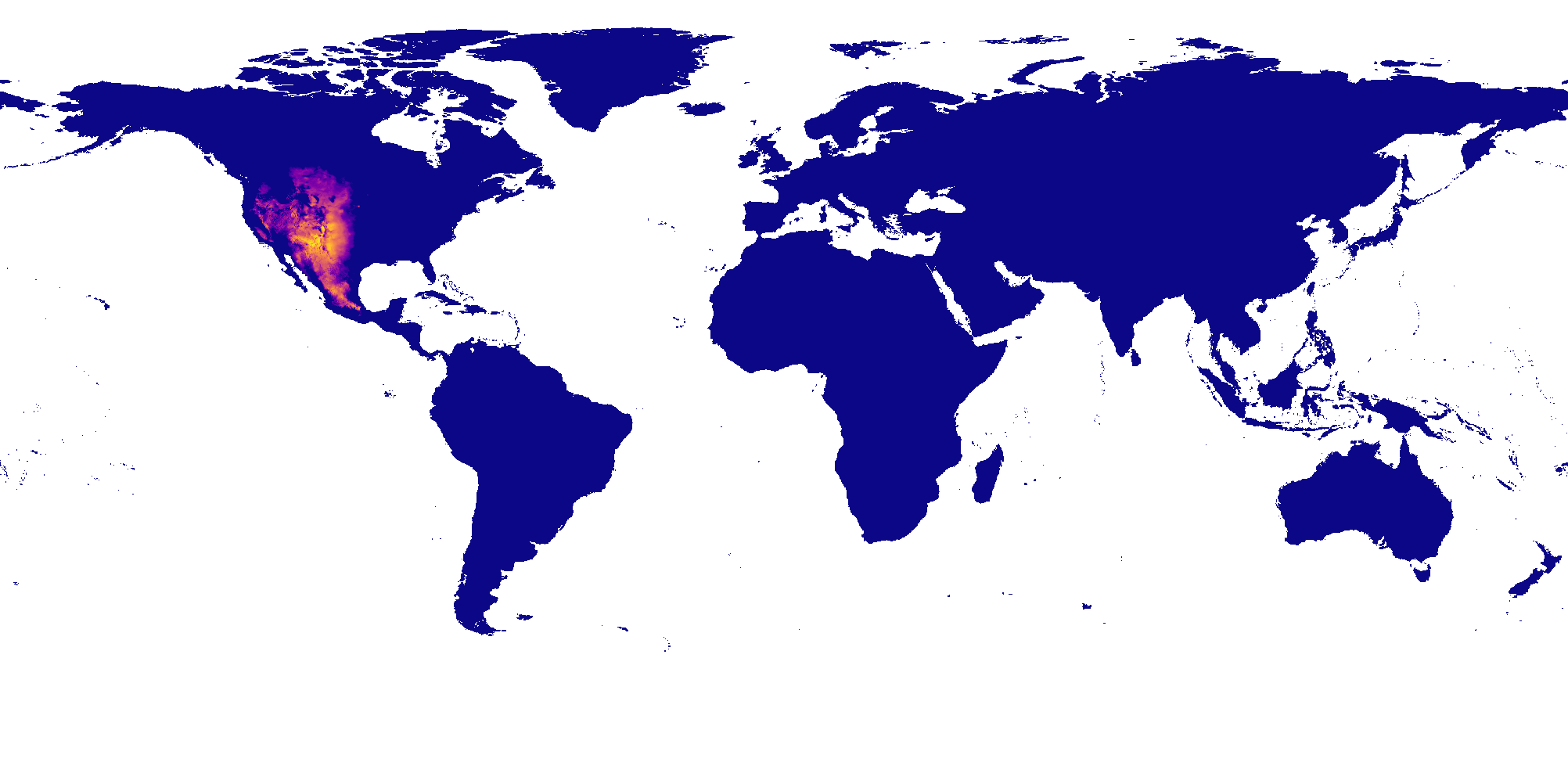}
    \end{minipage}
    \begin{minipage}{0.31\textwidth}
        \includegraphics[width=\linewidth]{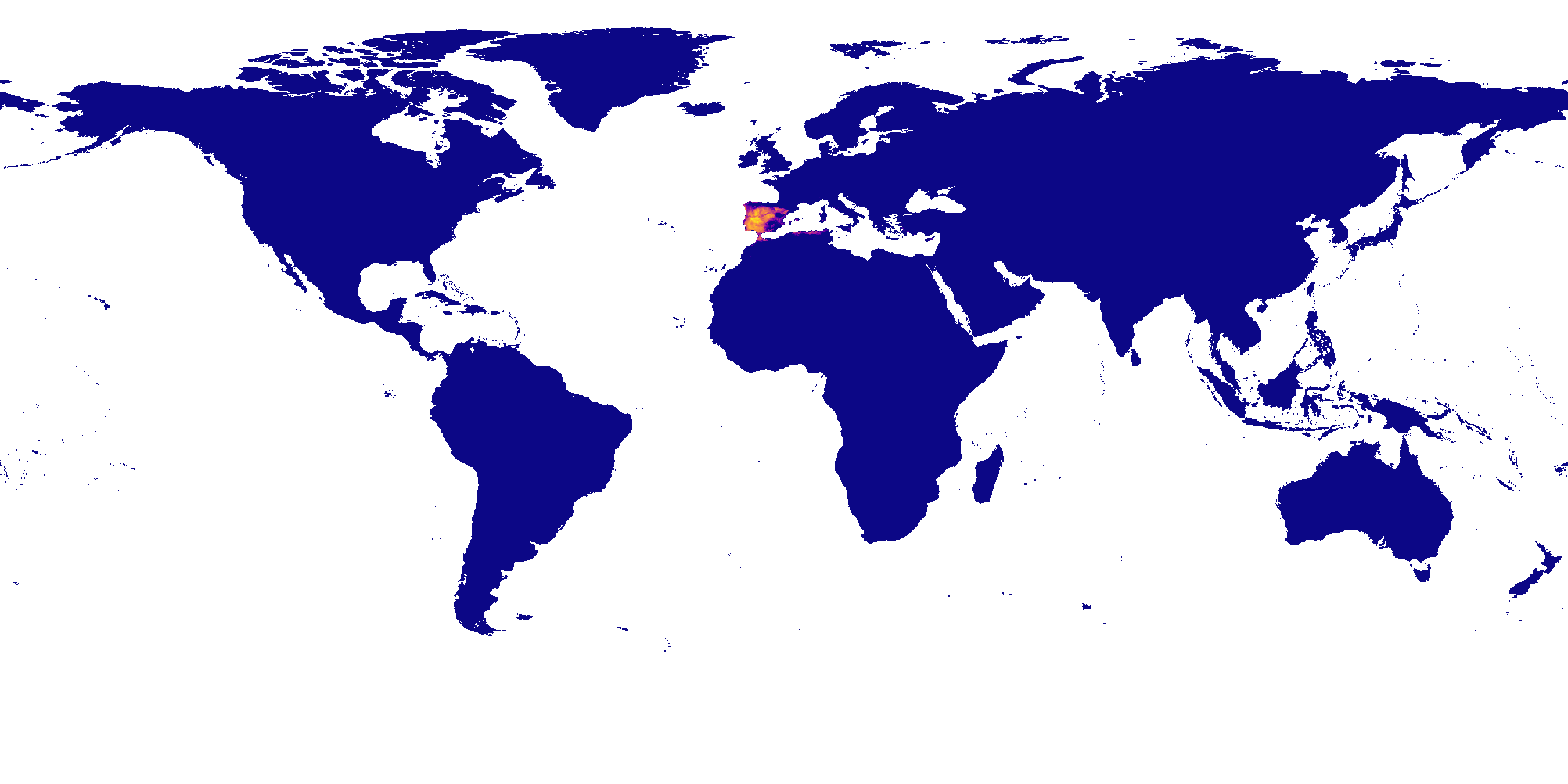}
    \end{minipage}

    \rotatebox{90}{\hspace{-15pt}\parbox{20mm}{\small\centering\texttt{one shot}}}
    \begin{minipage}{0.31\textwidth}
        \includegraphics[width=\linewidth]{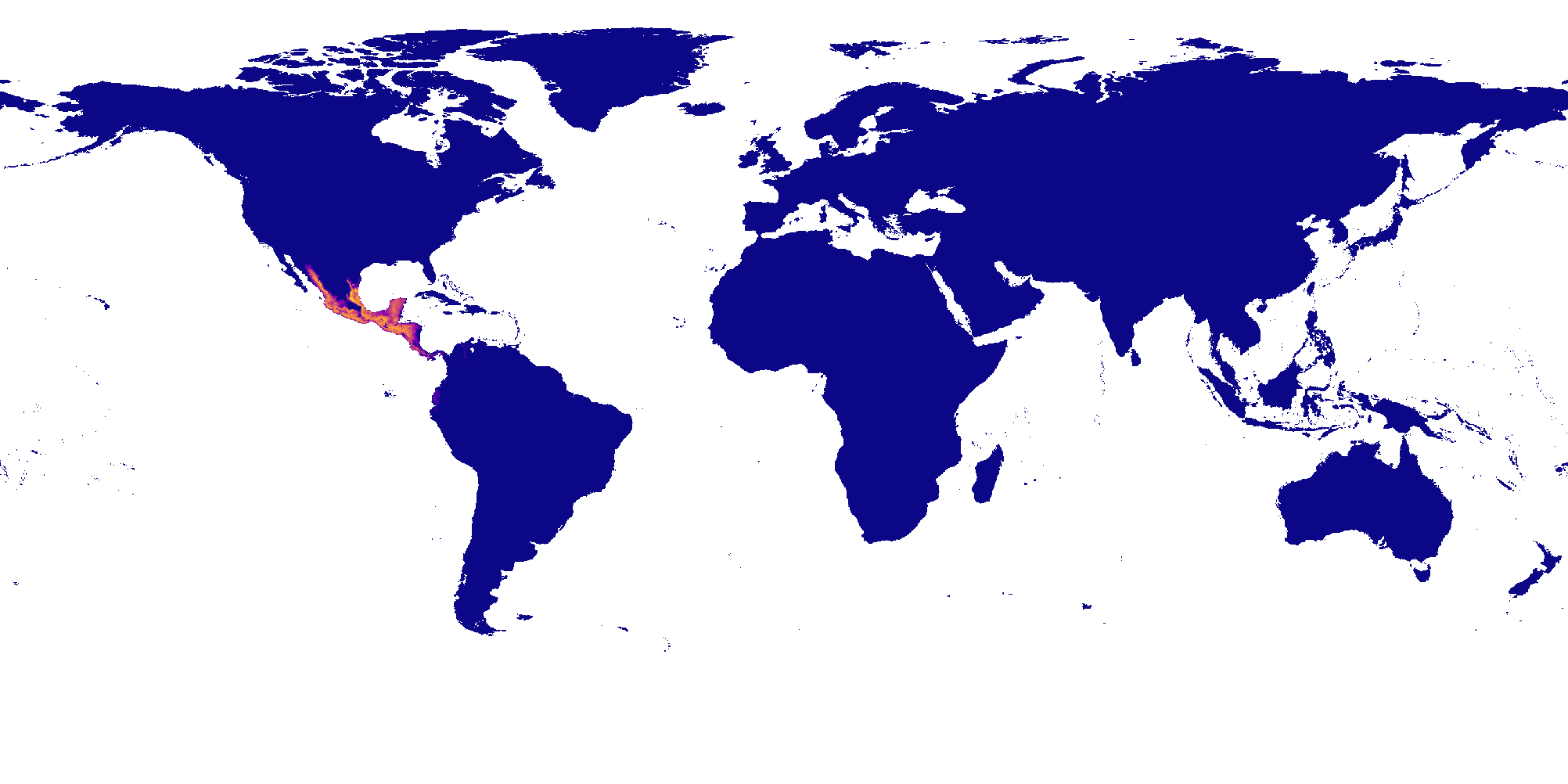} %
    \end{minipage}
    \begin{minipage}{0.31\textwidth}
        \includegraphics[width=\linewidth]{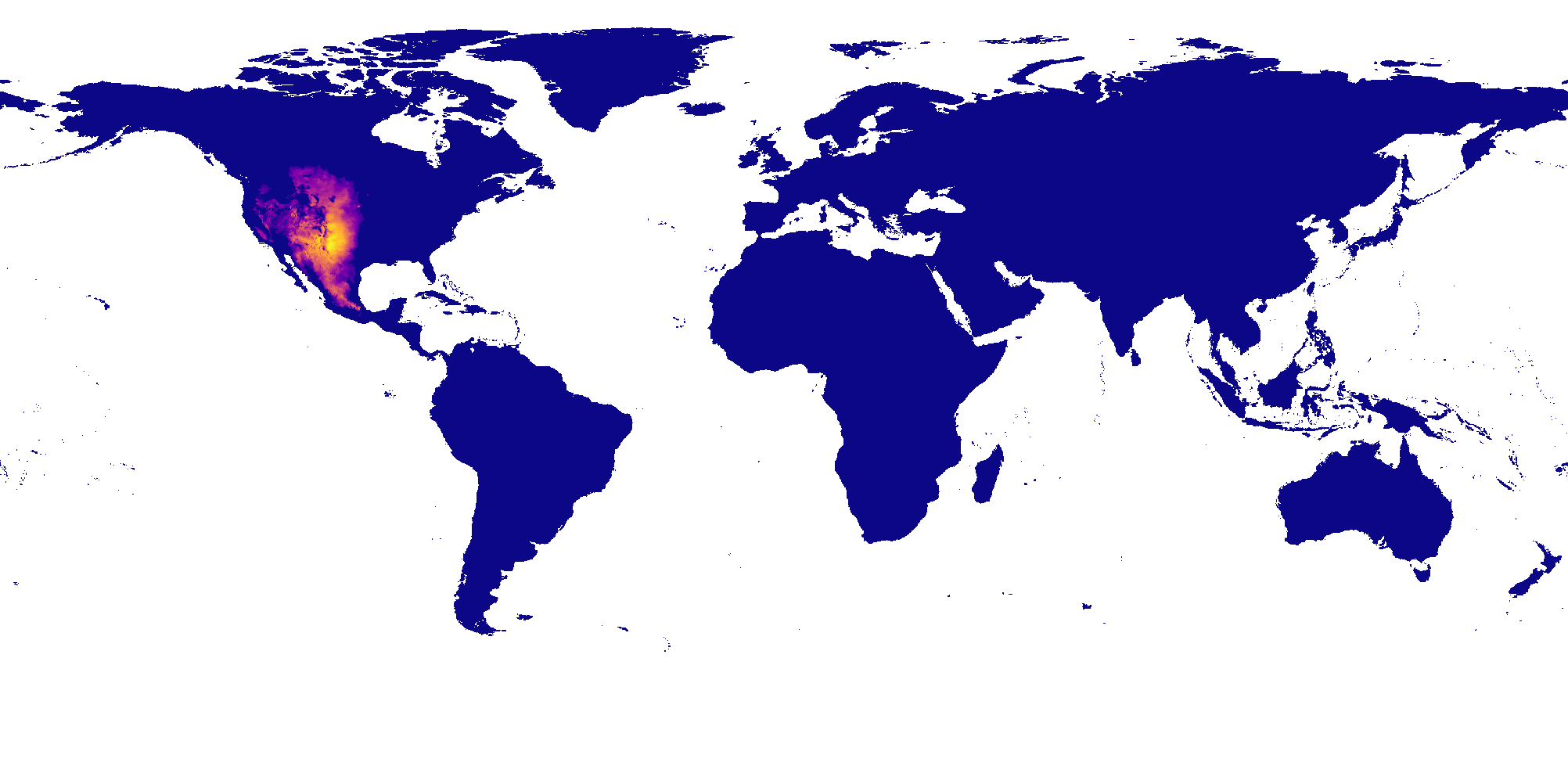} %
    \end{minipage}
    \begin{minipage}{0.31\textwidth}
        \includegraphics[width=\linewidth]{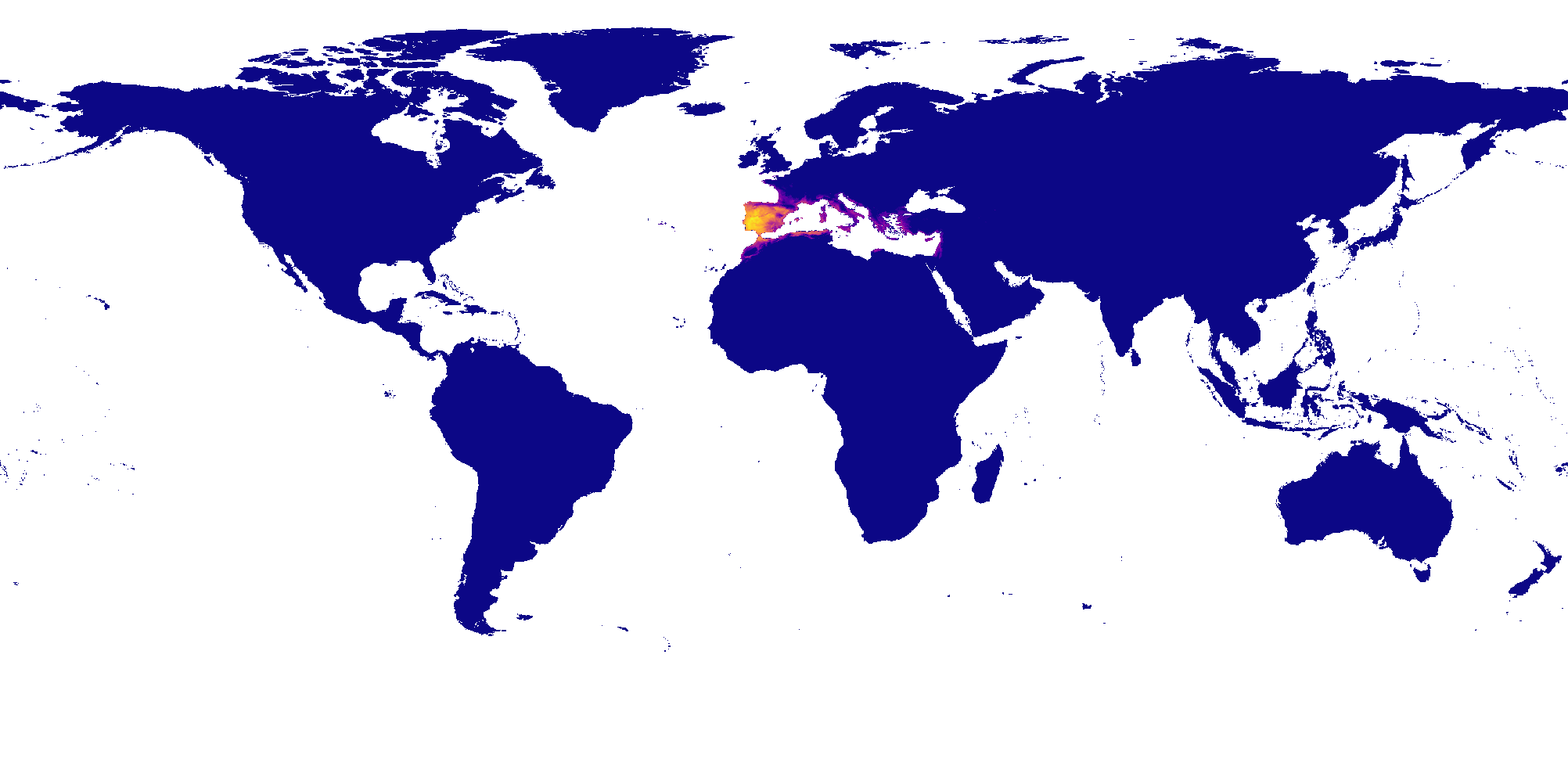} %
    \end{minipage}

    \rotatebox{90}{\hspace{-15pt}\parbox{20mm}{\small\centering\texttt{five shot}}}
    \begin{minipage}{0.31\textwidth}
        \includegraphics[width=\linewidth]{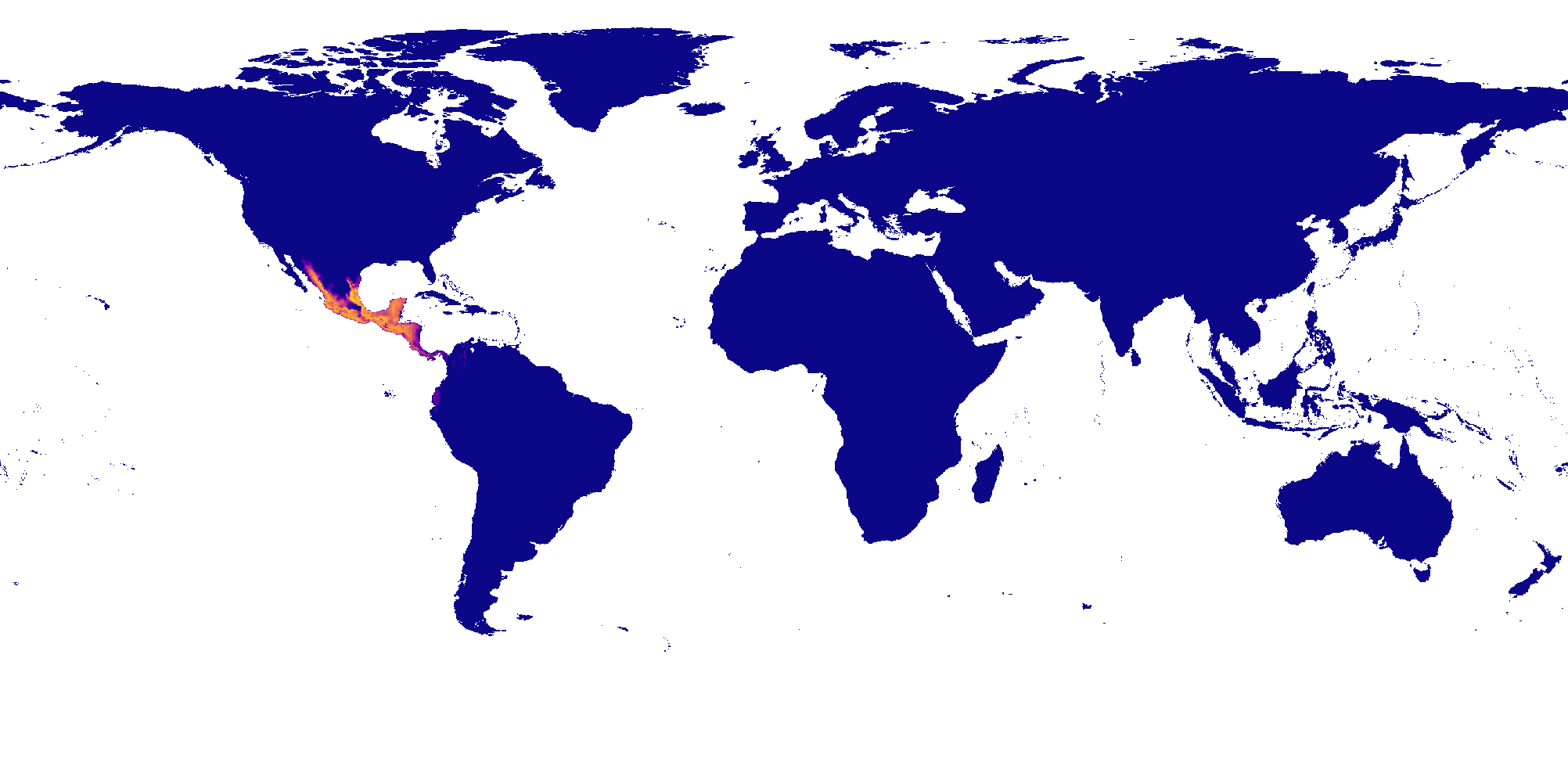}
    \end{minipage}
    \begin{minipage}{0.31\textwidth}
        \includegraphics[width=\linewidth]{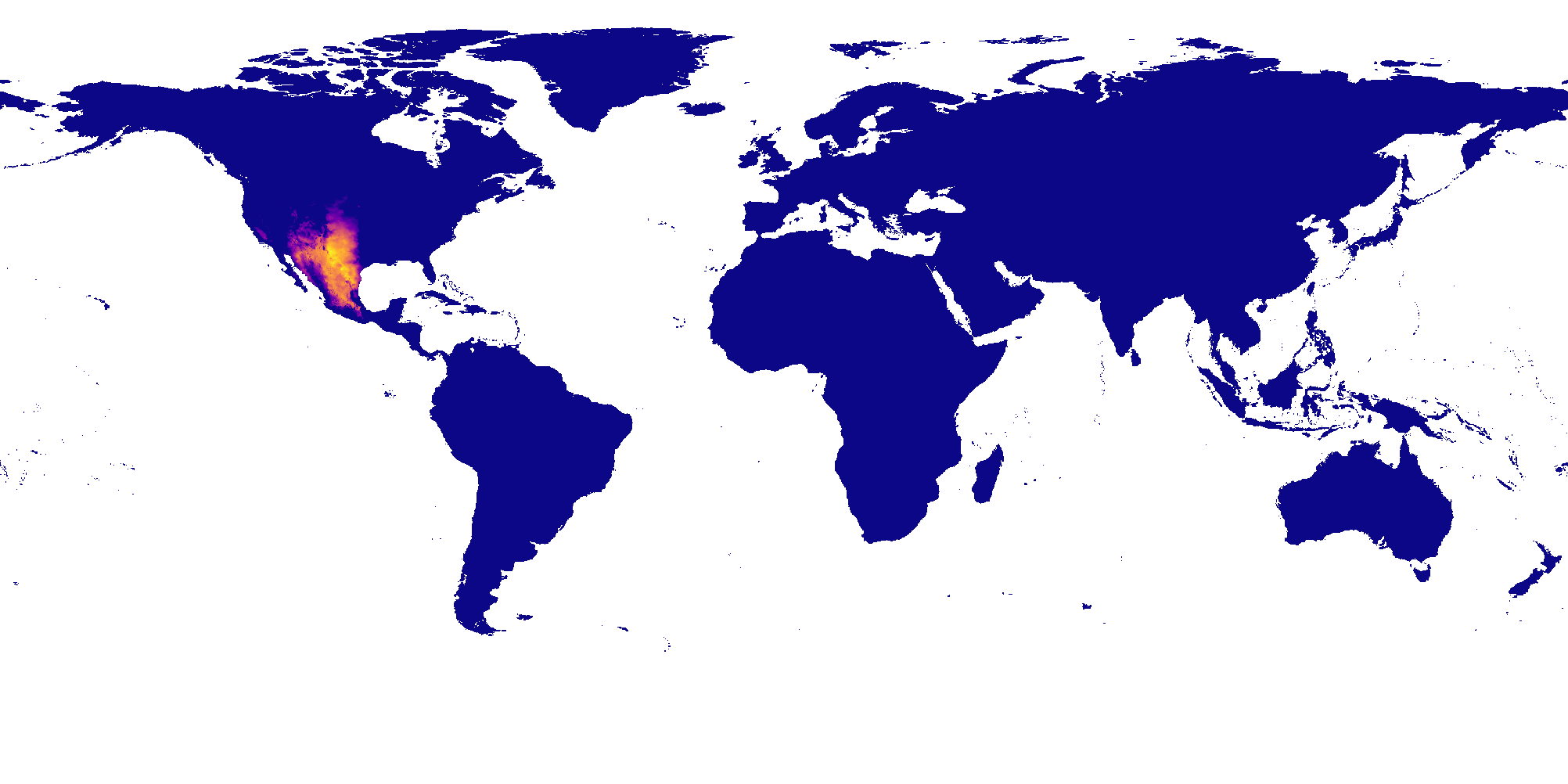}
    \end{minipage}
    \begin{minipage}{0.31\textwidth}
        \includegraphics[width=\linewidth]{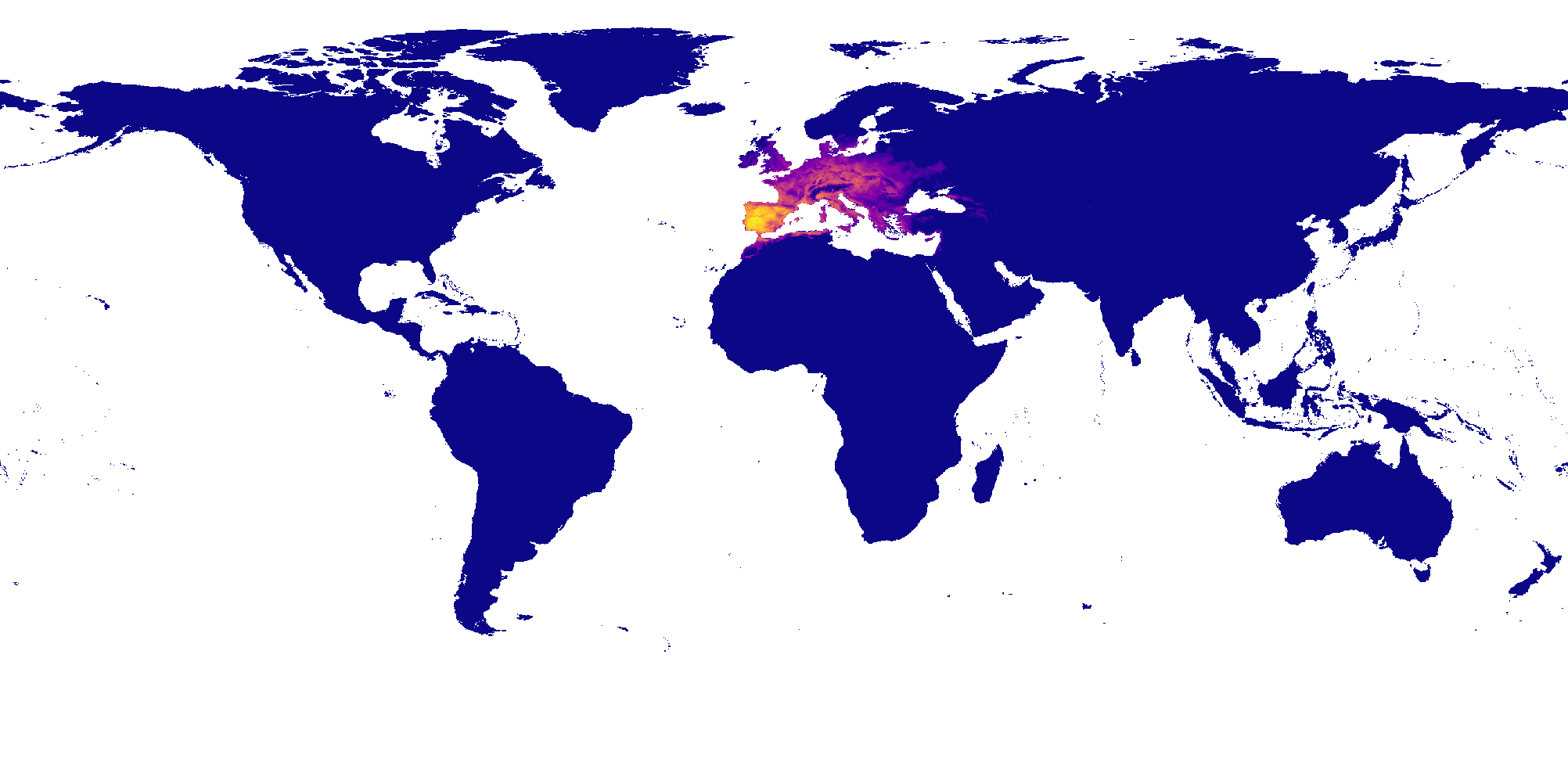}
    \end{minipage}

        \rotatebox{90}{\hspace{-15pt}\parbox{20mm}{\small\centering\texttt{ten shot}}}
    \begin{minipage}{0.31\textwidth}
        \includegraphics[width=\linewidth]{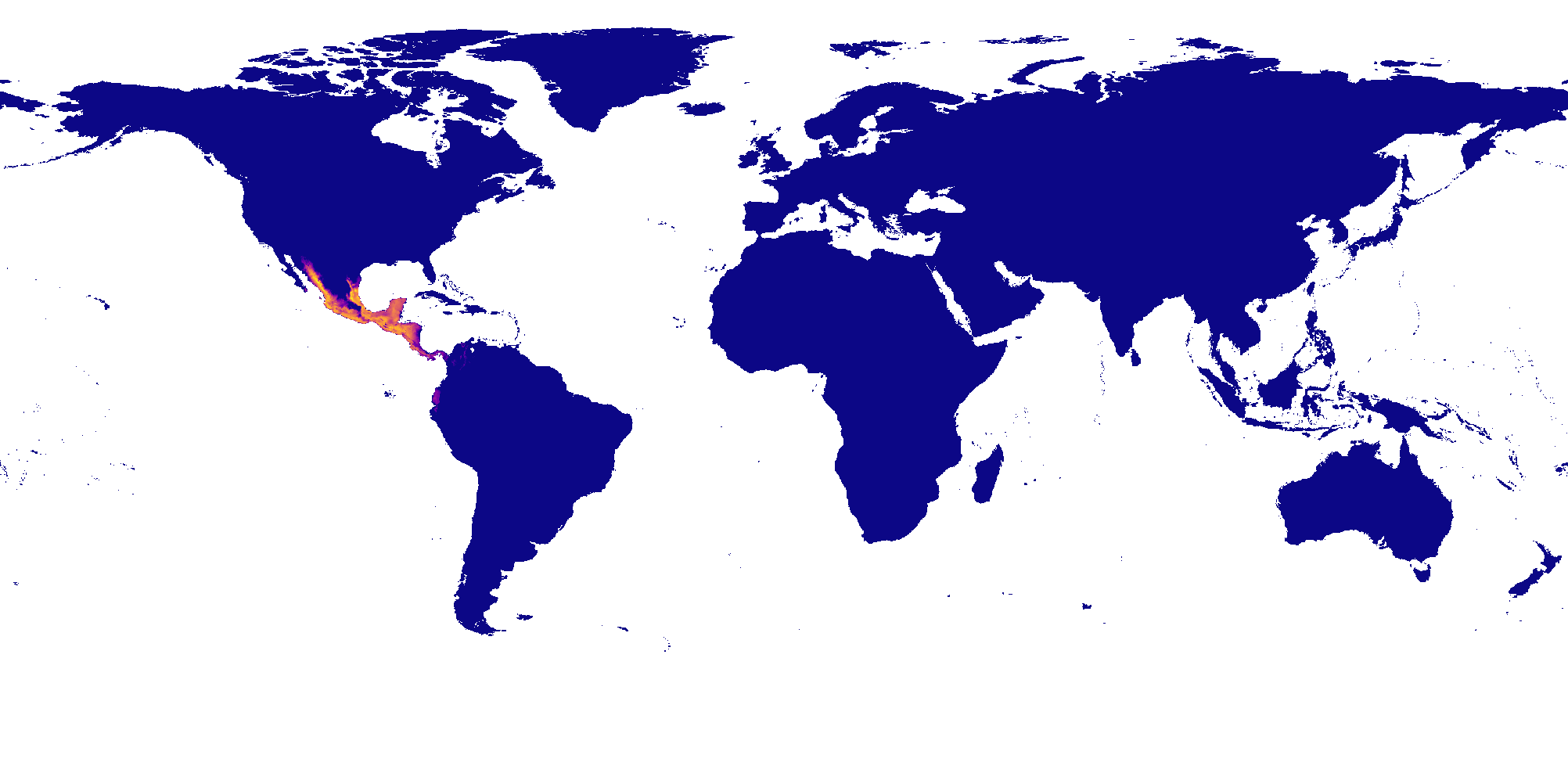}
    \end{minipage}
    \begin{minipage}{0.31\textwidth}
        \includegraphics[width=\linewidth]{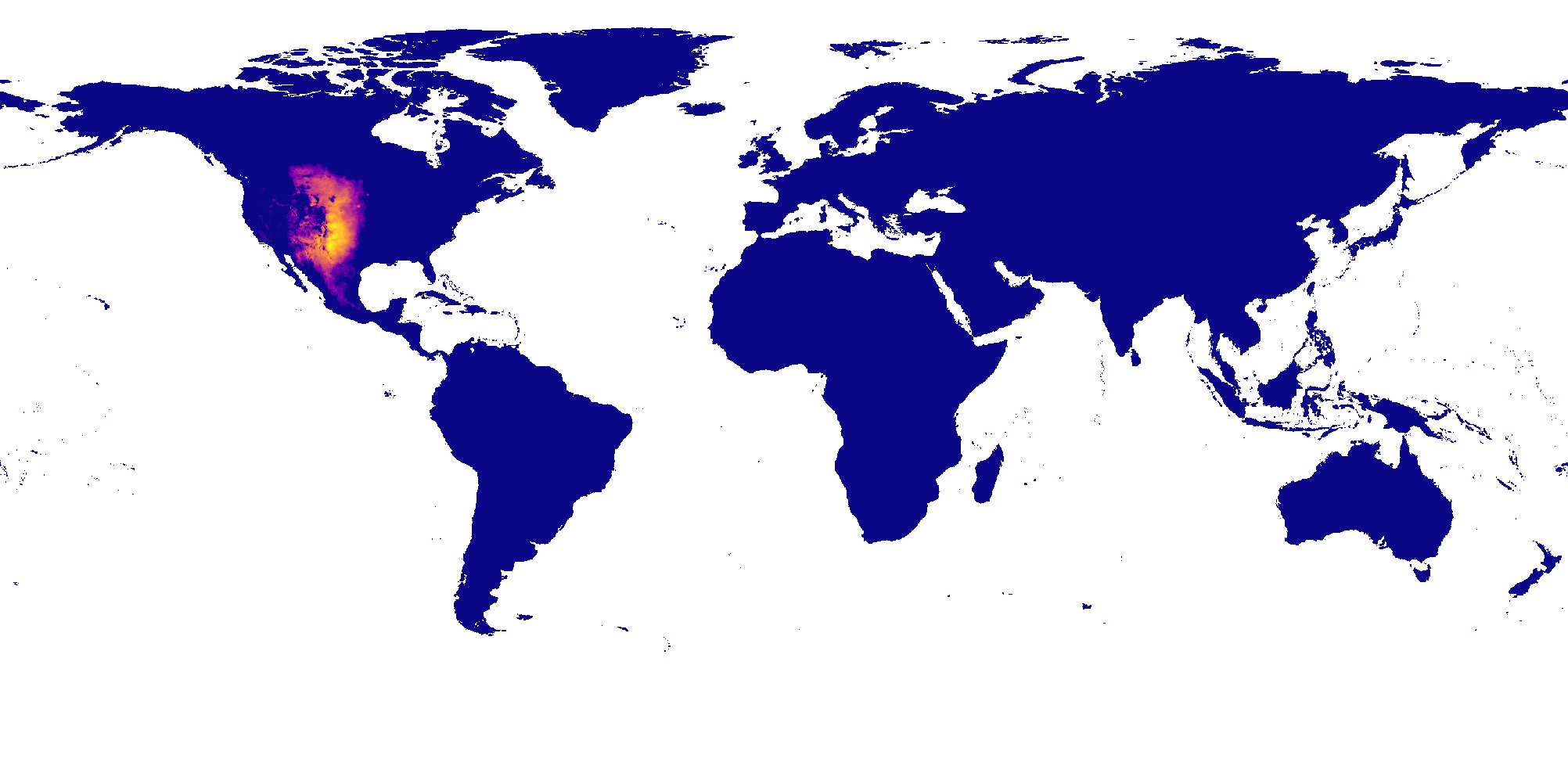}
    \end{minipage}
    \begin{minipage}{0.31\textwidth}
        \includegraphics[width=\linewidth]{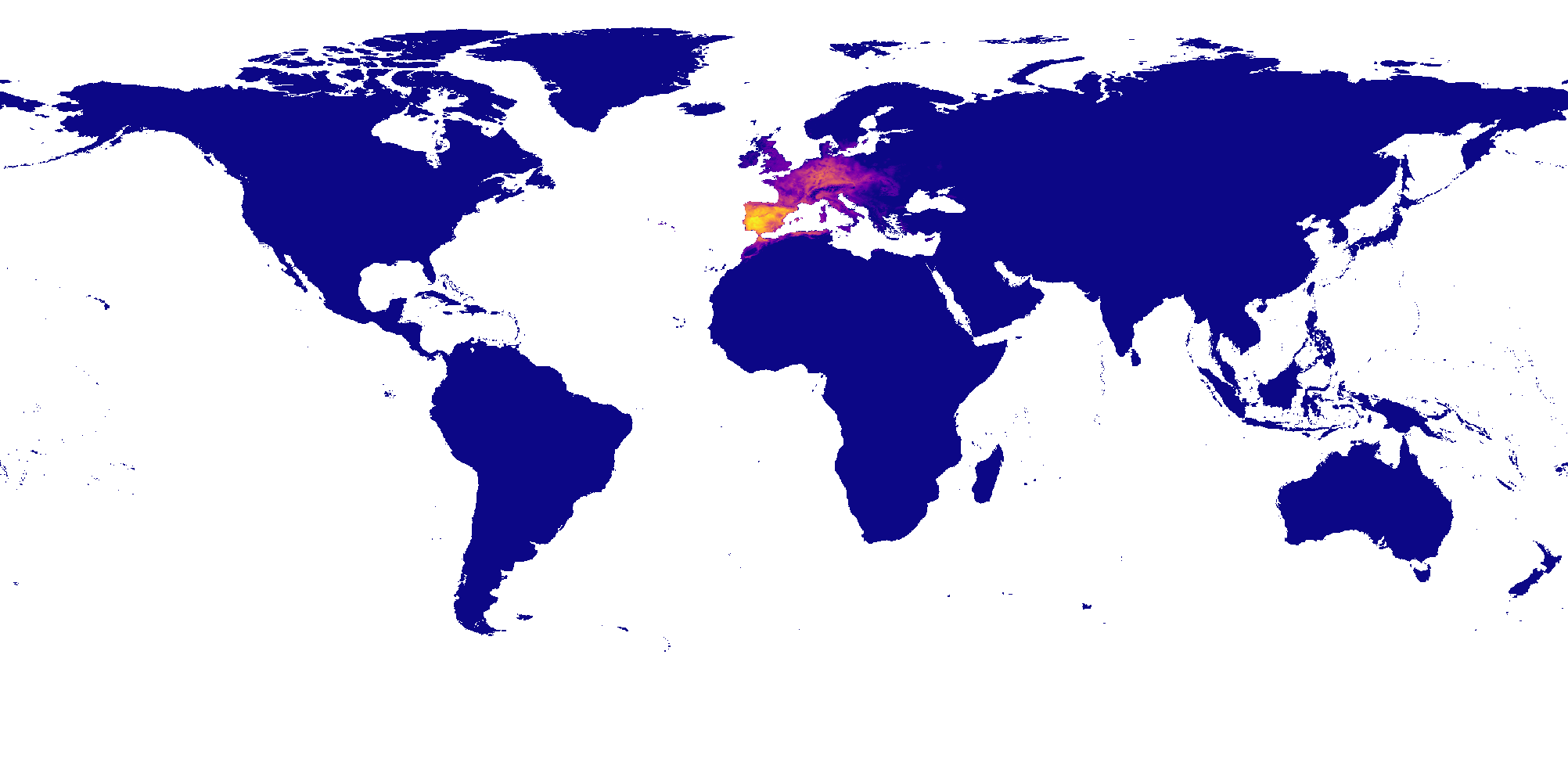}
    \end{minipage}

        \rotatebox{90}{\hspace{-15pt}\parbox{20mm}{\small\centering\texttt{Expert Range}}}
    \begin{minipage}{0.31\textwidth}
        \includegraphics[width=\linewidth]{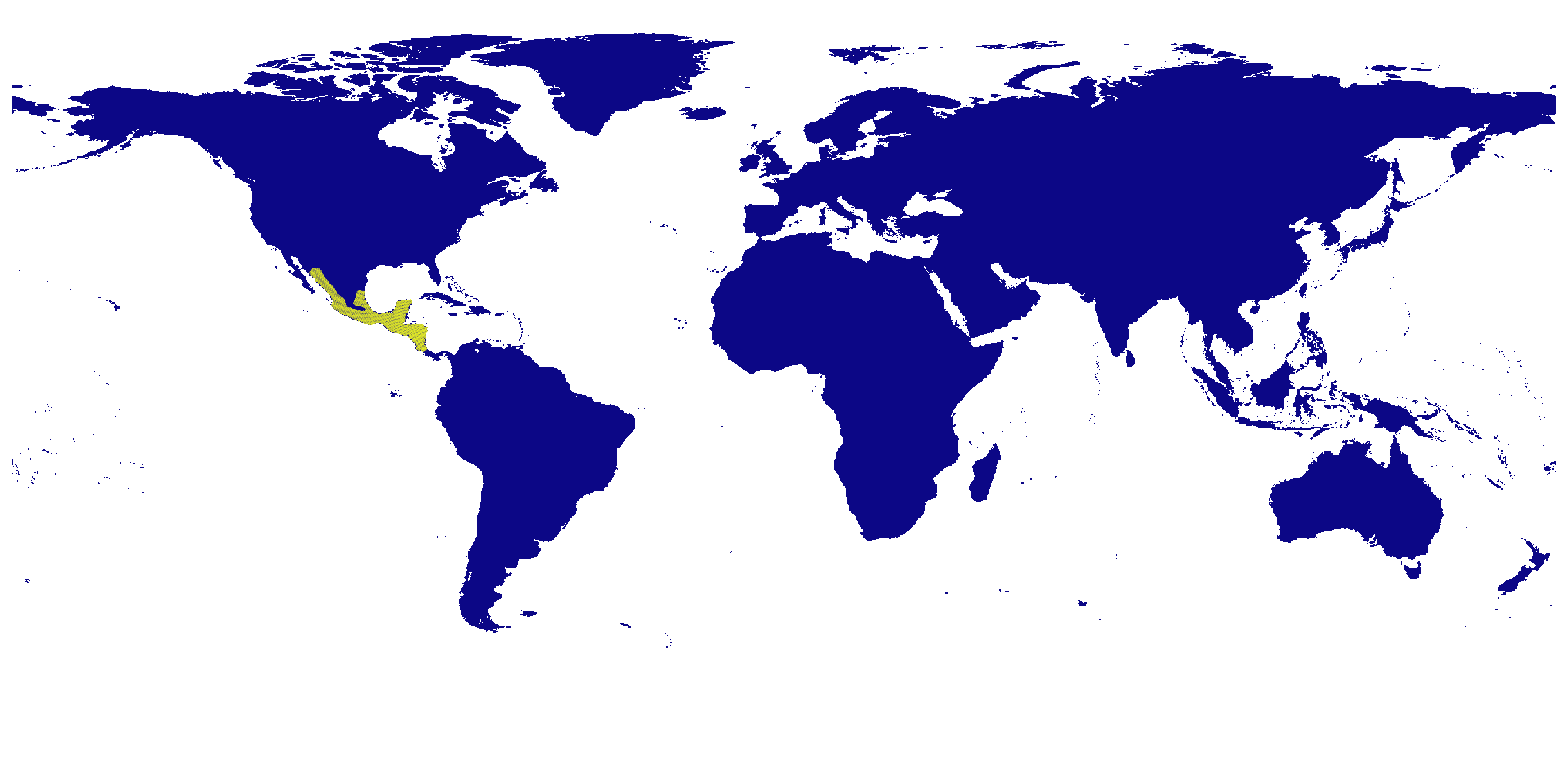}
        \centering{\small \texttt{Northern Yellow-shouldered Bat}}
    \end{minipage}
    \begin{minipage}{0.31\textwidth}
        \includegraphics[width=\linewidth]{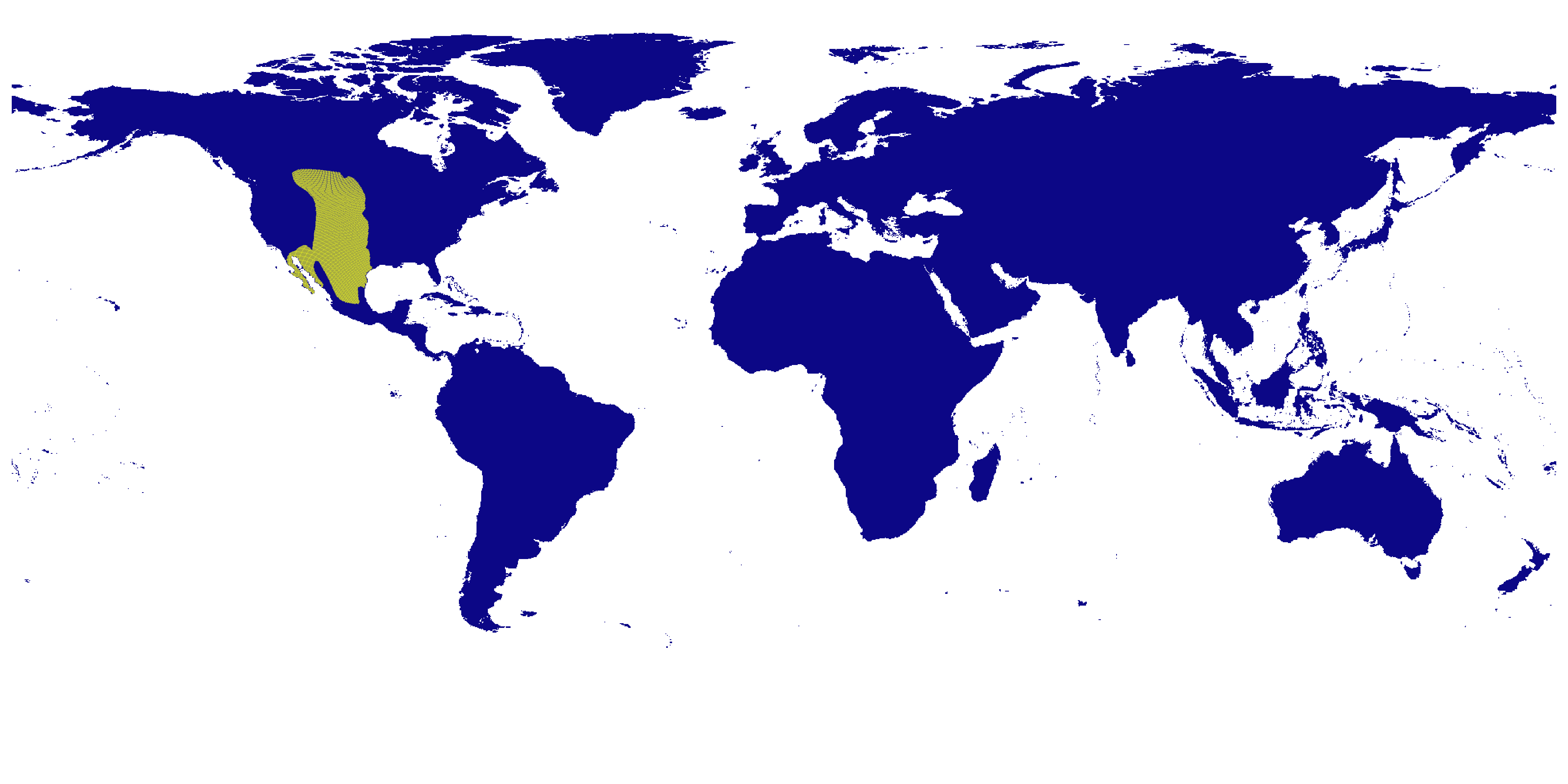}
        \centering{\small \texttt{Lark Bunting}}
    \end{minipage}
    \begin{minipage}{0.31\textwidth}
        \includegraphics[width=\linewidth]{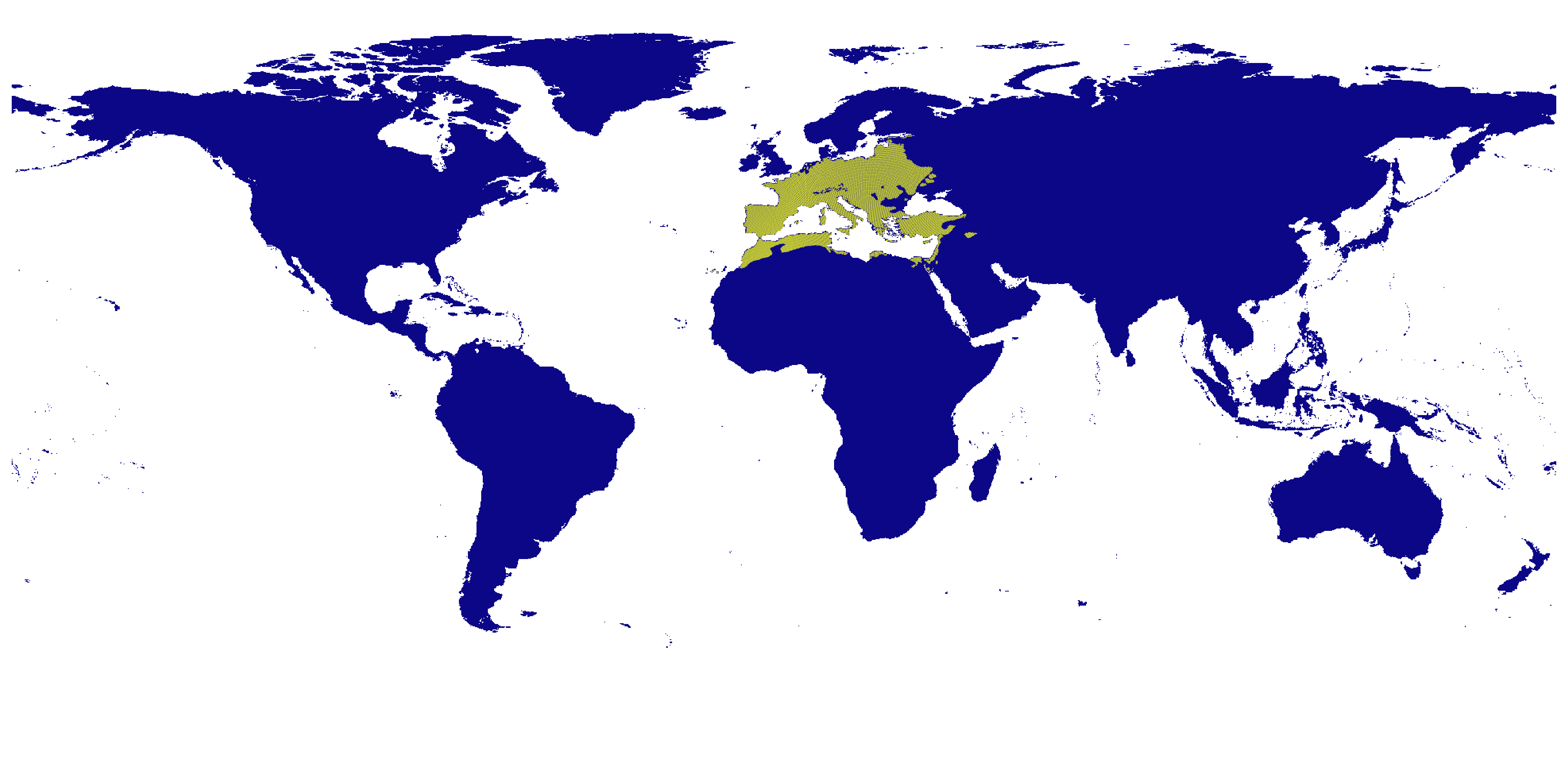}
        \centering{\small \texttt{European Serin}}
    \end{minipage}
    
    \caption{\textbf{Visualization of Zero-Shot and Few-Shot Range Estimates.} Range estimates obtained from text only (zero-shot) for three different species is shown on the top row. Range estimates with one, five, and ten observations are shown in rows two, three, and four respectively, while the ground truth range shown at the bottom row along with the name of the species. Zero-shot predictions were made from the `range' text. Zoom in for details.    
    }
    \label{fig:qualitative_low_shot}
    \vspace{-10pt}
\end{figure}

In Figure~\ref{fig:qualitative_low_shot} we show the range estimates obtained from text (zero-shot setting) as well as a few examples (few-shot setting corresponding to Table~\ref{tab:main-results} from the main paper) for three different species, namely the \texttt{Northern Yellow-shouldered Bat}, \texttt{Lark Bunting}, and \texttt{European Serin} along with the associated range maps curated by experts. 

\begin{figure}
    \centering
    \rotatebox{90}{\hspace{-15pt}\parbox{20mm}{\small\centering\texttt{Striated Babbler}}}
    \begin{minipage}{0.31\textwidth}
        \includegraphics[width=\linewidth]{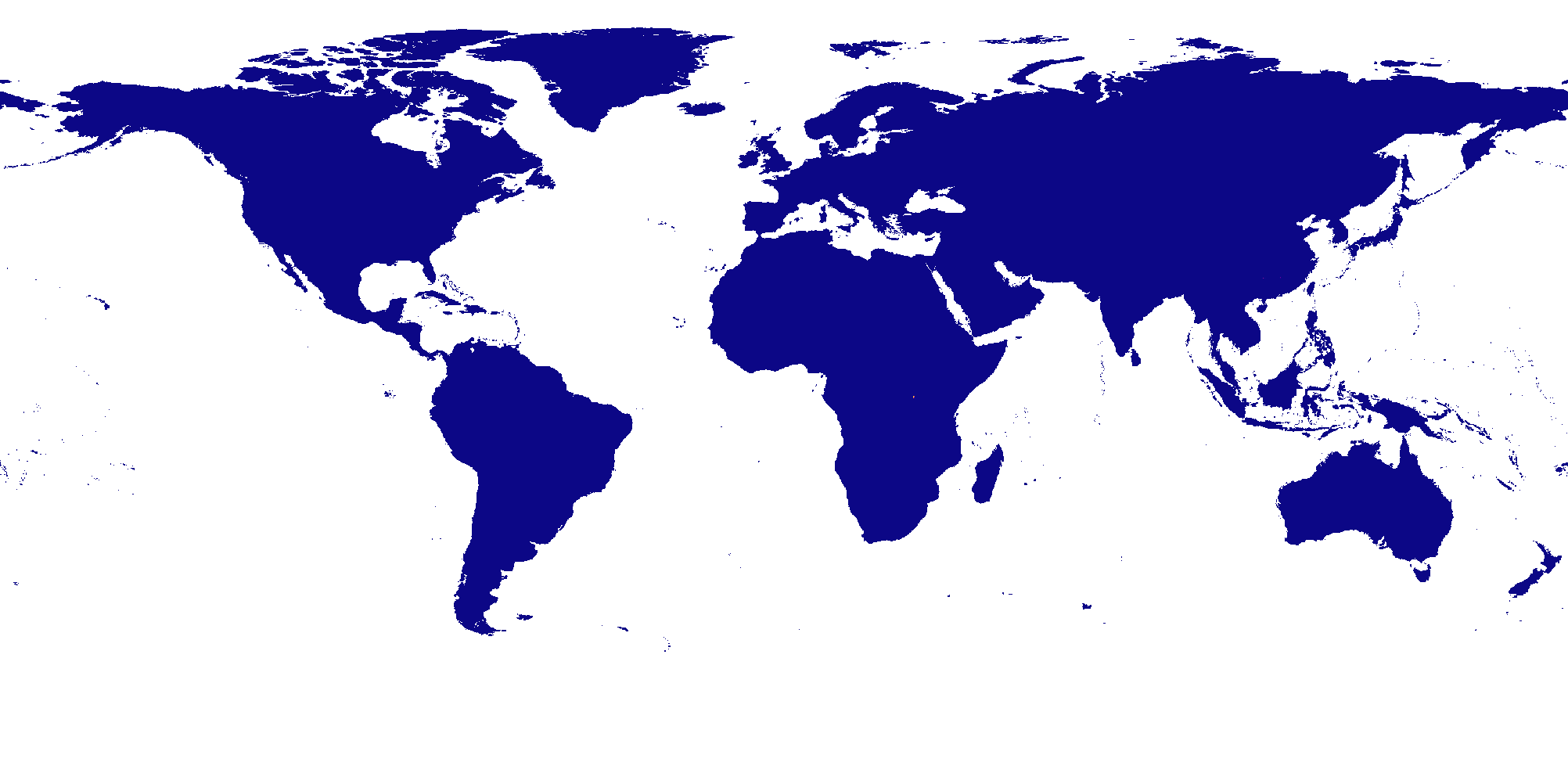}
    \end{minipage}
    \begin{minipage}{0.31\textwidth}
        \includegraphics[width=\linewidth]{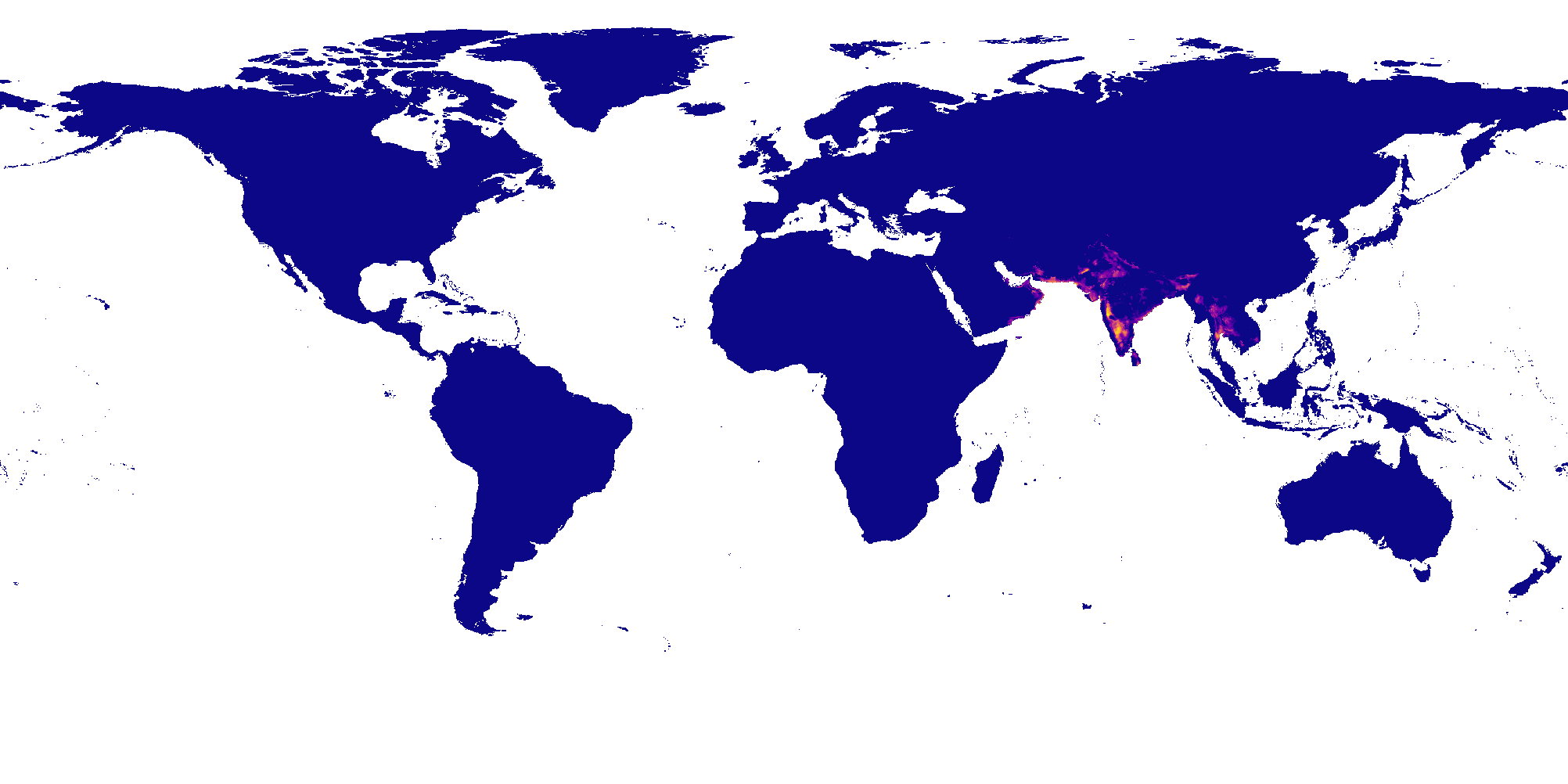}
    \end{minipage}
    \begin{minipage}{0.31\textwidth}
        \includegraphics[width=\linewidth]{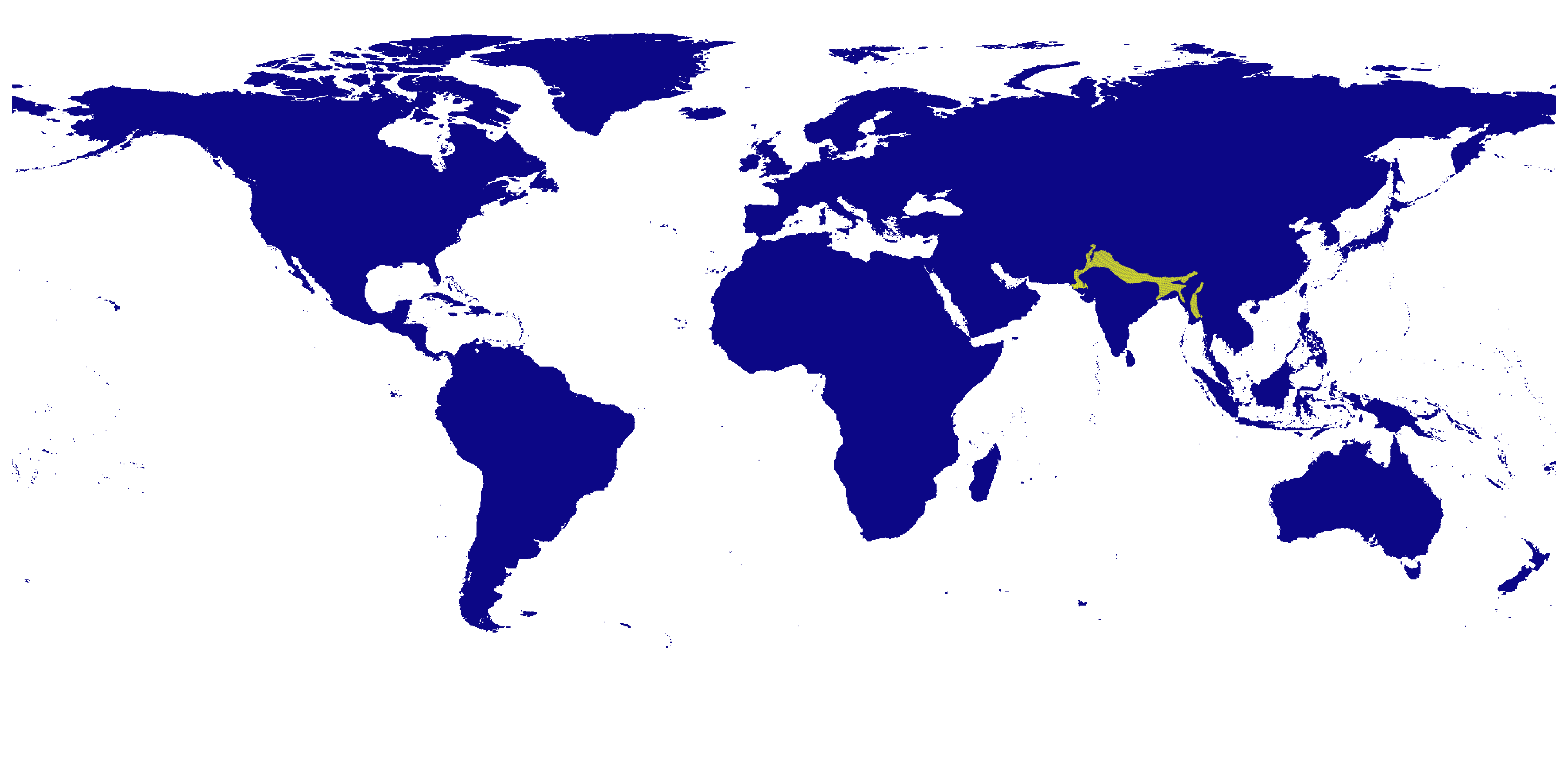}
    \end{minipage}

    \rotatebox{90}{\hspace{-15pt}\parbox{20mm}{\small\centering\texttt{Striped Sticky Frog}}}
    \begin{minipage}{0.31\textwidth}
        \includegraphics[width=\linewidth]{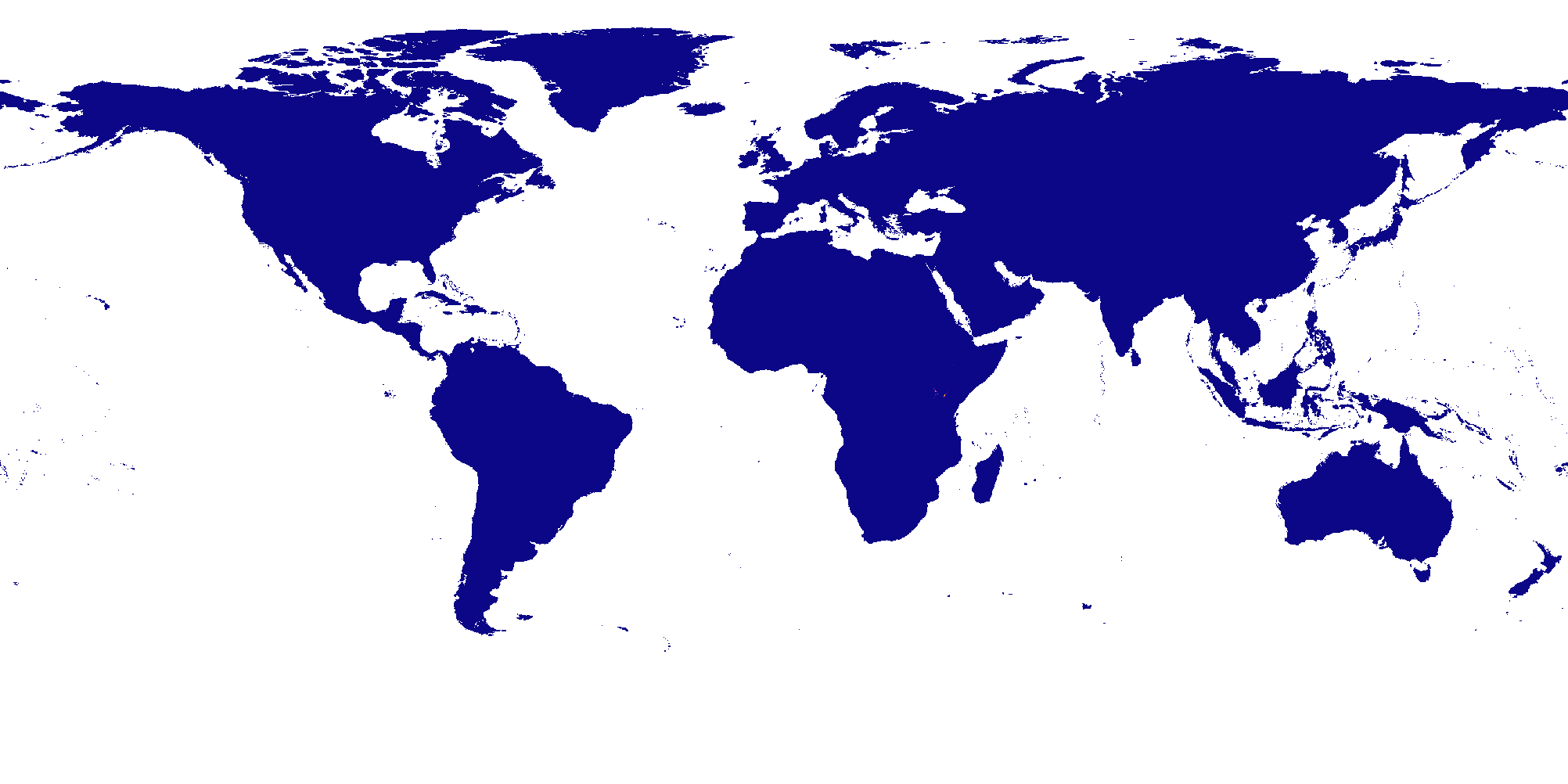} %
    \end{minipage}
    \begin{minipage}{0.31\textwidth}
        \includegraphics[width=\linewidth]{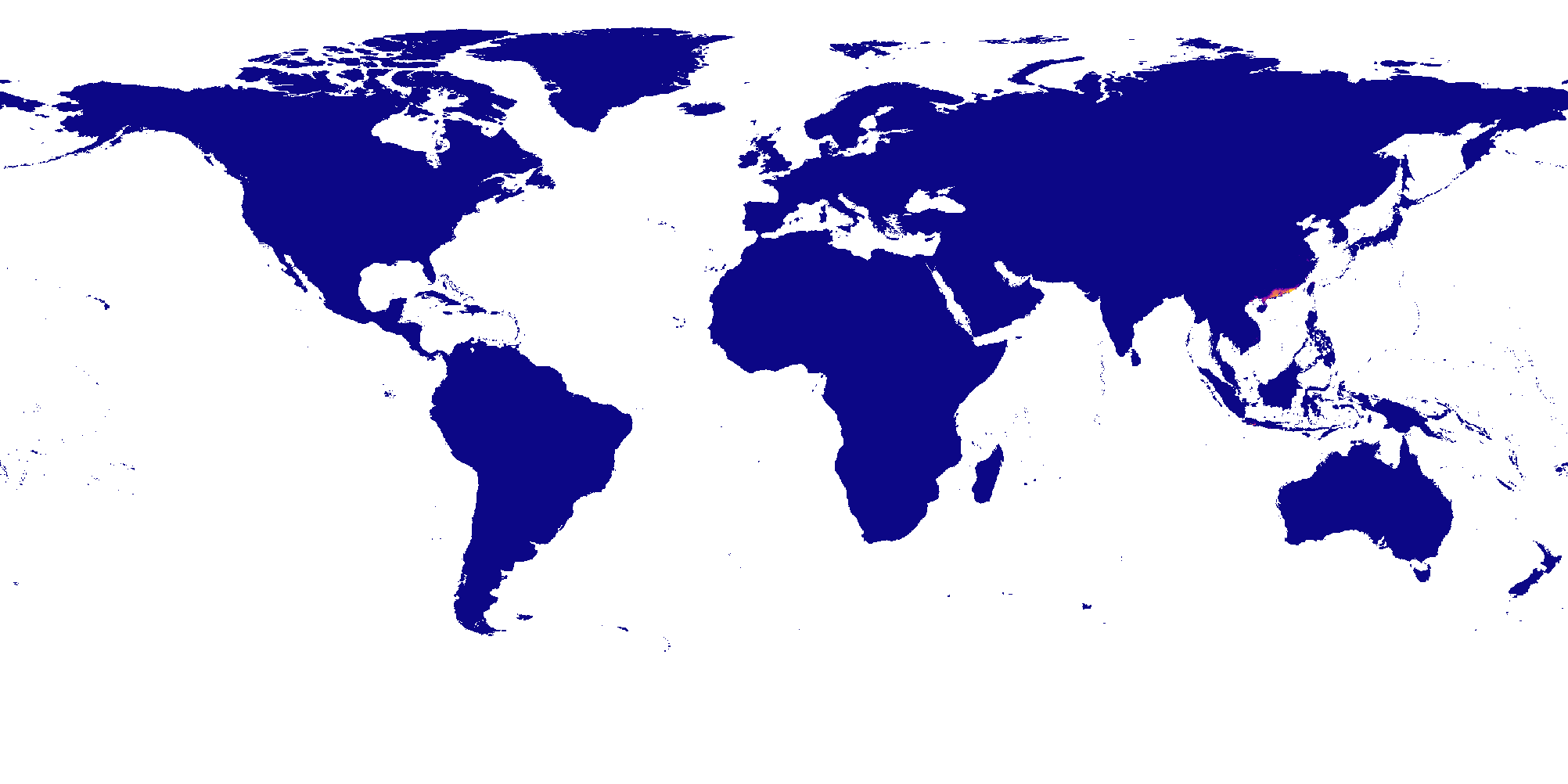} %
    \end{minipage}
    \begin{minipage}{0.31\textwidth}
        \includegraphics[width=\linewidth]{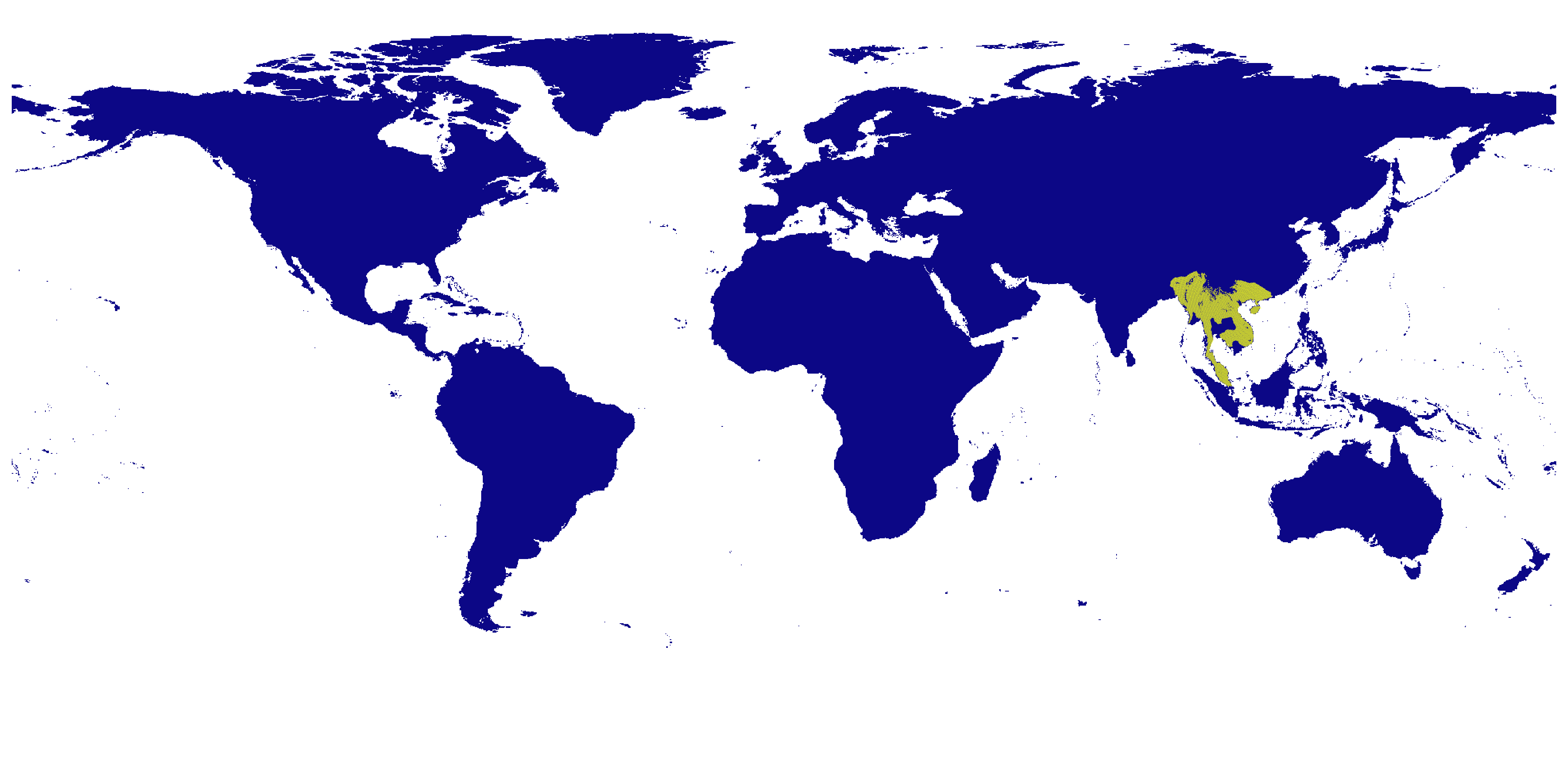} %
    \end{minipage}

    \rotatebox{90}{\hspace{-15pt}\parbox{20mm}{\small\centering\texttt{Common Hawk-Cuckoo}}}
    \begin{minipage}{0.31\textwidth}
        \includegraphics[width=\linewidth]{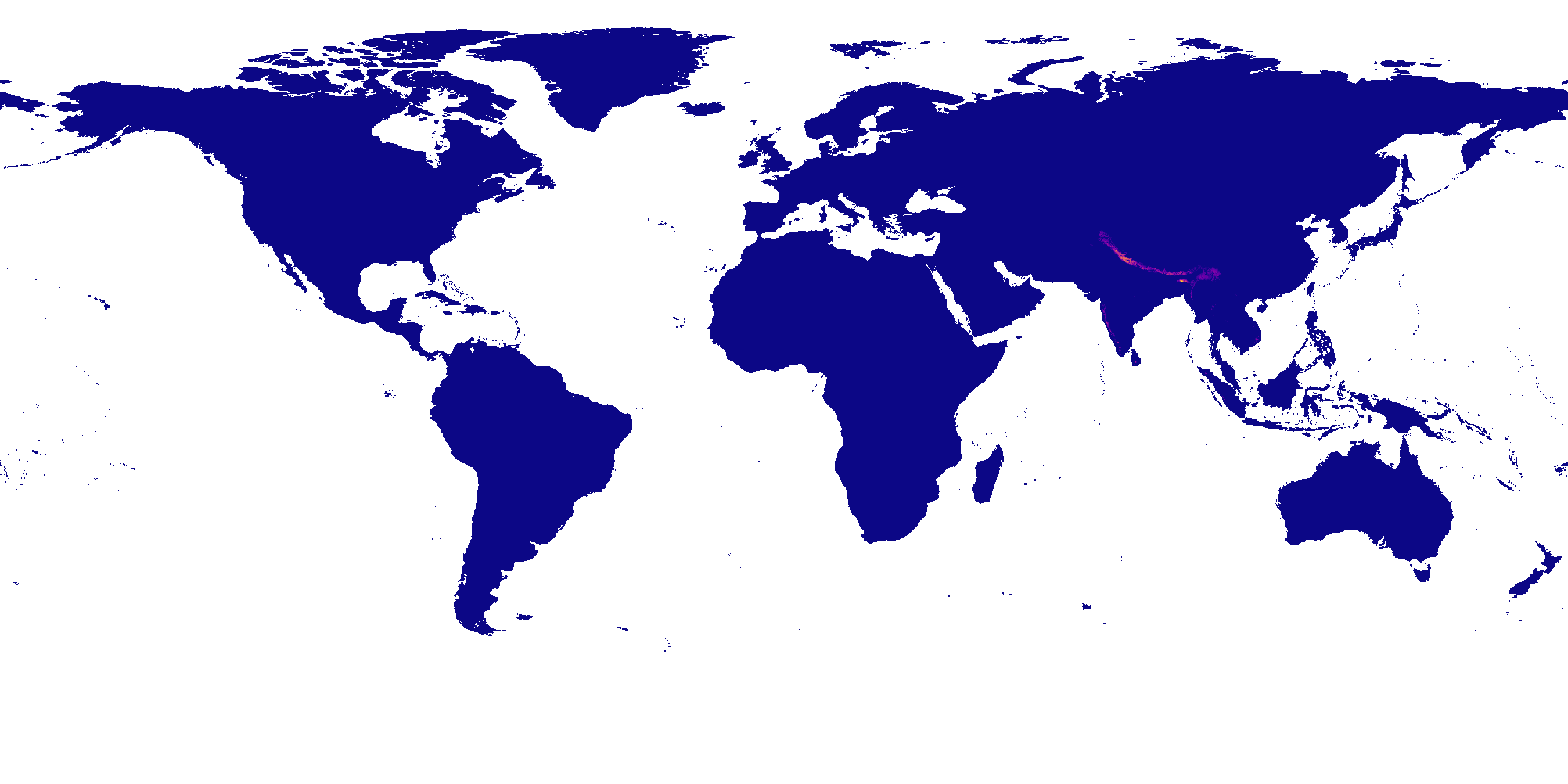}
    \end{minipage}
    \begin{minipage}{0.31\textwidth}
        \includegraphics[width=\linewidth]{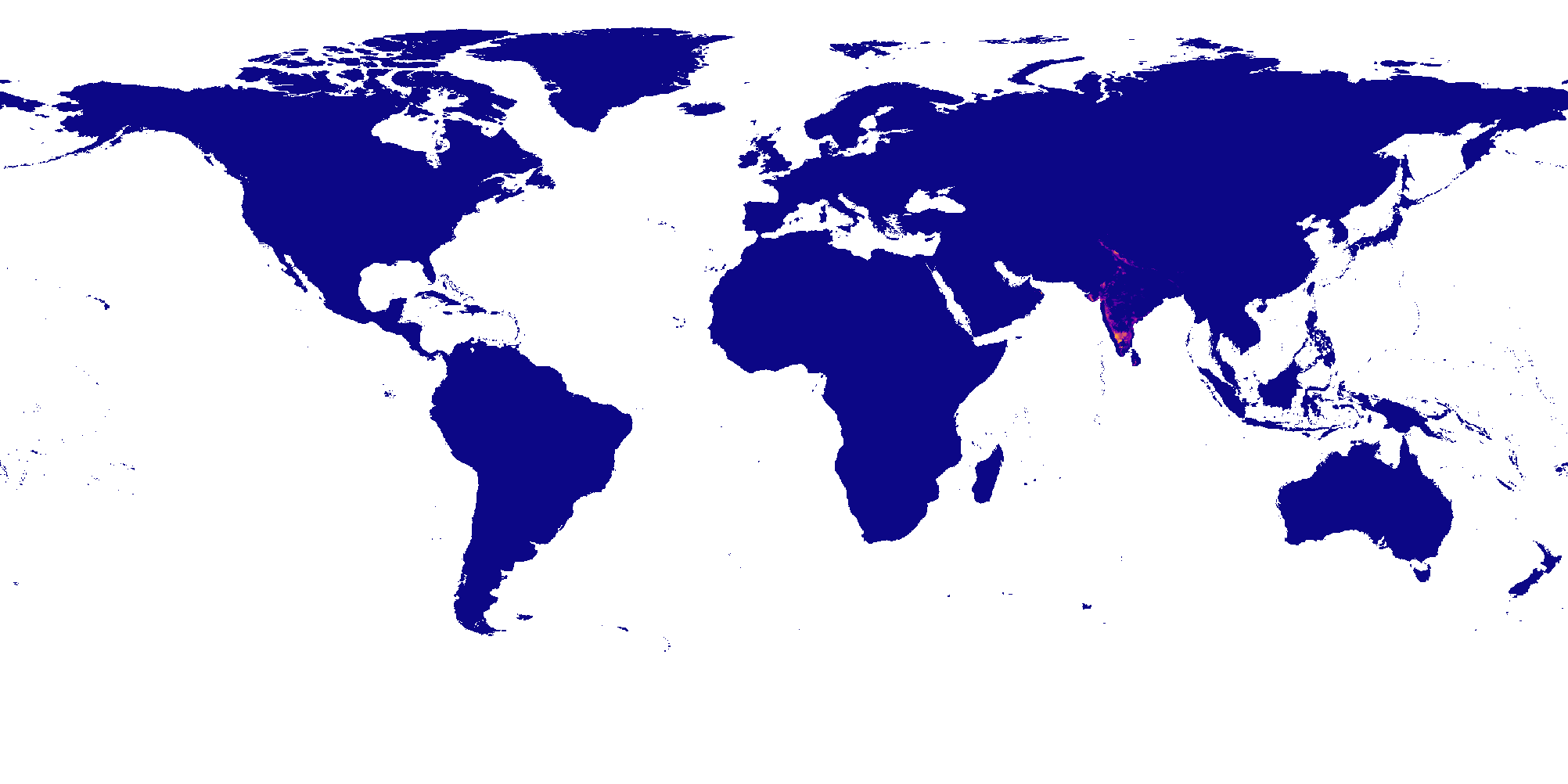}
    \end{minipage}
    \begin{minipage}{0.31\textwidth}
        \includegraphics[width=\linewidth]{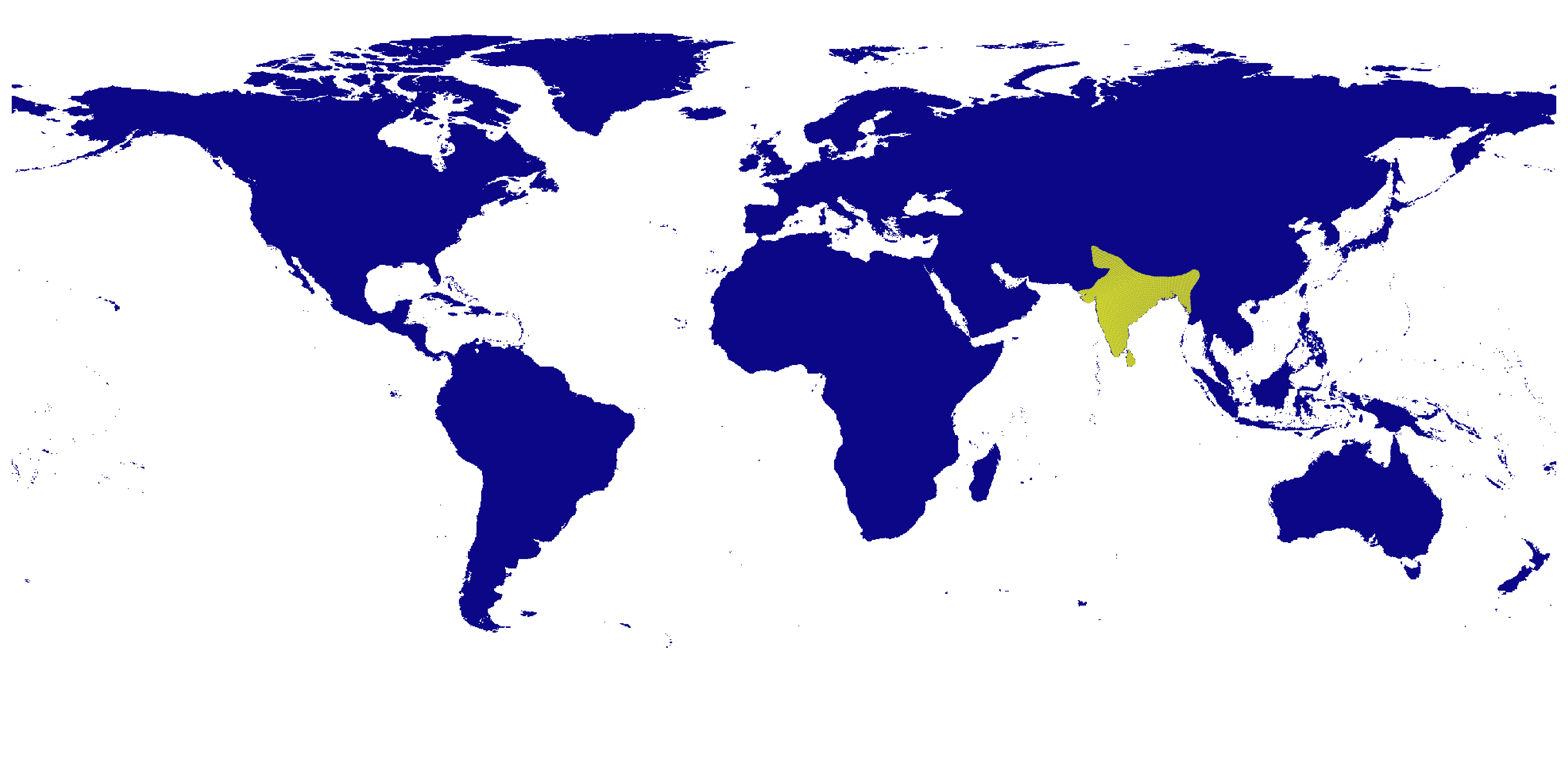}
    \end{minipage}

        \rotatebox{90}{\hspace{-15pt}\parbox{20mm}{\small\centering\texttt{Cape Griffon}}}
    \begin{minipage}{0.31\textwidth}
        \includegraphics[width=\linewidth]{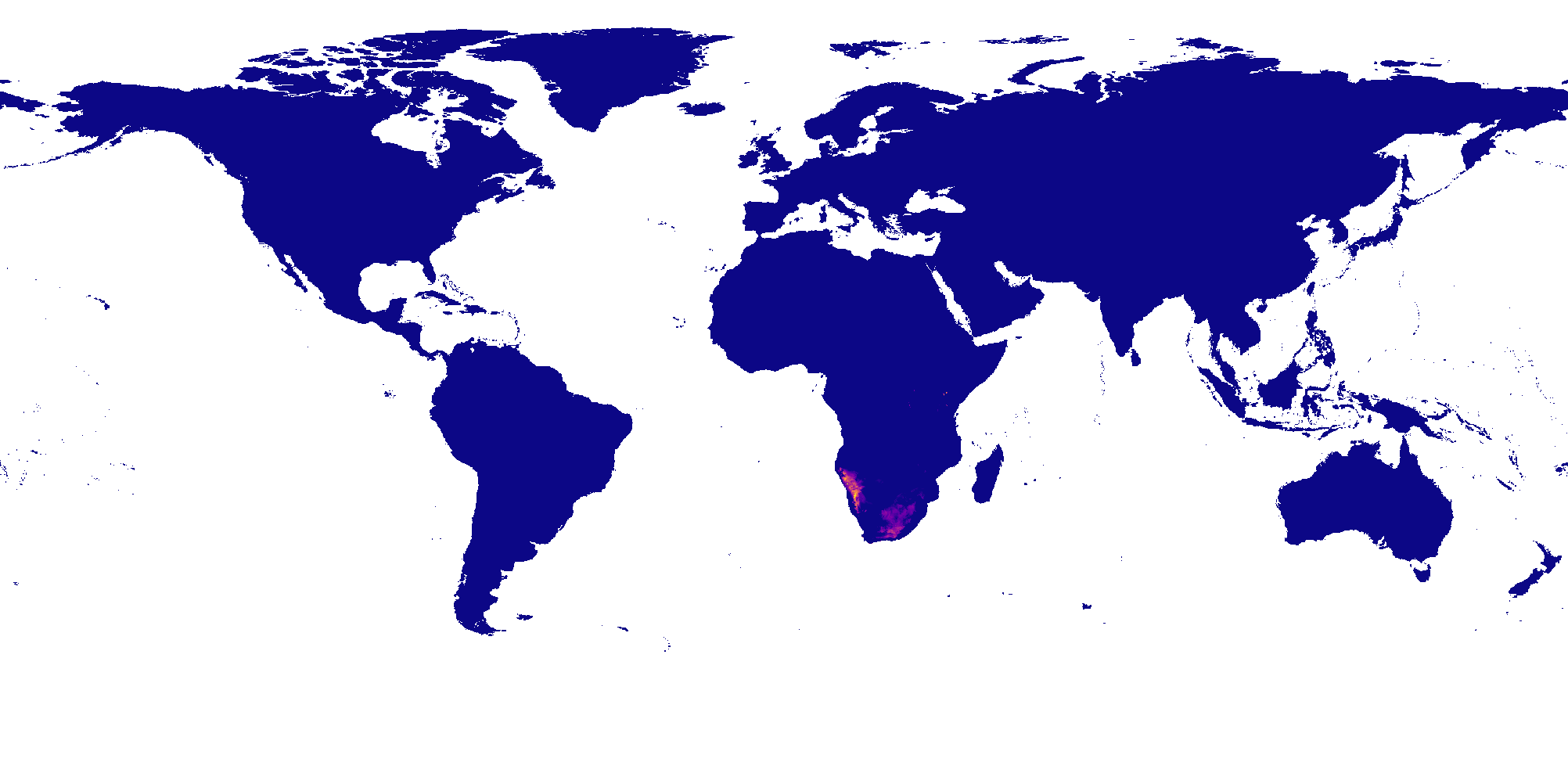}
    \end{minipage}
    \begin{minipage}{0.31\textwidth}
        \includegraphics[width=\linewidth]{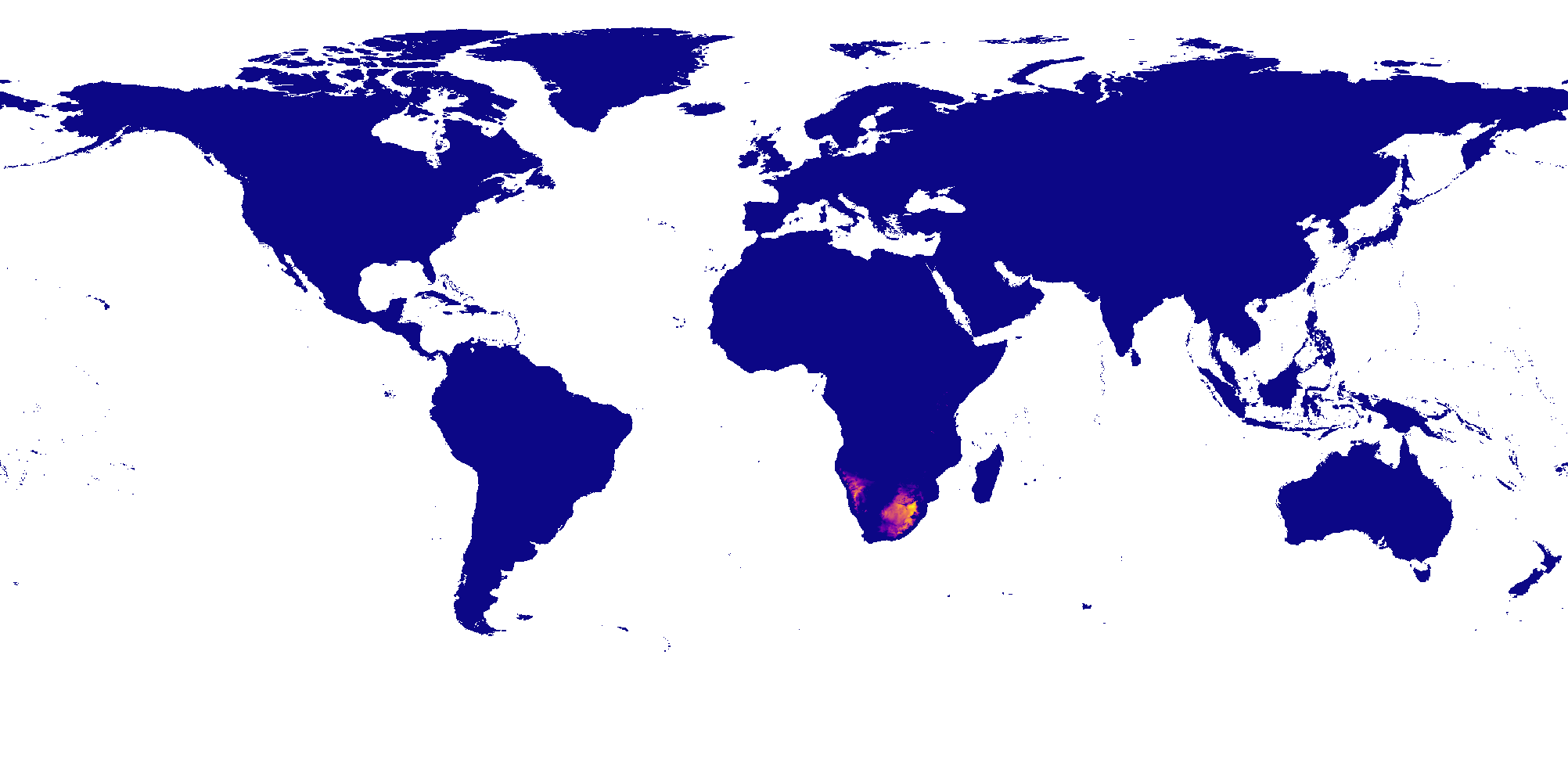}
    \end{minipage}
    \begin{minipage}{0.31\textwidth}
        \includegraphics[width=\linewidth]{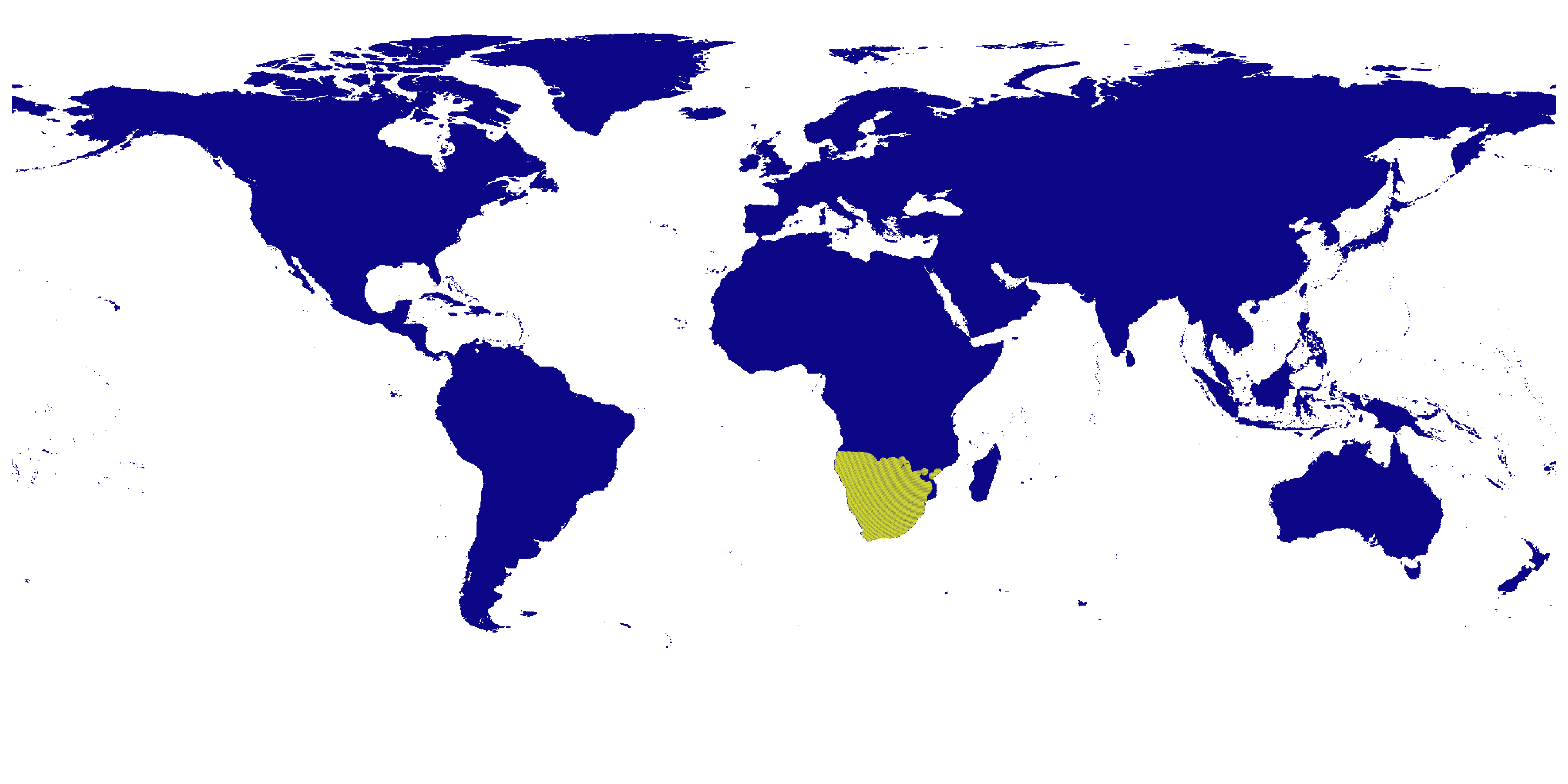}
    \end{minipage}

        \rotatebox{90}{\hspace{-15pt}\parbox{20mm}{\small\centering\texttt{Madagascar Hoopoe}}}
    \begin{minipage}{0.31\textwidth}
        \includegraphics[width=\linewidth]{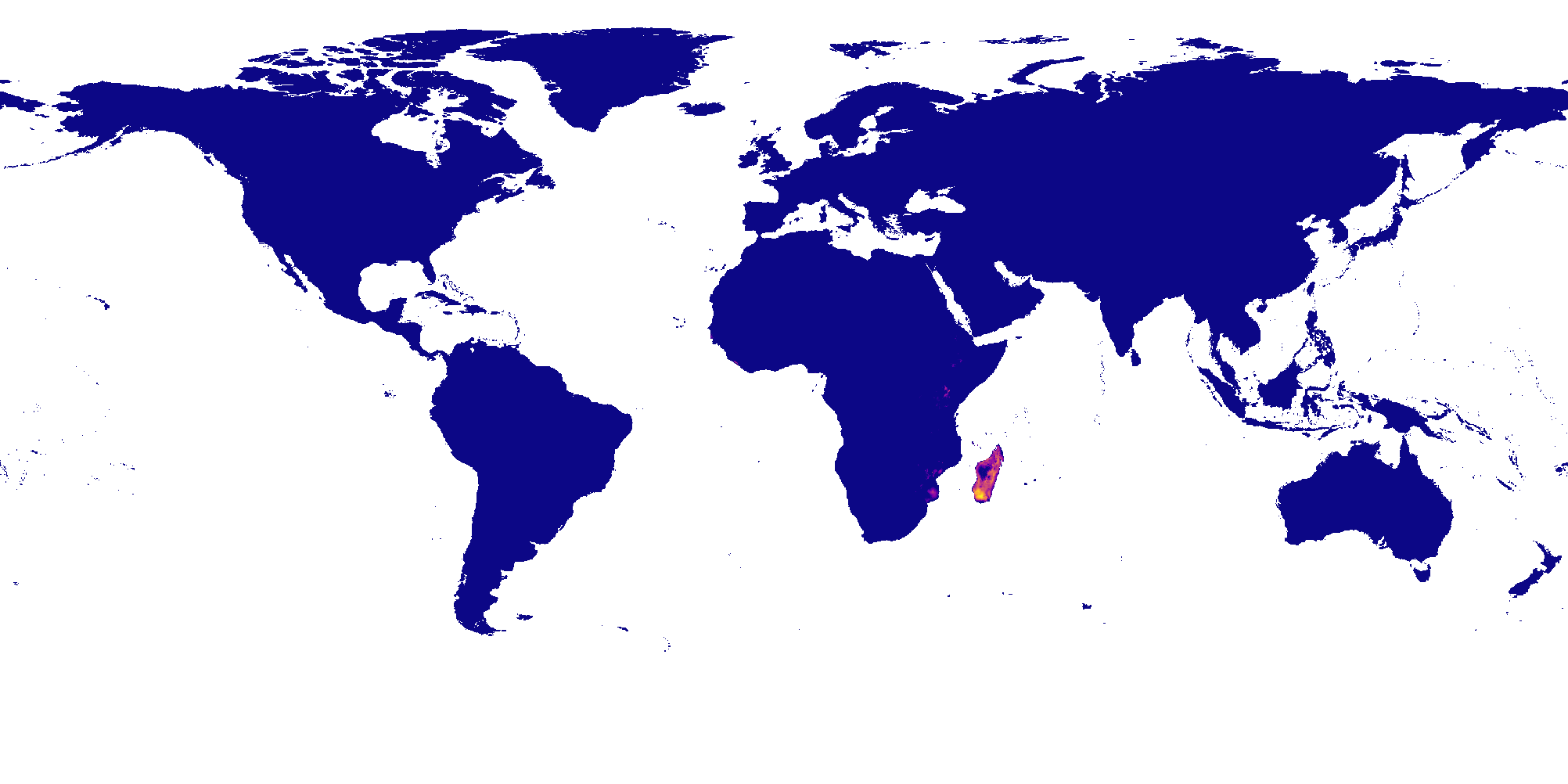}
    \end{minipage}
    \begin{minipage}{0.31\textwidth}
        \includegraphics[width=\linewidth]{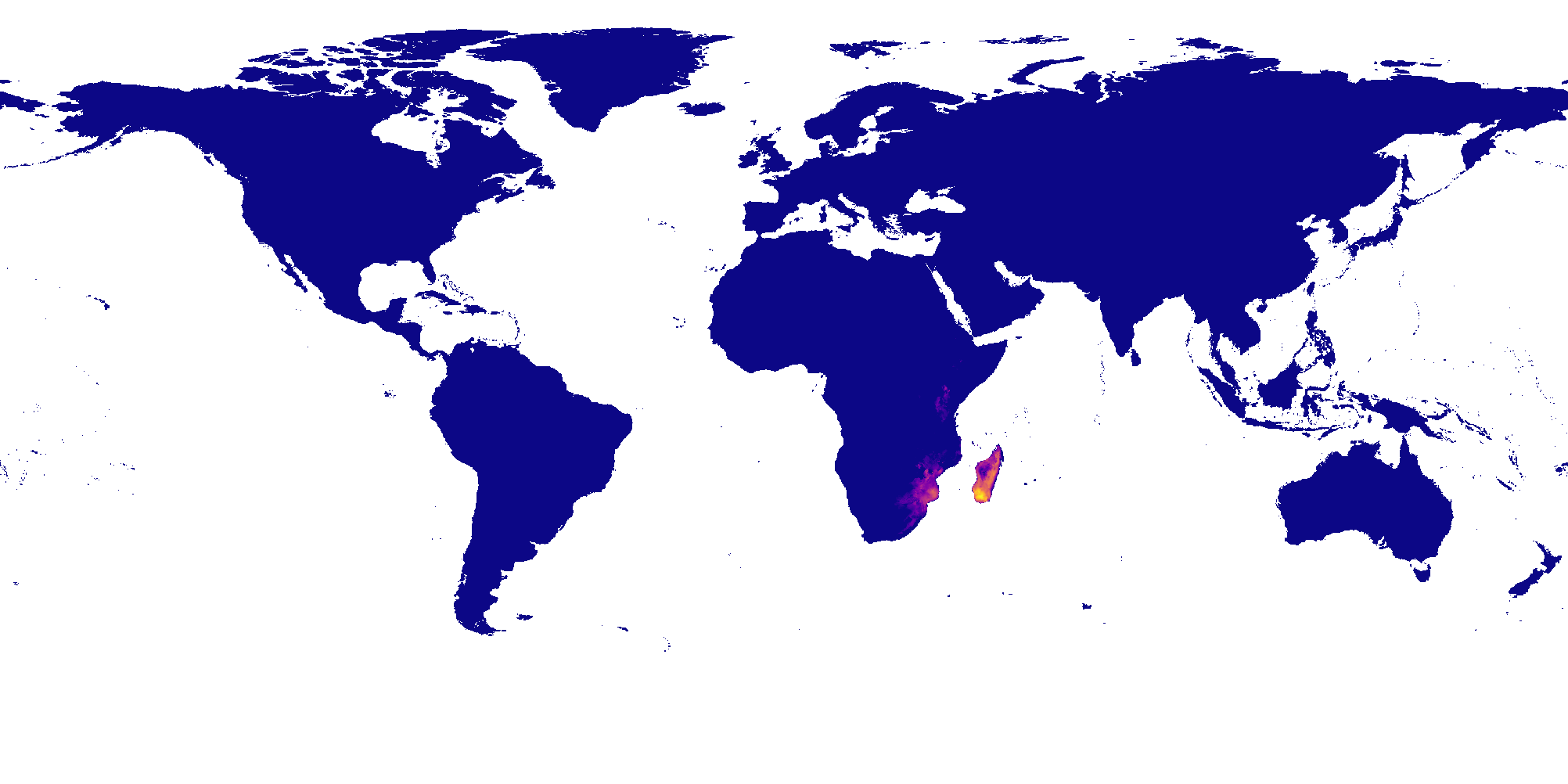}
    \end{minipage}
    \begin{minipage}{0.31\textwidth}
        \includegraphics[width=\linewidth]{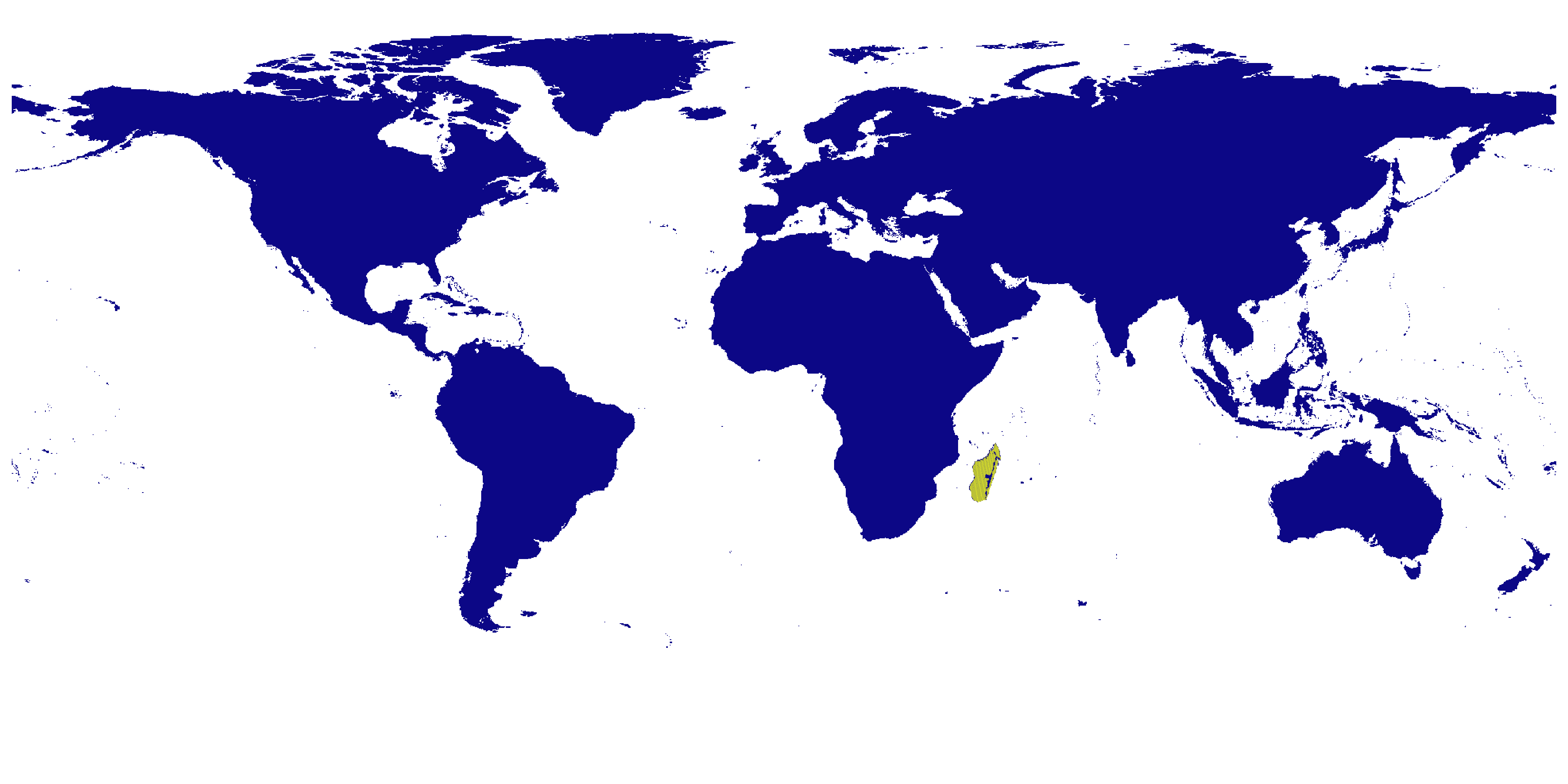}
    \end{minipage}

        \rotatebox{90}{\hspace{-15pt}\parbox{20mm}{\small\centering\texttt{Raucous Toad}}}
    \begin{minipage}{0.31\textwidth}
        \includegraphics[width=\linewidth]{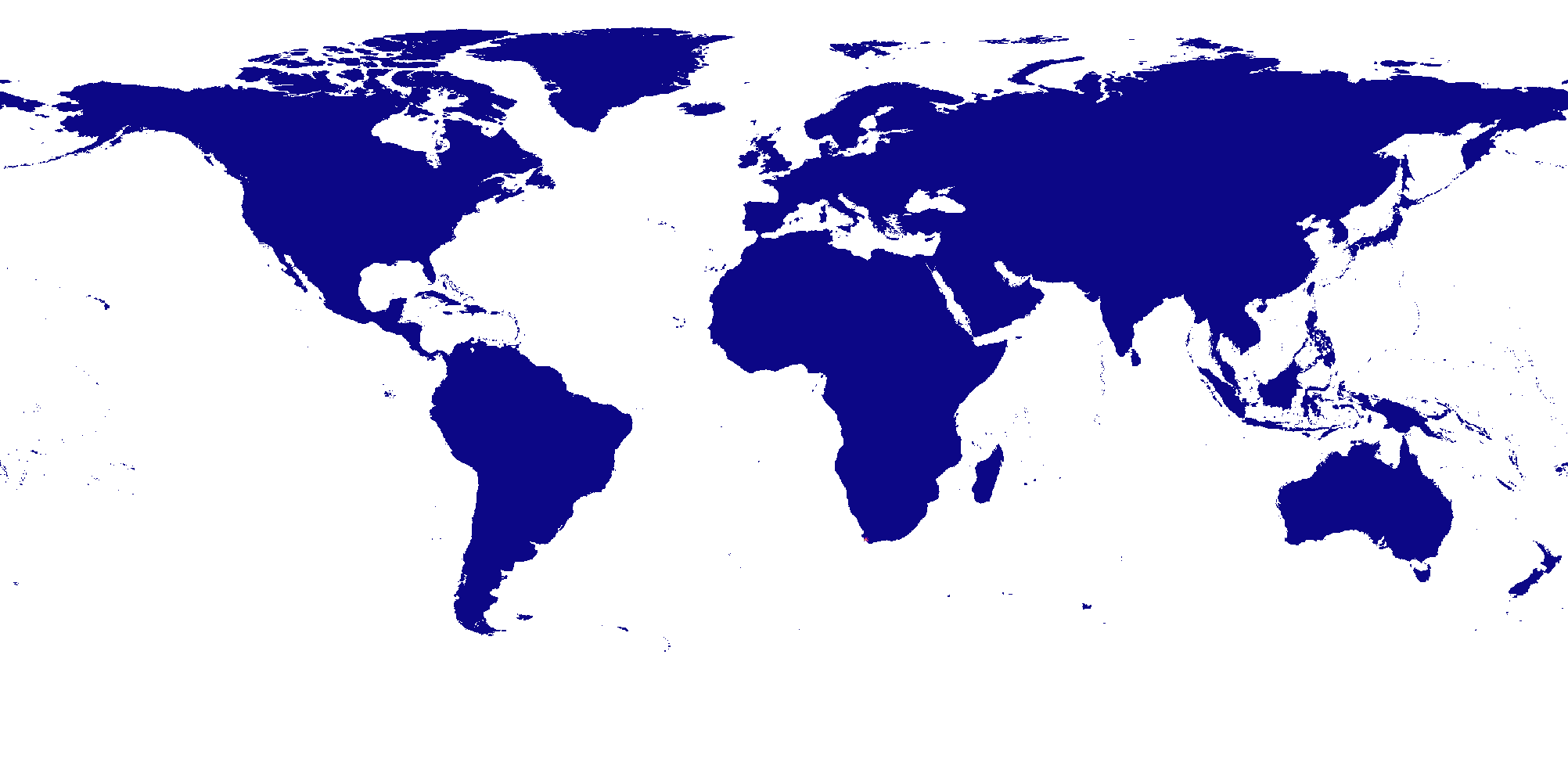}
        \centering{\small {Habitat text}}
    \end{minipage}
    \begin{minipage}{0.31\textwidth}
        \includegraphics[width=\linewidth]{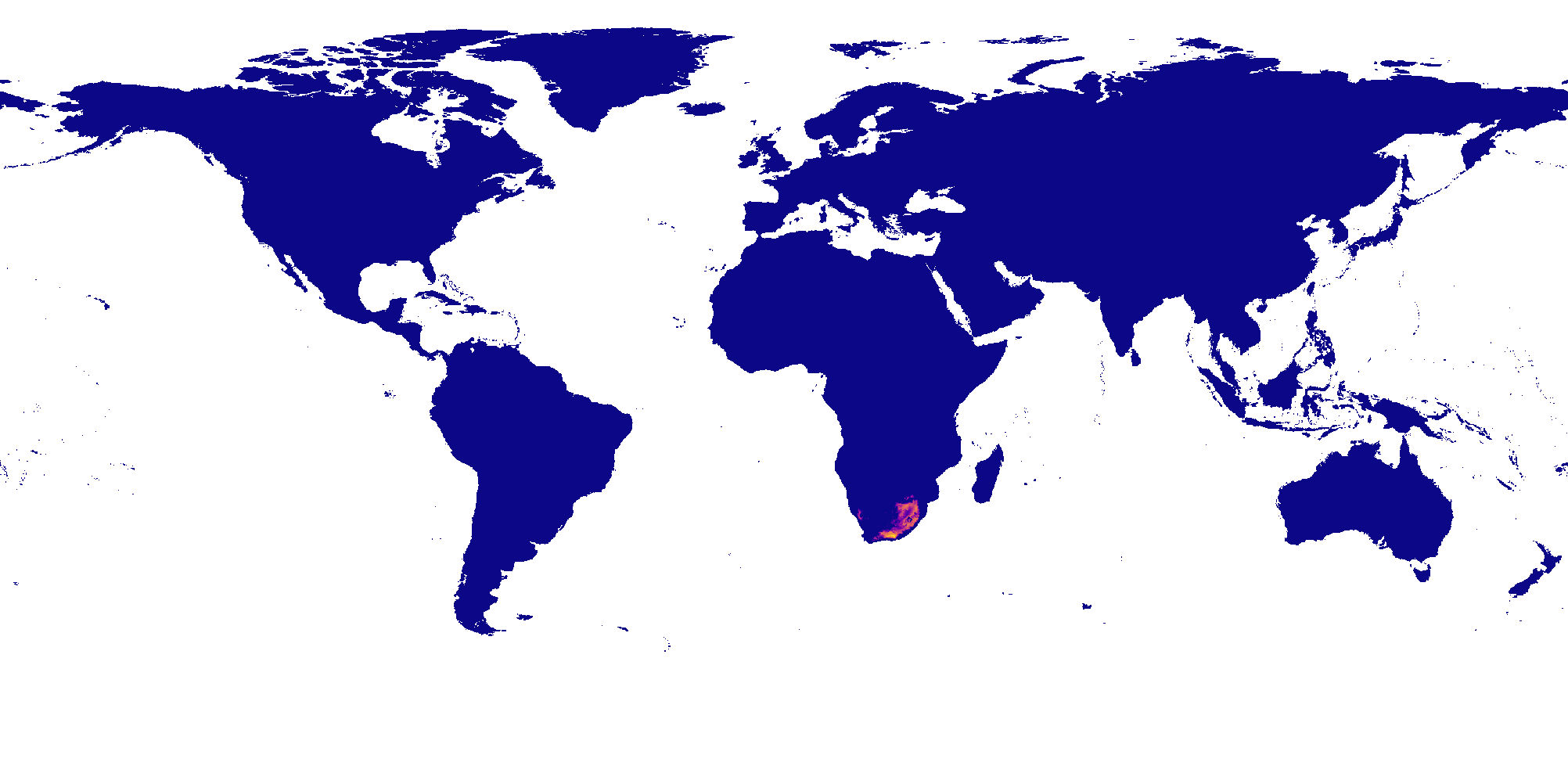}
        \centering{\small {Range text}}
    \end{minipage}
    \begin{minipage}{0.31\textwidth}
        \includegraphics[width=\linewidth]{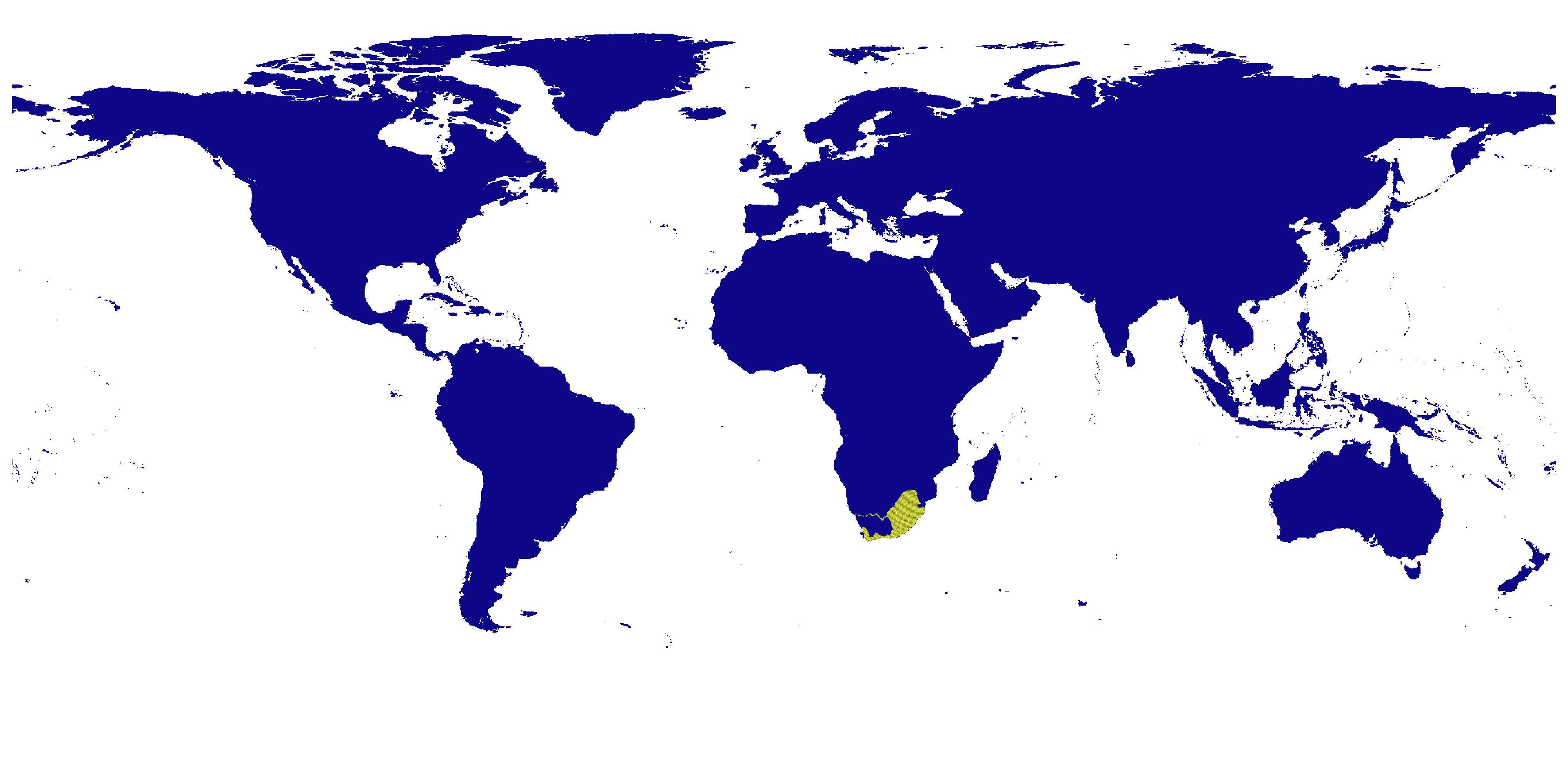}
        \centering{\small Expert Derived Range}
    \end{minipage}
    
    \caption{\textbf{Additional Zero-Shot Range Estimation.} Here we show the ‘Habitat’ (left) and ‘Range’ (center) text zero-shot range maps with associated expert derived range maps (right), for a variety of species. While the ‘Range’ text provides a strong prior in all cases with 
high probability assigned to regions within the expert derived range, the ‘Habitat’ text does not always do this, with the \texttt{Striated Babbler}, \texttt{Striped Sticky Frog}, and the \texttt{Raucous Toad} providing no strong prior. Zoom in to see details.}
    \label{fig:more_zero_shot}
\end{figure}

In Figure~\ref{fig:more_zero_shot} we show the range estimates obtained from both habitat and range text embeddings (zero-shot setting) alongside the associated range maps curated by experts for six additional species, namely the \texttt{Striated Babbler}, \texttt{Striped Sticky Frog}, \texttt{Common Hawk-Cuckoo}, \texttt{Cape Griffon}, \texttt{Madagascar Hoopoe} and \texttt{Raucous Toad}.

\begin{figure}[t]
    \centering

    \noindent
    \begin{minipage}[t]{0.47\textwidth}
        \centering
        \textbf{Habitat Text}
    \end{minipage}
    \hspace{0.04\textwidth}
    \begin{minipage}[t]{0.47\textwidth}
        \centering
        \textbf{Range Text}
    \end{minipage}

    \begin{minipage}[t]{0.47\textwidth}
        \vspace{0pt} %
        \includegraphics[trim={0 4cm 0 0},clip,width=\linewidth]{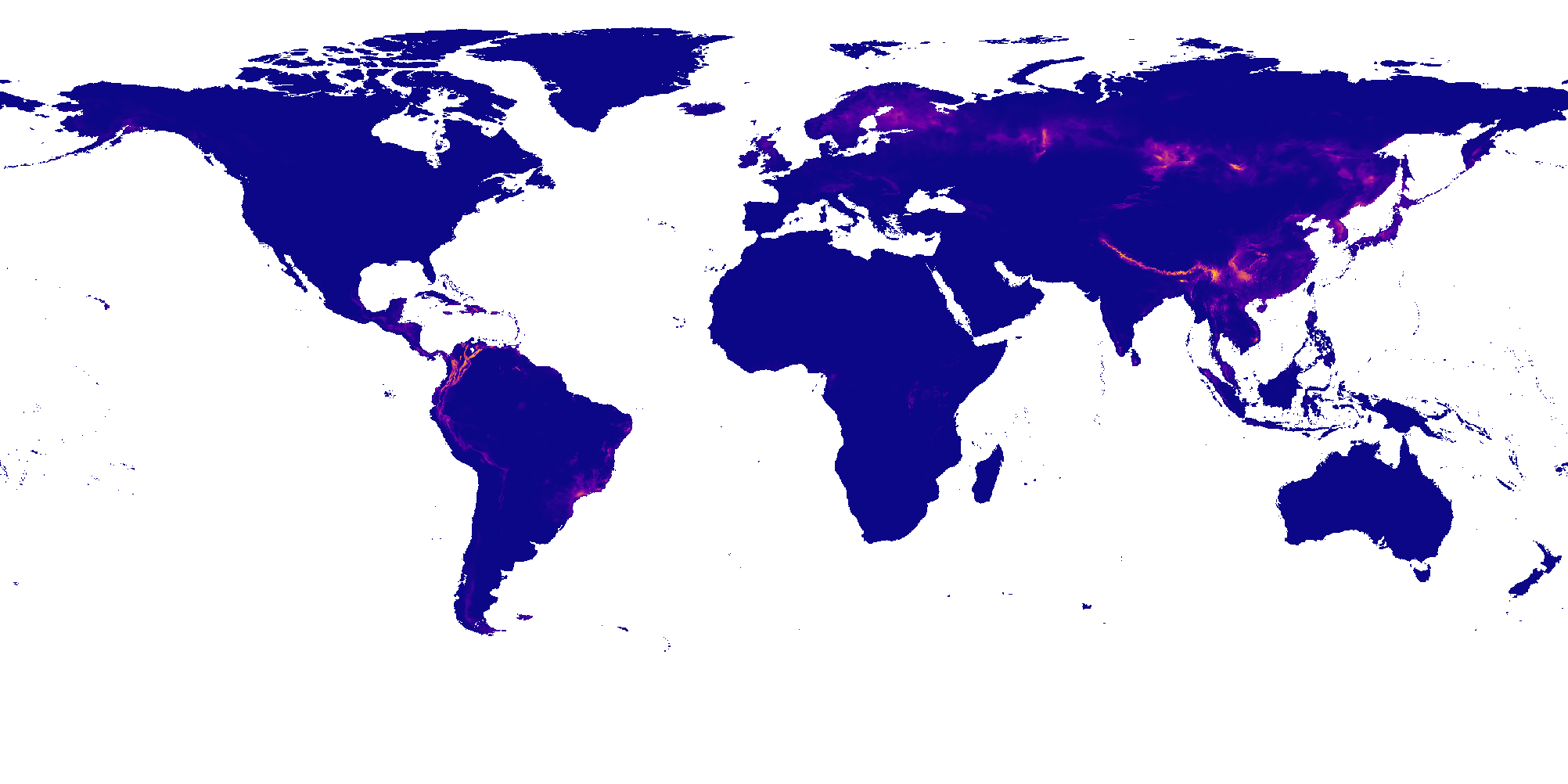}
        \subcaption*{\scriptsize The species inhabits undergrowth of coniferous forests with shrubs and bamboo, and can also be seen in parks and along roads.}
    \end{minipage}
    \hspace{0.04\textwidth}
    \begin{minipage}[t]{0.47\textwidth}
        \vspace{0pt}
        \includegraphics[trim={0 4cm 0 0},clip,width=\linewidth]{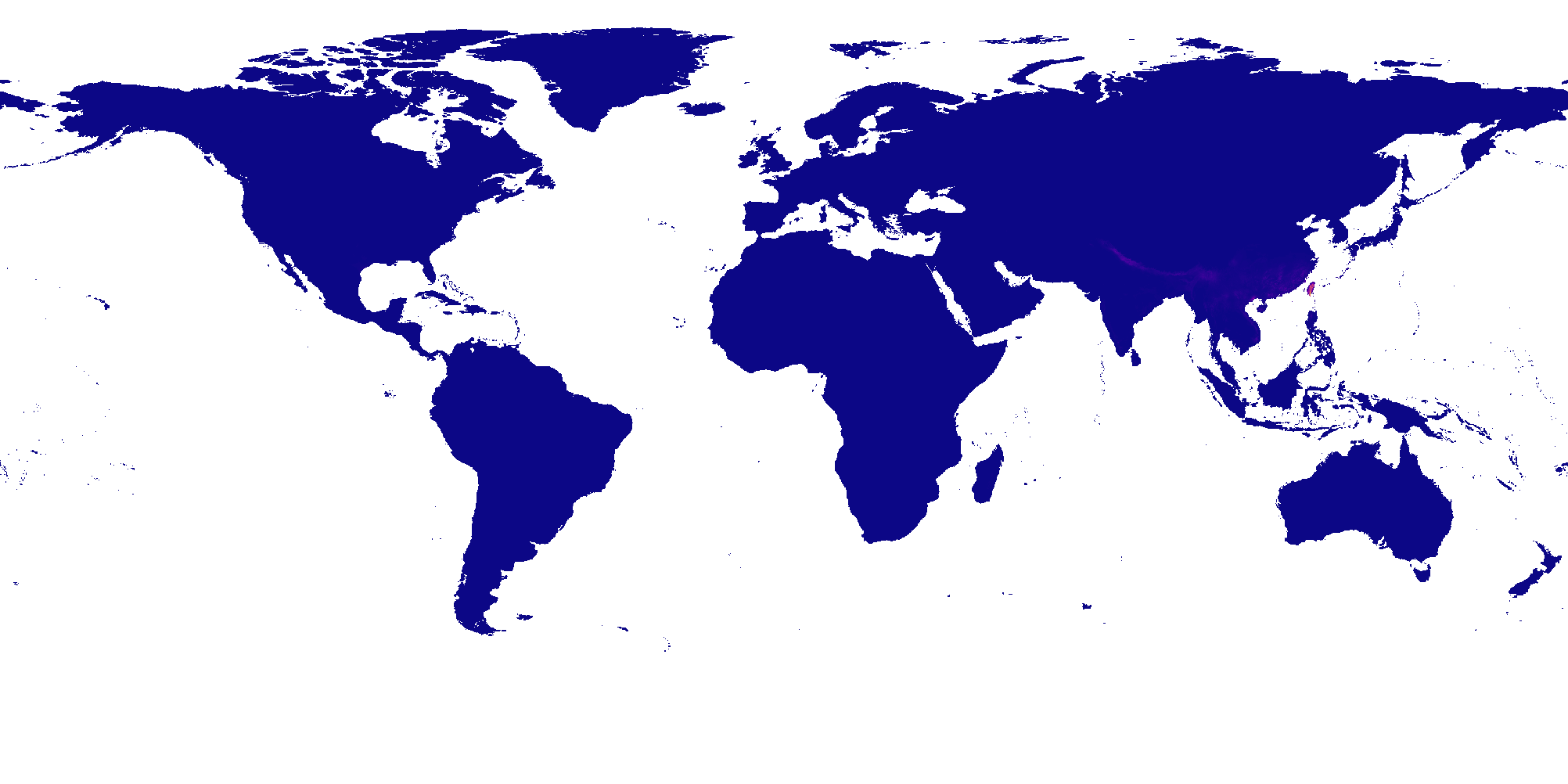}
        \subcaption*{\scriptsize The collared bush robin is endemic to Taiwan,}
    \end{minipage}

    \par\bigskip
    \noindent
    \begin{minipage}[t]{0.47\textwidth}
        \vspace{0pt}
        \includegraphics[trim={0 4cm 0 0},clip,width=\linewidth]{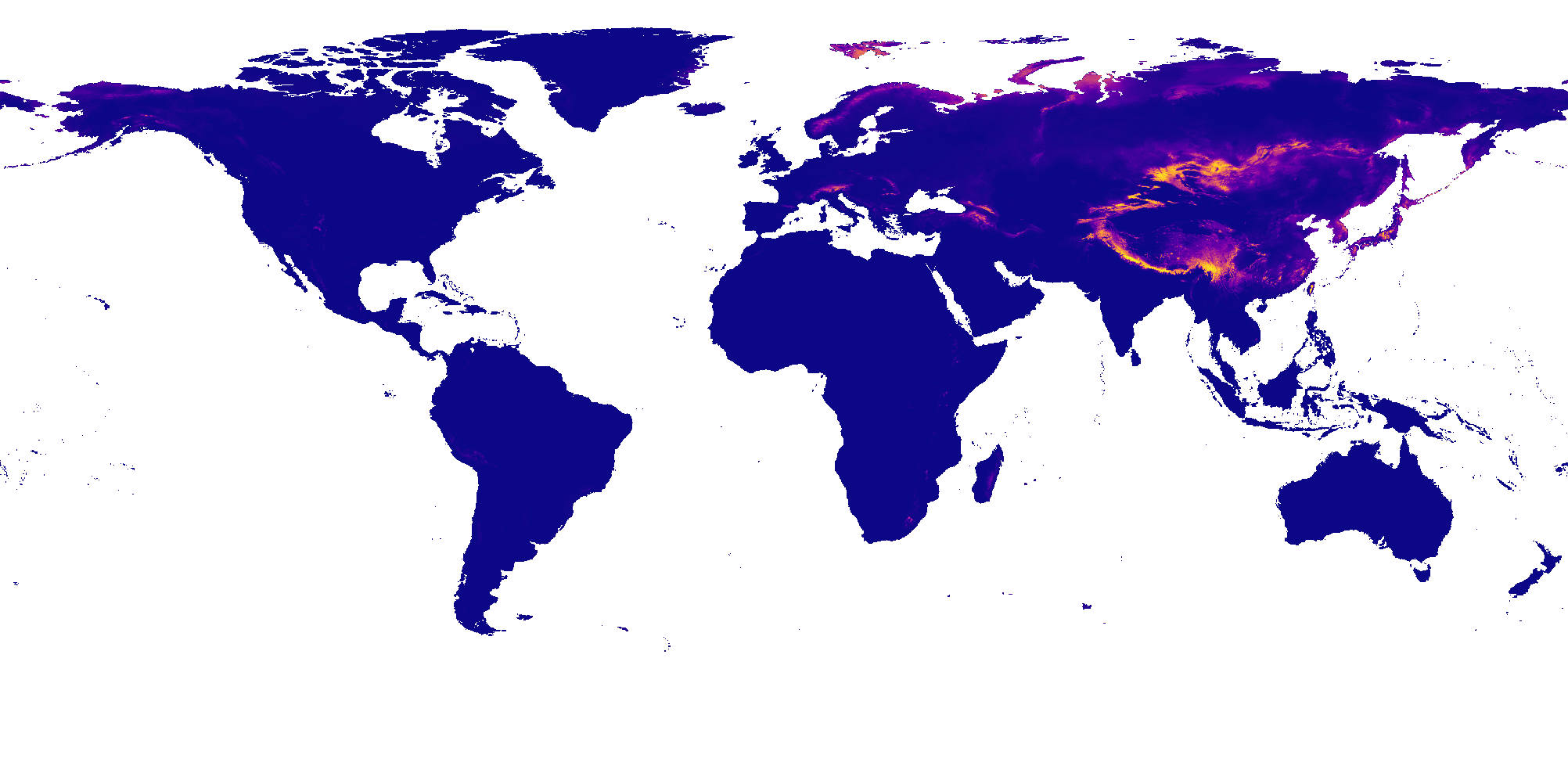}
        \subcaption*{\scriptsize It typically resides at elevations of 2,000 to 2,800m (6,600 to 9,200 ft) and sometimes above the tree line, descending to lower elevations in winter.}
    \end{minipage}
    \hspace{0.04\textwidth}
    \begin{minipage}[t]{0.47\textwidth}
        \vspace{0pt}
        \includegraphics[trim={0 4cm 0 0},clip,width=\linewidth]{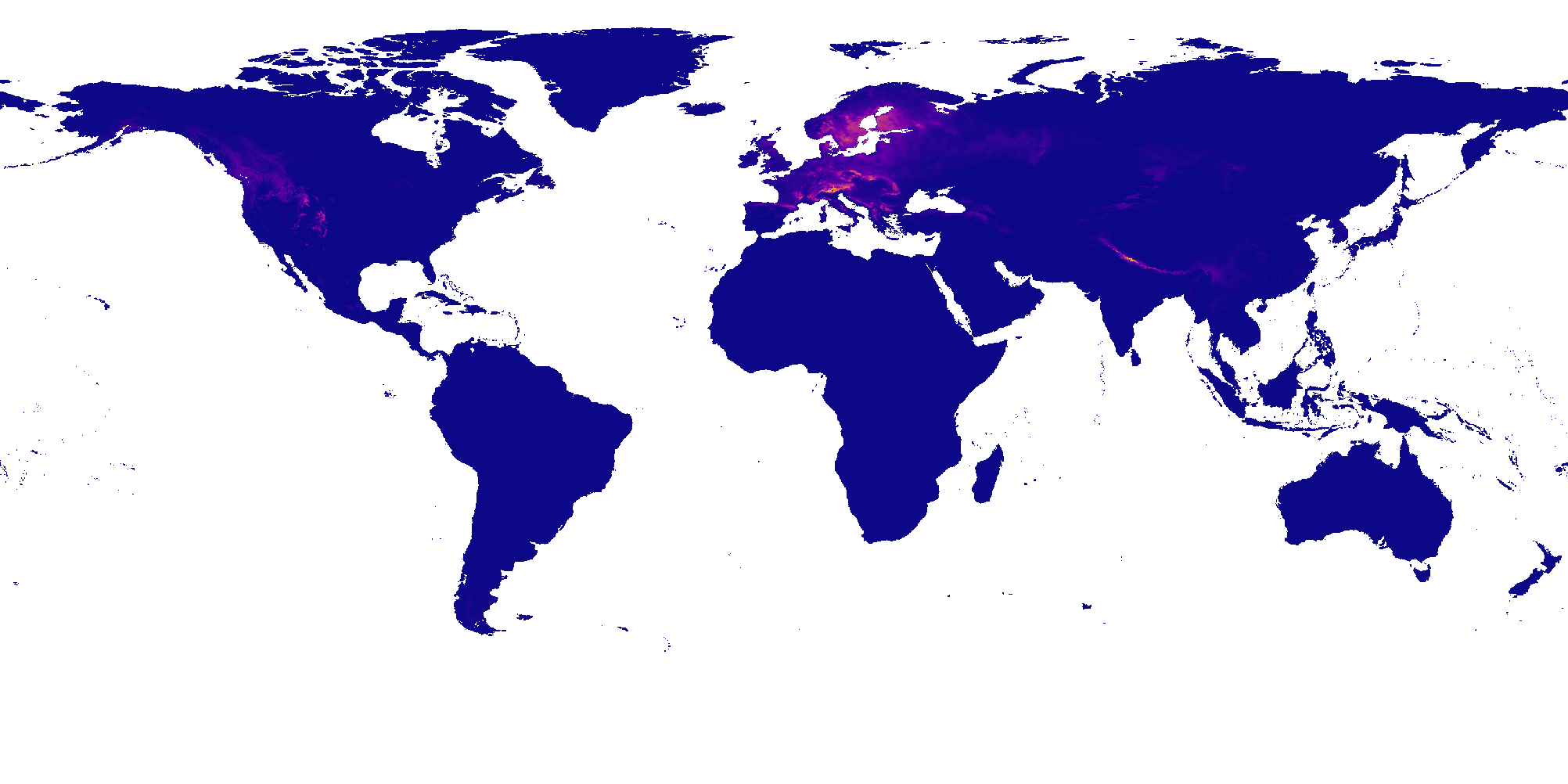}
        \subcaption*{\scriptsize living in montane and subalpine forests.}
    \end{minipage}

    \par\bigskip
    \noindent
    \begin{minipage}[t]{0.47\textwidth}
        \vspace{0pt}
        \includegraphics[trim={0 4cm 0 0},clip,width=\linewidth]{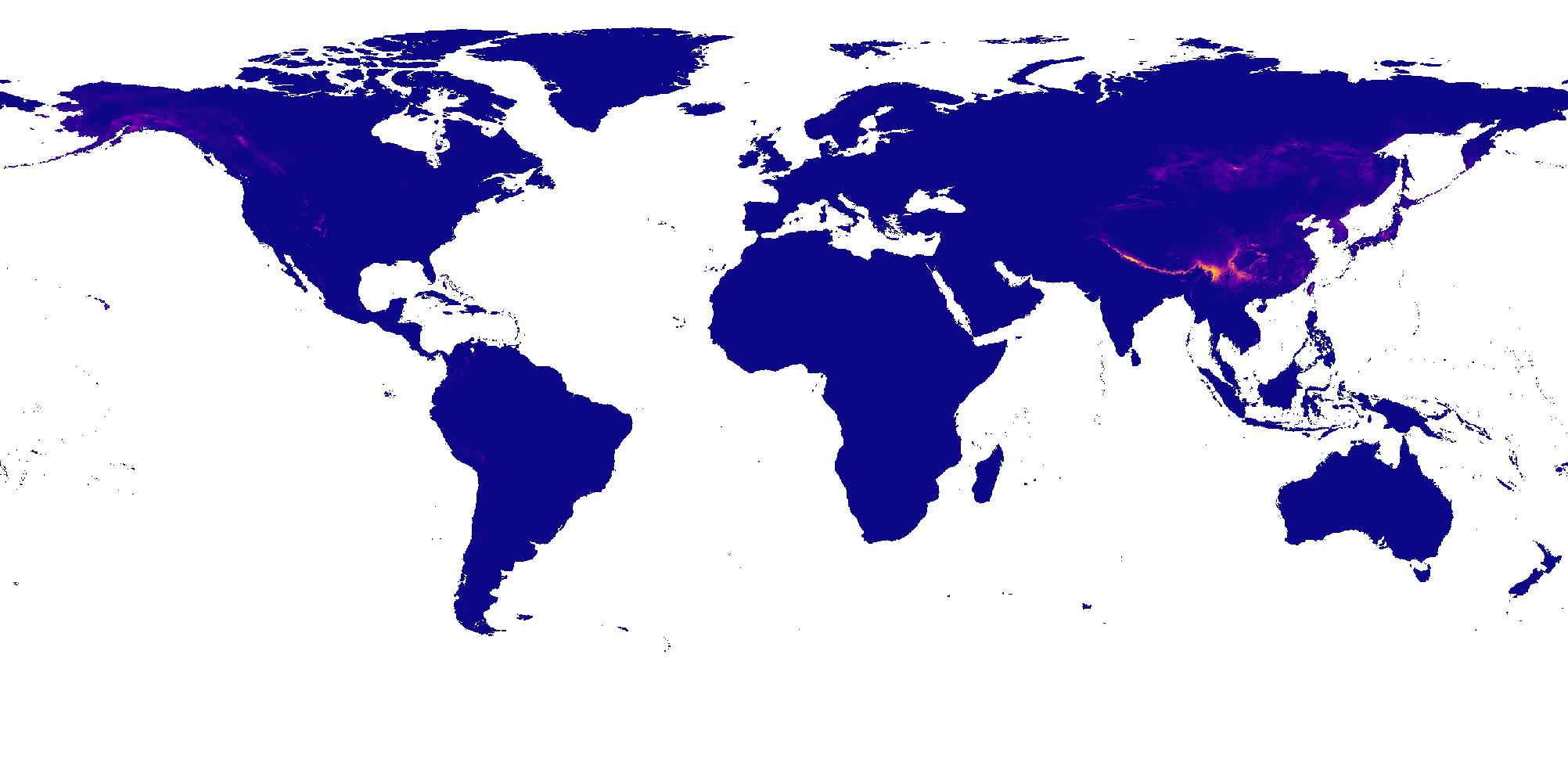}
        \subcaption*{\scriptsize The species inhabits undergrowth of coniferous forests with shrubs and bamboo, and can also be seen in parks and along roads. It typically resides at elevations of 2,000 to 2,800m (6,600 to 9,200 ft) and sometimes above the tree line, descending to lower elevations in winter.}
    \end{minipage}
    \hspace{0.04\textwidth}
    \begin{minipage}[t]{0.47\textwidth}
        \vspace{0pt}
        \includegraphics[trim={0 4cm 0 0},clip,width=\linewidth]{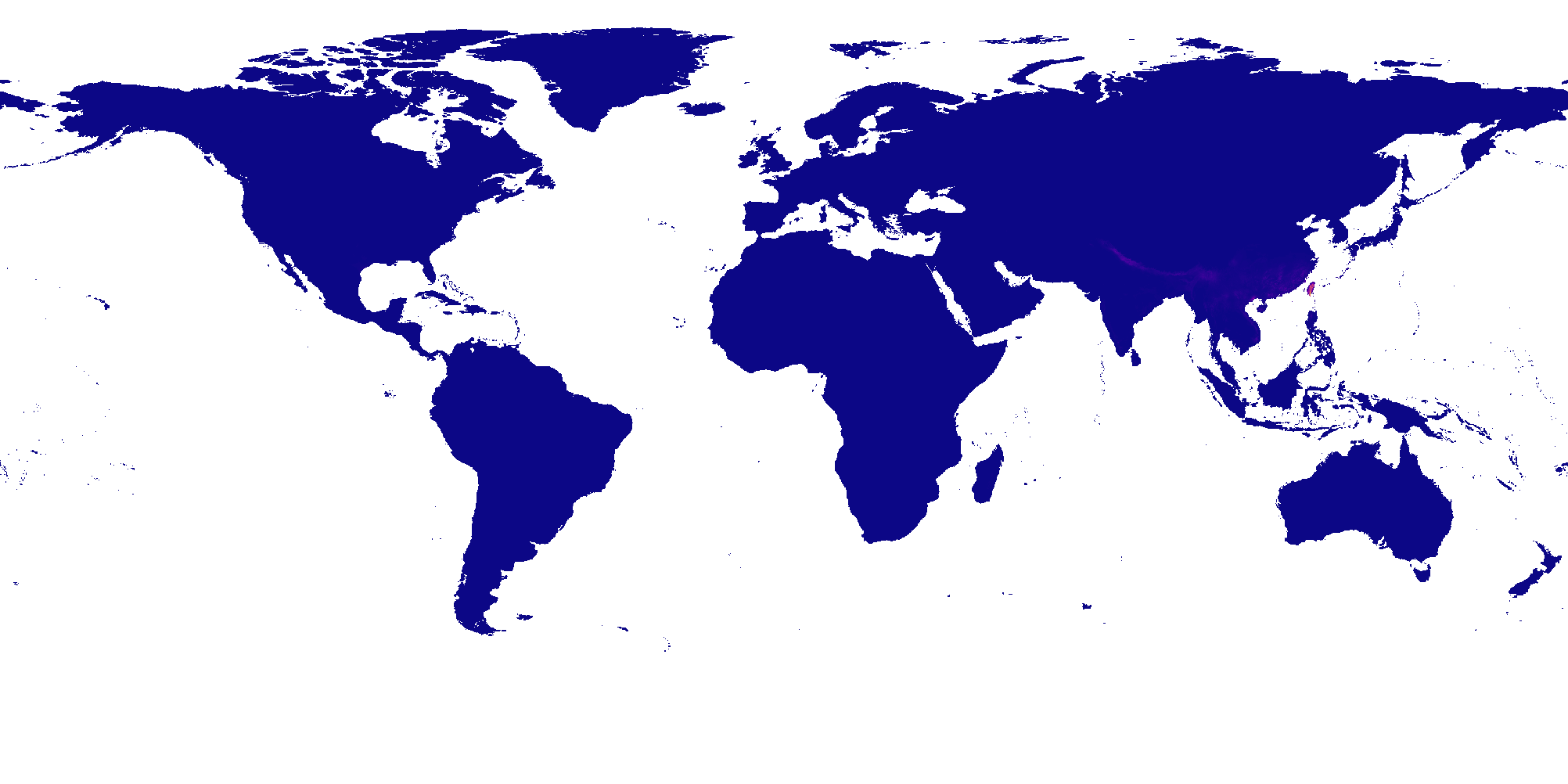}
        \subcaption*{\scriptsize The collared bush robin is endemic to Taiwan, living in montane and subalpine forests.}
    \end{minipage}

    \caption{\textbf{Zero-Shot Range Estimation From Parts of Text.} Here we show the zero-shot  predicted ranges when different parts of the habitat text (left column) and range text (right column) are given to our model for the \texttt{Collared Bush Robin}, which is found in Taiwan. 
    From top to bottom, the rows show the predicted range for the first part of the text, the second part of the text, and  the entire text, \ie the first and second parts concatenated.  
    For the habitat text we see that the first part of the text loosely identifies several forested areas but not Taiwan, while the second part loosely identifies several mountainous areas including Taiwan. Combining these parts reduces the false positives in the range map produced while still correctly including Taiwan. For the range text, we again see that the second part of the text identifies some mountainous areas, while the first part seems to locate Taiwan effectively. Together the range is correctly limited to the inland highland areas of Taiwan. Please zoom in to see more detail.}
    \vspace{-8pt}
    \label{fig:zero_shot_w_parts_of_text}
\end{figure}

In Figure~\ref{fig:zero_shot_w_parts_of_text} we show how zero-shot range estimates for the \texttt{Collared Bush Robin} change as we use different parts of a piece of text. 

\begin{table}[h]
\caption{\textbf{Evaluation Text Oracles}. We can get a tighter upper bound on performance by evaluating our oracle model with the habitat and range text summaries rather than species tokens. These summaries are not seen in the training data.}
  \vspace{5pt}
  \centering
  \resizebox{0.6\textwidth}{!}{
  \begin{tabular}{l|c|c|c}
        Method & +Eval Sp. & IUCN & S\&T \\\hline
        LE-SINR (Habitat Text) & & 0.32 & 0.52\\
        LE-SINR (Range Text) & & 0.53 & 0.64\\
        LE-SINR (Habitat Text) & \checkmark & \color{gray}0.39 & \color{gray}0.61\\
        LE-SINR (Range Text) & \checkmark & \color{gray}0.60 & \color{gray}0.69\\
  \end{tabular}
  }
  \label{tab:text_oracle}    
  \vspace{-10pt}
\end{table}

\section{Dataset Examples}
Below we show `range', `habitat' text for several species used for our zero-shot range estimation experiments. These were generated by using Llama3 language model to summarize the Wikipedia text into content that describes the range, \ie the geographical extent where the species is found, including names of countries, continents, and geographic regions, as well as its habitat, which indicates specific environmental conditions and types of habitats that a species thrives in such as descriptions of climate, vegetation, topography, soil types, food resources, etc. 

\begin{enumerate}[leftmargin=20pt]
\item \texttt{Gray Kingbird (Tyrannus dominicensis)}
\begin{itemize}
\item \textit{Range}: The gray kingbird is found in the southeast USA, Colombia, and Venezuela, with two recognized subspecies: T. d. dominicensis and T. d. vorax. It breeds from the extreme southeast of the United States, mainly in Florida, as well as Central America, and through the West Indies south to Venezuela, Trinidad and Tobago, the Guianas, and Colombia. Northern populations are migratory, wintering on the Caribbean coast of Central America and northern South America.
\item \textit{Habitat}: The gray kingbird favors tall trees and shrubs, including the edges of savanna and marshes. It is found in increasing numbers in the state of Florida, and is more often found inland though it had been previously restricted to the coast.
\end{itemize}
\item \texttt{Cape Weaver (Ploceus capensis)}
\begin{itemize}
    \item \textit{Range}: The Cape weaver is endemic to South Africa, Lesotho, and Eswatini, occurring across much of the area excluding the Kalahari Desert from the Orange River in the Northern Cape south to the Cape of Good Hope, then east to northern KwaZulu Natal, and inland almost to Bloemfontein in the Free State.
    \item \textit{Habitat}: The Cape weaver occurs in open grassland, lowland fynbos, coastal thicket, and farmland, so long as there is permanent water and trees. In the more arid, hotter regions, it is restricted to upland areas and never occurs in forest.
\end{itemize}
\item \texttt{Plate-billed Mountain-Toucan (Andigena laminirostris)}
\begin{itemize}
    \item \textit{Range}: The plate-billed mountain toucan is found in the western foothills of the Andes of western Ecuador and far southwestern Colombia, specifically from Pita Canyon (Narino) in southwestern Colombia and south to the northwestern border of Morona-Santiago Province, in Ecuador.
    \item \textit{Habitat}: The species inhabits the humid forest and edges of the temperate forest of the lateral slope of the Andes Mountains, featuring abundant epiphytes, bromeliads, and mosses. The forests receive an average of 14 feet of rainfall per year, and the canopy ranges from 6 to 10 meters high. Their altitudinal range is between 1600 and 2600 meters above sea level.
\end{itemize}
\item \texttt{Harlequin Racerunner (Plica umbra)}
\begin{itemize}
    \item \textit{Range}: The blue-lipped tree lizard or harlequin racerunner (Plica umbra) is found in South America, specifically in Colombia, Venezuela, Guyana, Suriname, French Guiana, Brazil, Bolivia, Peru, and Ecuador.
    \item \textit{Habitat}: The species inhabits tropical rainforests, savannas, and dry forests, typically at elevations below 500 meters. It prefers areas with dense vegetation, rocky outcrops, and sandy or clay soils.
\end{itemize}

\item \texttt{Dusky Rattlesnake (Crotalus triseriatus)}
\begin{itemize}
    \item \textit{Range}: The Mexican dusky rattlesnake (Crotalus triseriatus) is found in Mexico, along the southern edge of the Mexican Plateau in the highlands of the Transverse Volcanic Cordillera, including the states of Jalisco, México, Michoacán, Morelos, Nayarit, Puebla, Tlaxcala, and Veracruz.
    \item \textit{Habitat}: Crotalus triseriatus occurs in pine-oak forest, boreal forest, coniferous forest, and bunchgrass grasslands. On Volcán Orizaba, it is found at very high altitudes, with the species being found within the zone where the snow line comes down to about 4,572 m (15,000 ft), and green plants can be found up to 4,573 m (15,003 ft).
\end{itemize}
\item \texttt{Western Ghats Flying Lizard (Draco dussumieri)}
\begin{itemize}
    \item \textit{Range}: The species is found principally in the Western Ghats and some other hill forests of Southern India, including Karnataka, Kerala, Tamil Nadu, Goa, and Maharashtra. It is also reported from some parts of the Eastern Ghats in Andhra Pradesh.
    \item \textit{Habitat}: The southern flying lizard is almost entirely arboreal, found on trees in forests and adjoining palm plantations. It climbs trees in search of insect prey on the trunks and leaps off when it reaches the top to land on adjoining trees. The species is active during the day after it has warmed up in the early morning sun.
\end{itemize}
\end{enumerate}

\clearpage

\section*{NeurIPS Paper Checklist}

\begin{enumerate}

\item {\bf Claims}
    \item[] Question: Do the main claims made in the abstract and introduction accurately reflect the paper's contributions and scope?
    \item[] Answer: \answerYes{} %
    \item[] Justification: Section~\ref{sec:results} provide detailed experimental evidence quantifying the accuracy of predicted range maps using expert-derived range maps for species in the S\&T and IUCN lists. We show the value of language supervision by demonstrating zero-shot range estimation from descriptions of habitat and range preferences. We also demonstrate the value of learning from language with superior performance over the baseline SINR model trained with observations only in the few-shot setting


\item {\bf Limitations}
    \item[] Question: Does the paper discuss the limitations of the work performed by the authors?
    \item[] Answer: \answerYes{} %
    \item[] Justification: Section~\ref{sec:limitations} describes some of the limitations of our work, which include the need for a more thorough evaluation of the few-shot experiments, inheriting biases from language models and text data on the internet, and the lack of precision required for high-stakes use cases in conservation and planning.
       \item[] Guidelines:
    \begin{itemize}
        \item The answer NA means that the paper has no limitation while the answer No means that the paper has limitations, but those are not discussed in the paper. 
        \item The authors are encouraged to create a separate "Limitations" section in their paper.
        \item The paper should point out any strong assumptions and how robust the results are to violations of these assumptions (\eg independence assumptions, noiseless settings, model well-specification, asymptotic approximations only holding locally). The authors should reflect on how these assumptions might be violated in practice and what the implications would be.
        \item The authors should reflect on the scope of the claims made, \eg if the approach was only tested on a few datasets or with a few runs. In general, empirical results often depend on implicit assumptions, which should be articulated.
        \item The authors should reflect on the factors that influence the performance of the approach. For example, a facial recognition algorithm may perform poorly when image resolution is low or images are taken in low lighting. Or a speech-to-text system might not be used reliably to provide closed captions for online lectures because it fails to handle technical jargon.
        \item The authors should discuss the computational efficiency of the proposed algorithms and how they scale with dataset size.
        \item If applicable, the authors should discuss possible limitations of their approach to address problems of privacy and fairness.
        \item While the authors might fear that complete honesty about limitations might be used by reviewers as grounds for rejection, a worse outcome might be that reviewers discover limitations that aren't acknowledged in the paper. The authors should use their best judgment and recognize that individual actions in favor of transparency play an important role in developing norms that preserve the integrity of the community. Reviewers will be specifically instructed to not penalize honesty concerning limitations.
    \end{itemize}

\item {\bf Theory Assumptions and Proofs}
    \item[] Question: For each theoretical result, does the paper provide the full set of assumptions and a complete (and correct) proof?
    \item[] Answer: \answerNA{} %
    \item[] Justification: The paper does not include theoretical results.

    \item {\bf Experimental Result Reproducibility}
    \item[] Question: Does the paper fully disclose all the information needed to reproduce the main experimental results of the paper to the extent that it affects the main claims and/or conclusions of the paper (regardless of whether the code and data are provided or not)?
    \item[] Answer: \answerYes{} %
    \item[] Justification: The work closely follows the experimental setup, source code, and training data of SINR using iNaturalist, both of which are publicly available. The novel part is the incorporation of Wikipedia text and their short summaries, which are provided in the Supplementary Material. We have described the model architecture and hyperparameters for training in our paper, and will publicly release the dataset and the evaluation framework for the zero-shot and few-shot experiments upon publication of the paper. 
   \item[] Guidelines:
    \begin{itemize}
        \item The answer NA means that the paper does not include experiments.
        \item If the paper includes experiments, a No answer to this question will not be perceived well by the reviewers: Making the paper reproducible is important, regardless of whether the code and data are provided or not.
        \item If the contribution is a dataset and/or model, the authors should describe the steps taken to make their results reproducible or verifiable. 
        \item Depending on the contribution, reproducibility can be accomplished in various ways. For example, if the contribution is a novel architecture, describing the architecture fully might suffice, or if the contribution is a specific model and empirical evaluation, it may be necessary to either make it possible for others to replicate the model with the same dataset, or provide access to the model. In general. releasing code and data is often one good way to accomplish this, but reproducibility can also be provided via detailed instructions for how to replicate the results, access to a hosted model (\eg in the case of a large language model), releasing of a model checkpoint, or other means that are appropriate to the research performed.
        \item While NeurIPS does not require releasing code, the conference does require all submissions to provide some reasonable avenue for reproducibility, which may depend on the nature of the contribution. For example
        \begin{enumerate}
            \item If the contribution is primarily a new algorithm, the paper should make it clear how to reproduce that algorithm.
            \item If the contribution is primarily a new model architecture, the paper should describe the architecture clearly and fully.
            \item If the contribution is a new model (\eg a large language model), then there should either be a way to access this model for reproducing the results or a way to reproduce the model (\eg with an open-source dataset or instructions for how to construct the dataset).
            \item We recognize that reproducibility may be tricky in some cases, in which case authors are welcome to describe the particular way they provide for reproducibility. In the case of closed-source models, it may be that access to the model is limited in some way (\eg to registered users), but it should be possible for other researchers to have some path to reproducing or verifying the results.
        \end{enumerate}
    \end{itemize}

\item {\bf Open access to data and code}
   \item[] Question: Does the paper provide open access to the data and code, with sufficient instructions to faithfully reproduce the main experimental results, as described in supplemental material?
    \item[] Answer: \answerYes{} %
    \item[] Justification: We include a link to a publicly available github in the intro that contains instructions and code needed to reproduce the experimental results. We also provide pre-trained weights of the models used in this work as well as the training and evaluation data.
    \item[] Guidelines:
    \begin{itemize}
        \item The answer NA means that paper does not include experiments requiring code.
        \item Please see the NeurIPS code and data submission guidelines (\url{https://nips.cc/public/guides/CodeSubmissionPolicy}) for more details.
        \item While we encourage the release of code and data, we understand that this might not be possible, so “No” is an acceptable answer. Papers cannot be rejected simply for not including code, unless this is central to the contribution (\eg for a new open-source benchmark).
        \item The instructions should contain the exact command and environment needed to run to reproduce the results. See the NeurIPS code and data submission guidelines (\url{https://nips.cc/public/guides/CodeSubmissionPolicy}) for more details.
        \item The authors should provide instructions on data access and preparation, including how to access the raw data, preprocessed data, intermediate data, and generated data, etc.
        \item The authors should provide scripts to reproduce all experimental results for the new proposed method and baselines. If only a subset of experiments are reproducible, they should state which ones are omitted from the script and why.
        \item At submission time, to preserve anonymity, the authors should release anonymized versions (if applicable).
        \item Providing as much information as possible in supplemental material (appended to the paper) is recommended, but including URLs to data and code is permitted.
    \end{itemize}

\item {\bf Experimental Setting/Details}
    \item[] Question: Does the paper specify all the training and test details (\eg data splits, hyperparameters, how they were chosen, type of optimizer, etc.) necessary to understand the results?
    \item[] Answer: \answerYes{} %
    \item[] Justification: We follow the same evaluation splits and metrics as previous work, which is publicly available. The training data is also identical, except for the zero-shot and few-shot settings where the evaluation species are held out.
\item {\bf Experiment Statistical Significance}
    \item[] Question: Does the paper report error bars suitably and correctly defined or other appropriate information about the statistical significance of the experiments?
    \item[] Answer: \answerNo{} %
    \item[] Justification: We did not report error bars, but the MAP numbers are the result of averaging AP across hundreds or thousands of species for S\&T and IUCN, respectively, and the confidence intervals for MAP are extremely narrow ($\approx$0.1 MAP). The improvements from language supervision in the zero-shot and few-shot settings in Figure~\ref{fig:few_shot_results} are 10 to 15 points MAP, so the improvements are statistically significant.
      \item[] Guidelines:
    \begin{itemize}
        \item The answer NA means that the paper does not include experiments.
        \item The authors should answer "Yes" if the results are accompanied by error bars, confidence intervals, or statistical significance tests, at least for the experiments that support the main claims of the paper.
        \item The factors of variability that the error bars are capturing should be clearly stated (for example, train/test split, initialization, random drawing of some parameter, or overall run with given experimental conditions).
        \item The method for calculating the error bars should be explained (closed form formula, call to a library function, bootstrap, etc.)
        \item The assumptions made should be given (\eg Normally distributed errors).
        \item It should be clear whether the error bar is the standard deviation or the standard error of the mean.
        \item It is OK to report 1-sigma error bars, but one should state it. The authors should preferably report a 2-sigma error bar than state that they have a 96\% CI, if the hypothesis of Normality of errors is not verified.
        \item For asymmetric distributions, the authors should be careful not to show in tables or figures symmetric error bars that would yield results that are out of range (\eg negative error rates).
        \item If error bars are reported in tables or plots, The authors should explain in the text how they were calculated and reference the corresponding figures or tables in the text.
    \end{itemize}

\item {\bf Experiments Compute Resources}
    \item[] Question: For each experiment, does the paper provide sufficient information on the computer resources (type of compute workers, memory, time of execution) needed to reproduce the experiments?
    \item[] Answer: \answerYes{} %
    \item[] Justification: Section~\ref{sec:training_details} describes the computational resources needed for training.
    \item[] Guidelines:
    \begin{itemize}
        \item The answer NA means that the paper does not include experiments.
        \item The paper should indicate the type of compute workers CPU or GPU, internal cluster, or cloud provider, including relevant memory and storage.
        \item The paper should provide the amount of compute required for each of the individual experimental runs as well as estimate the total compute. 
        \item The paper should disclose whether the full research project required more compute than the experiments reported in the paper (\eg preliminary or failed experiments that didn't make it into the paper). 
    \end{itemize}
    
\item {\bf Code Of Ethics}
    \item[] Question: Does the research conducted in the paper conform, in every respect, with the NeurIPS Code of Ethics \url{https://neurips.cc/public/EthicsGuidelines}?
    \item[] Answer: \answerYes{}{} %
    \item[] Justification: We have conformed to ethics code to the best of our ability and knowledge.
    \item[] Guidelines:
    \begin{itemize}
        \item The answer NA means that the authors have not reviewed the NeurIPS Code of Ethics.
        \item If the authors answer No, they should explain the special circumstances that require a deviation from the Code of Ethics.
        \item The authors should make sure to preserve anonymity (\eg if there is a special consideration due to laws or regulations in their jurisdiction).
    \end{itemize}

\item {\bf Broader Impacts}
    \item[] Question: Does the paper discuss both potential positive societal impacts and negative societal impacts of the work performed?
    \item[] Answer: \answerYes{} %
    \item[] Justification: Section~\ref{sec:limitations} discusses the broader impacts. 
    \item[] Guidelines:
    \begin{itemize}
        \item The answer NA means that there is no societal impact of the work performed.
        \item If the authors answer NA or No, they should explain why their work has no societal impact or why the paper does not address societal impact.
        \item Examples of negative societal impacts include potential malicious or unintended uses (\eg disinformation, generating fake profiles, surveillance), fairness considerations (\eg deployment of technologies that could make decisions that unfairly impact specific groups), privacy considerations, and security considerations.
        \item The conference expects that many papers will be foundational research and not tied to particular applications, let alone deployments. However, if there is a direct path to any negative applications, the authors should point it out. For example, it is legitimate to point out that an improvement in the quality of generative models could be used to generate deepfakes for disinformation. On the other hand, it is not needed to point out that a generic algorithm for optimizing neural networks could enable people to train models that generate Deepfakes faster.
        \item The authors should consider possible harms that could arise when the technology is being used as intended and functioning correctly, harms that could arise when the technology is being used as intended but gives incorrect results, and harms following from (intentional or unintentional) misuse of the technology.
        \item If there are negative societal impacts, the authors could also discuss possible mitigation strategies (\eg gated release of models, providing defenses in addition to attacks, mechanisms for monitoring misuse, mechanisms to monitor how a system learns from feedback over time, improving the efficiency and accessibility of ML).
    \end{itemize}
    
\item {\bf Safeguards}
    \item[] Question: Does the paper describe safeguards that have been put in place for responsible release of data or models that have a high risk for misuse (\eg pretrained language models, image generators, or scraped datasets)?
    \item[] Answer: \answerNA{} %
    \item[] Justification: While we do scrape text from the internet, the datasets do not contain images, and are less extensive focusing on a few thousand Wikipedia articles covering a wide range of species in natural world. Thus we believe it poses low safety risks such as those associated with unsafe images or text. 
    \begin{itemize}
        \item The answer NA means that the paper poses no such risks.
        \item Released models that have a high risk for misuse or dual-use should be released with necessary safeguards to allow for controlled use of the model, for example by requiring that users adhere to usage guidelines or restrictions to access the model or implementing safety filters. 
        \item Datasets that have been scraped from the Internet could pose safety risks. The authors should describe how they avoided releasing unsafe images.
        \item We recognize that providing effective safeguards is challenging, and many papers do not require this, but we encourage authors to take this into account and make a best faith effort.
    \end{itemize}
\item {\bf Licenses for existing assets}
    \item[] Question: Are the creators or original owners of assets (\eg code, data, models), used in the paper, properly credited and are the license and terms of use explicitly mentioned and properly respected?
    \item[] Answer: \answerYes{} %
    \item[] Justification: Our benchmark builds on data from iNaturalist and Wikipedia. The iNaturalist data is already publicly available from the original SINR paper, as we noted in our paper. Meanwhile, Wikipedia grants permission to copy, distribute and/or modify Wikipedia's text under the terms of the Creative Commons Attribution-Share Alike 4.0 International License. We will include these details in the public release of our benchmark.
      \item[] Guidelines:
    \begin{itemize}
        \item The answer NA means that the paper does not use existing assets.
        \item The authors should cite the original paper that produced the code package or dataset.
        \item The authors should state which version of the asset is used and, if possible, include a URL.
        \item The name of the license (\eg CC-BY 4.0) should be included for each asset.
        \item For scraped data from a particular source (\eg website), the copyright and terms of service of that source should be provided.
        \item If assets are released, the license, copyright information, and terms of use in the package should be provided. For popular datasets, \url{paperswithcode.com/datasets} has curated licenses for some datasets. Their licensing guide can help determine the license of a dataset.
        \item For existing datasets that are re-packaged, both the original license and the license of the derived asset (if it has changed) should be provided.
        \item If this information is not available online, the authors are encouraged to reach out to the asset's creators.
    \end{itemize}

\item {\bf New Assets}
    \item[] Question: Are new assets introduced in the paper well documented and is the documentation provided alongside the assets?
    \item[] Answer: \answerYes{} %
    \item[] Justification: Documentation for the datasets and models is provided in the github repository.

\item {\bf Crowdsourcing and Research with Human Subjects}
    \item[] Question: For crowdsourcing experiments and research with human subjects, does the paper include the full text of instructions given to participants and screenshots, if applicable, as well as details about compensation (if any)? 
    \item[] Answer: \answerNA{} %
    \item[] Justification: The paper does not involve crowdsourcing nor research with human subjects. 
\item {\bf Institutional Review Board (IRB) Approvals or Equivalent for Research with Human Subjects}
    \item[] Question: Does the paper describe potential risks incurred by study participants, whether such risks were disclosed to the subjects, and whether Institutional Review Board (IRB) approvals (or an equivalent approval/review based on the requirements of your country or institution) were obtained?
    \item[] Answer: \answerNA{} %
    \item[] Justification: The paper does not involve crowdsourcing nor research with human subjects. 
\end{enumerate}

\end{document}